%% file: ms.tex
\documentclass[iop]{emulateapj}
\slugcomment{{\sc Submitted to ApJ: 26 June 2018}}
\usepackage{amssymb,amsmath,bm}
\usepackage{graphicx}
\usepackage[english]{babel}
\usepackage{natbib}
\usepackage{adjustbox}
\usepackage{tabularx}
\usepackage[section]{placeins}
\usepackage[colorlinks=true, linkcolor=blue,urlcolor=blue,citecolor=blue]{hyperref}

\begin{document}
\title{Merging Cluster Collaboration: A Panchromatic Atlas of Radio Relic Mergers} 
\shorttitle{Atlas of Radio Relic Mergers}

\author{
N. Golovich$^{\ast}$\altaffilmark{1,2},
W. A. Dawson\altaffilmark{1},
D. M. Wittman\altaffilmark{2,3}\\
R. J. van Weeren\altaffilmark{4,5},
F. Andrade-Santos\altaffilmark{4},
M. J. Jee\altaffilmark{2,6},
B. Benson\altaffilmark{2},
F. de Gasperin\altaffilmark{5,7},
T. Venturi\altaffilmark{8},
A. Bonafede\altaffilmark{8,9},
D. Sobral\altaffilmark{5,10},
G. A. Ogrean\altaffilmark{11,12},
B. C. Lemaux\altaffilmark{2},
M. Brada\v{c}\altaffilmark{2},
M. Br\"{u}ggen\altaffilmark{7},
A. Peter\altaffilmark{13,14,15}
}
\altaffiltext{1}{Lawrence Livermore National Laboratory, 7000 East Avenue, Livermore, CA 94550, USA}
\altaffiltext{2}{Department of Physics, University of California, One Shields Avenue, Davis, CA 95616, USA}
\altaffiltext{3}{Instituto de Astrof\'{\i}sica e Ci\^{e}ncias do Espa\c{c}o, Universidade de Lisboa, Lisbon, Portugal}
\altaffiltext{4}{Harvard-Smithsonian Center for Astrophysics, 60 Garden Street, Cambridge, MA 02138, USA}
\altaffiltext{5}{Leiden Observatory, Leiden University, P.O. Box 9513, 2300 RA Leiden, the Netherlands}
\altaffiltext{6}{Department of Astronomy, Yonsei University, 50 Yonsei-ro, Seodaemun-gu, Seoul, South Korea}
\altaffiltext{7}{Hamburger Sternwarte, Universit\"at Hamburg, Gojenbergsweg 112, 21029 Hamburg, Germany}
\altaffiltext{9}{Dipartimento di Fisica e Astronomia, Universit\'a di Bologna, via P. Gobetti 93/2, 40129 Bologna, Italy}
\altaffiltext{8}{INAF Istituto di Radioastronomia, via P. Gobetti 101, 40129 Bologna, Italy}
\altaffiltext{10}{Department of Physics, Lancaster University, Lancaster, LA1 4YB, UK}
\altaffiltext{11}{Kavli Institute for Particle Astrophysics and Cosmology, Stanford University, 452 Lomita Mall, Stanford, CA 94305, USA}
\altaffiltext{12}{Hubble Fellow}
\altaffiltext{13}{Department of Astronomy, The Ohio State University, 140 W. 18th Avenue, Columbus, OH 43210, USA}
\altaffiltext{14}{Center for Cosmology and AstroParticle Physics, The Ohio State University, 191 W. Woodruff Avenue, Columbus, OH 43210, USA}
\altaffiltext{15}{Department of Physics, The Ohio State University, 191 W. Woodruff Avenue, Columbus, Ohio 43210, USA}

\email{$^{\ast}$golovich1@llnl.gov}

\shortauthors{Golovich et al.}

\label{firstpage}

\begin{abstract}
\citet{Golovich:2017b} presents an optical imaging and spectroscopic survey of 29 radio relic merging galaxy clusters. In this paper, we study this survey to identify substructure and quantify the dynamics of the mergers. Using a combined photometric and spectroscopic approach, we identify the minimum number of substructures in each system to describe the galaxy populations and estimate the line of sight velocity difference between likely merging subclusters. We find that the line-of-sight velocity components of the mergers are typically small compared with the maximum three dimensional relative velocity (usually $<1000$ km s$^{-1}$ and often consistent with zero). This suggests that the merger axes of these systems are generally in or near the plane of the sky matching findings in magneto-hydrodynamical simulations. In 28 of the 29 systems we identify substructures in the galaxy population aligned with the radio relic(s) and presumed associated merger induced shock. From this ensemble, we identify eight systems to include in a ``gold'' sample that is prime for further observation, modeling, and simulation study. Additional papers will present weak lensing mass maps and dynamical modeling for each merging system, ultimately leading to new insight into a wide range of astrophysical phenomena at some of the largest scales in the universe.
\end{abstract}



\defcitealias{Golovich:2017b}{Paper 1}

\section{Introduction}\label{sec:intro}

Galaxy clusters have long served as astrophysical laboratories for the study of dark matter (DM), the intra-cluster medium (ICM), and galaxies, including their interactions and evolution. In the past decade, galaxy cluster {\it mergers} have been recognized as particularly fruitful laboratories because of the immense gravitational potential energy released into astrophysical interactions ($\sim10^{57}$ erg). Cluster mergers have been used to study the nature of DM \citep{Markevitch04,Clowe06,Randall:2008} and matter/antimatter asymmetry \citep{steigman08} as well as longstanding astronomical questions such as the origin of cosmic rays \citep[e.g.,][]{Bell:1978,Giler:1980,Gabici:2003,Brunetti:2014} and of the red sequence; i.e, do mergers quench, stimulate, or have little effect on star formation and AGN activity \citep[e.g.,][]{Miller:2003,Poggianti:2004,Chung:2009,Stroe:2017,Mansheim:2017b, Sobral:2015}? Yet, most observed mergers are still poorly understood even in terms of basic dynamical properties such as time since first pericenter (which we will refer to simply as {\it age}), relative velocity at first pericenter, and viewing angle. In addition, mergers evolve over billions of years and must be pieced together from instantaneous snapshots of a variety of systems, with simulations providing a framework for fitting them together. 

The Merging Cluster Collaboration\footnote{\url{www.mergingclustercollaboration.org}} has undertaken a comprehensive study---combining panchromatic observations and modeling of an ensemble of merging systems, with the following goals:

\begin{itemize}
\item improved constraints on DM self interactions
\item improved modeling of shocks, particle acceleration, and turbulence in the ICM
\item temporal and spatial description of galaxy evolution of member galaxies
\end{itemize}

Each of these goals requires a dynamical model of the merger including geometry, temporal information, and relative velocities of the merging subclusters and outward propagating shocks. Such modeling necessitates observations including optical photometry and spectroscopy to map the mass and galaxy distributions; i.e., the non-collisional components. Presented in \citet[][hereafter Paper 1]{Golovich:2017b}, these observations are more valuable when augmented with radio and X-ray data, which offer complimentary information regarding the ICM, which imprints the merger history for more than a Gyr after core passage \citep[see e.g., the merging cluster El Gordo;][]{Menanteau:2012,Ng:2015}.

A basic understanding of major mergers begins with the composition of the subclusters, which is typically $\sim85\%$ DM, $\sim13\%$ gas, and $\sim2\%$ galaxies by mass \citep{Vikhlinin:2006, Sun:2009, Sanderson:2013, Gonzalez:2013}. As the subclusters approach and pass through pericenter the gas halos exchange momentum while the approximately non-collisional galaxies and DM continue well past pericenter. Thus, a separation arises between the gas, which remains closer to the center of mass, and the outbound galaxies and DM; mergers clearly exhibiting this separation are said to be {\it dissociative} \citep{Dawson11}. The dissociative phase of a merging system is eventually erased as the subclusters pass through subsequent core crossing phases and dynamical friction causes each component to relax and create a new, more massive cluster. In this paper we will demonstrate the utility of dissociative mergers as efficient astrophysical laboratories for the study of the constituent galaxies, DM halos, and intra-cluster gas. 

Dark matter interaction models are generating interest in the context of galaxy cores and low mass halos \citep{Tulin:2017}.  Merging clusters provide a complementary probe by serving as large DM colliders: any momentum exchange between the DM halos could result in those halos lagging the outbound galaxies. Such lags will be visible only to the extent that the merger is transverse; the classic transverse merger, called the Bullet Cluster \citep{Markevitch04,Clowe06}, provided direct evidence for the existence of DM and has yielded the tightest upper limit on DM self-interaction \citep[$\sigma_{DM}<0.7$ cm$^2$ g$^{-1}$]{Randall:2008} to date based on the DM collider concept \citep[see, however,][who argue that the Bullet Cluster is not as constraining as previously thought]{Robertson:2017}. \citet{Kim:2017} also showed that the orientation of the lag depends on the merger phase, the size of the lags are always small under hard-sphere-type scatter, and he lags are transient. Ensemble analyses have attempted to push the dark matter cross section limit lower \citep{Harvey:2015}; however, these results were based on single band imaging, and it was shown that the multi-band studies throughout the literature substantially relax this constraint \citep{Wittman:2017}. Furthermore, the \citet{Harvey:2015} analysis centers on a different scattering prescription than the large-angle scattering studies of \citet{Randall:2008, Robertson:2017, Kim:2017}. The \citet{Harvey:2015} modeling relies on a scattering that has no velocity dependence but highly constrained angular dependence. Alternative signals may be apparent in a skewed DM profile with a heavier tail toward the center of mass or even in an induced wobbling of the brightest cluster galaxy (BCG) position long after the merger commences \citep{Kim:2017, Harvey:2017}. We will discuss merging clusters as DM probes in more detail in \S\ref{sec:discussion}.

Member galaxies are largely non-collisional, but their trajectories undergo an impulsive period of acceleration as the subclusters overlap and the gravitational potential deepens. Because of this fact, galaxy velocity dispersion is notoriously biased as a mass-estimator for a few hundred million years after pericenter \citep[e.g.,][]{Pinkney:1996}, but this effect is strongest for mergers observed along the line of sight and near core-passage \citep{Takizawa:2010}. Galaxy spectroscopy is vital for modeling the dynamics of the merger as the galaxies serve as tracer particles for the underlying subcluster halo; however, these observations only offer insight into the line of sight motion. Within individual galaxies, there is still much uncertainty with regard to the change in star formation in merging clusters. The genesis of the red sequence of early type galaxies has been well established and studied in relation to redshift, environment, galaxy density, and merger history \citep{Gunn:1972, Dressler:1980, Aragon:1993,Kodama:1997,Stanford:1998, Gladders:1998}, but a direct understanding of the role of major mergers in this evolution has eluded astronomers to date. In particular, do changes occur in step with major mergers, or is it a continuous transformation more associated with the cluster environment on longer time scales? In this framework, comparison studies of emission lines upstream and downstream of merger shocks have yielded conflicting results with turbulence in the ICM possibly playing a role \citep[e.g.,][]{Stroe:2014a}.

Finally, due to pressure forces, the ICM undergoes immense change during a merger. Many merging clusters have been identified as such based on their projected two-dimensional X-ray surface brightness profiles, which often display bulk disturbances such as cold fronts as well as cluster scale supersonic shocks \citep[e.g.,][]{markevitch05, Russell:2010}. In systems with a merging cool-core remnant, iconic images such as that of the Bullet Cluster show clearly merging ``bullets'' with trailing wake features highlighting the stripped gas due to ram-pressure. Radio observations of merging clusters have shown that the underlying shocks in the ICM are often coincident with diffuse radio emission due to synchrotron radiation of electrons in the ICM as they interact with the shock-compressed magnetic fields in the cluster. Active research is underway in search of explanations of apparent discrepancies in the radio and X-ray inferences of particle acceleration at cluster shock interfaces \citep[e.g.,][]{Brunetti:2014, vanWeeren:2017}. 

A global understanding of the interactions of each component throughout the full timescale of the interaction requires intensive computation. The non-collisional DM and galaxies are often studied through gravity-only N-body simulations. Such simulations have been utilized from sub-galactic (kpc) to cosmological (Gpc) scales to study structure formation, halo mass functions, and DM density profiles with both cold dark matter (CDM) and various self-interacting dark matter (SIDM) prescriptions \citep[e.g.,][]{rocha2012,DarkSky}. Meanwhile, studies of galaxy evolution, active galactic nuclei (AGN), particle acceleration, and other baryonic physics require full magneto-hydro dynamics (MHD). These simulations have been carried out across a range of scales including cluster and cosmological scales \citep[e.g.,][]{Skillman:2013,Vogelsberger:2014, Vazza:2016, Crain:2015, Barnes:2017}. Simulations have progressed greatly as large-scale computation has become more accessible; however, they are hampered in terms of simulating realistic cluster mergers due to the vast parameter space of initial conditions relating to not only merger geometry but also subcluster ICM properties, formation histories, feedback, etc. 

In this paper we focus on observations and modeling of the galaxies and DM in 29 merging clusters by analyzing the optical photometry and spectroscopic of these 29 systems presented in \citetalias{Golovich:2017b}. Our results will substantially reduce the phase space of initial conditions for simulators as well as provide a dynamical understanding of each merging system including the number of merging subclusters, the relative line of sight motion, and the most likely merger scenario. We have developed a galaxy and redshift modeling technique through the analysis of several clusters \citep{Dawson:2015, Golovich:2016, vanWeeren:2017, Golovich:2017, Benson:2017}. The methods developed as successive clusters presented specific complications that further generalized our analysis technique. This paper is the culmination of this project and is organized as follows. In \S\ref{sec:sample}, we describe the rationale for our sample selection. We describe the ancillary X-ray and radio data utilized in this paper in \S\ref{sec:radio}. In \S\ref{sec:analysis}, we outline the data analysis methods applied in this paper. In \S\ref{sec:results_individual} we describe results for each cluster while in \S\ref{sec:results_sample} we summarize the findings of the sample as a whole. Finally, in \S\ref{sec:discussion} we discuss the implications of this initial look at the sample and offer our conclusions. Throughout, we emphasize the importance of a multi-wavelength study of merging galaxy clusters in order to gather insight into the merging process and to constrain physics therein. We assume a flat $\Lambda$CDM universe with $H_0 = 70\,\mathrm{km}\,\mathrm{s}^{ -1}\,\mathrm{Mpc}^{-1}$, $\Omega_M = 0.3$, and $\Omega_\Lambda = 0.7$. AB magnitudes are utilized throughout, and all distances are proper. 


\section{Radio Relic Sample: Goals and Selection}\label{sec:sample}

Each of the primary science goals addressed while observing, modeling, and simulating merging clusters of galaxies are time dependent. We desire a sample of mergers observed between first pericenter and first turnaround and with merger axis near the plane of the sky, so apparent offsets are maximized. Transverseness also facilitates our goals of (1) studying the effect of mergers on star formation and galaxy evolution, as galaxy membership is more easily assigned to a particular subcluster when the system is view transversely; and (2) studying shock physics and particle acceleration processes, as the shocks will be viewed mostly edge-on. This motivation in turn leads to the following selection considerations.

Radio relics are defined as peripheral, diffuse radio features covering regions several hundred kpc in projection. Most radio relics are unassociated with optical counterparts; although, recent observations have directly imaged a connection from a radio relic and associated shock to a radio jet from a cluster member AGN \citep{vanWeeren:2017}. We have adopted the detection of radio relics as our trigger for transverseness---as opposed to projected galaxy populations or detection of sharp X-ray shock features--- because radio data support further dynamical inferences. First, the position of the relic is closely linked to the time since pericenter because the shock is launched around the time of pericenter, at about the same speed as the corresponding subcluster in the center-of-mass frame \citep{springel2007}. The shock speed can vary somewhat thereafter depending on the details of the gas profile, but this argument still provides a valuable constraint on the time since pericenter as demonstrated by \citet{Ng:2015} in the case of El Gordo. Second, highly polarized relic emission provides strong evidence that the relic is viewed edge-on \citep{ensslin1998} which confines the merger axis to near the plane of the sky. In addition to the analytical models of \citet{ensslin1998}, the simulations of \citet{Skillman:2013} clearly show this effect. Low polarization, however, could result from disorganized magnetic fields even if the merger is transverse, so the constraint is one-sided. \citet{Ng:2015, Golovich:2016, Golovich:2017} also demonstrated the value of this constraint on models of El Gordo, MACS J1149.5+2223 (hereafter MACSJ1149), and ZwCl 0008.8+5215 (hereafter ZwCl0008), respectively. 

The main potential drawback of this selection is that it will disfavor the very youngest post-pericenter systems, which have not had time to generate radio relics. At the same time, this selection has the advantage of guaranteeing against the selection of pre-pericenter systems since the presence of a radio relic indicates a shockwave traveling in the ICM, which likely indicates a dissociative merger. 

The dynamical parameters allow a degeneracy between outbound and returning scenarios. Even if the viewing angle is known, the velocities and projected separations do not distinguish between these two scenarios, which may have markedly differing ages. The continued outward progression of the radio relic breaks this degeneracy, so the constraint from the relic position is valuable in a different way than the constraint from the polarization. In principle the ram-pressure slingshot \citep[see e.g. Abell 168:][]{Hallman:2004} can also distinguish between these two scenarios---gas beyond the DM suggests that the DM has already turned around---but this may be useful only in systems whose cool-core remnant is sufficiently robust to the extreme forces imparted upon it. Thus, the relic position constraints are the more practicable way to break the degeneracy further reinforcing the value of radio selection.

A further advantage of radio selection is that it seems to select for massive systems: masses of relic systems estimated with weak gravitational lensing include El Gordo at $\sim 3\times 10^{15} M_\odot$ \citep{Jee:2014}, CIZAJ2242 at $\sim 2\times 10^{15} M_\odot$ \citep{Jee:2014} as well as 1RXSJ0603 (hereafter 1RXSJ0604) and ZwCl0008 at $\sim 1\times 10^{15} M_\odot$ \citep{Jee:2015, Golovich:2017}. The high DM column density of these systems potentially makes them more sensitive to DM interactions. Furthermore, there is a positive relationship between radio relic power and cluster mass \citep{deGasperin:2014}.

The construction of a sample of merging clusters that satisfies these motivations is described in detail in \citetalias{Golovich:2017b}. An initial analysis of spectroscopic redshift distributions for each cluster was also presented in which the systems were well fit, with remarkable consistency, by a single Gaussian distribution, despite the clear on-going merger activity. This suggests that radio relic selection succeeded in identifying mergers occurring transverse to the line of sight. For further details, we refer readers to \citetalias{Golovich:2017b}. The 29 systems are listed in Table \ref{tab:sample}. 

\begin{table*}
\centering
\caption{The Merging Cluster Collaboration radio-selected sample.}
\input{table1.tex}
\label{tab:sample}

\end{table*}


\section{Radio and X-rays}\label{sec:radio}

\subsection{Radio Data}

The primary selection function for this ensemble of merging clusters is the presence of confirmed radio relics. The radio data utilized by the literature in order to confirm these features is quite diverse, and in this section we summarize this aspect of our ensemble. We also detail the archival X-ray data and discuss the additional advantages that X-ray surface brightness maps and radio relics offer when interpreting merger scenarios. 

In Table \ref{tab:radio}, we detail the radio and X-ray data that will be presented in this paper. The radio data are presented in the form of linearly spaced contours on many figures that follow. The wide field radio images were cut down to only display the radio relics using the {\tt ftcopy} function from ftools \citep{FTOOLS}. All data were shared generously by the authors of the various studies listed in Table \ref{tab:radio}. Readers are referred to those studies for details regarding observations, data reduction, and analyses.

\begin{table*}
\centering
\caption{Ancillary radio and X-ray data utilized in this paper}
\input{table5.tex}
\label{tab:radio}

\end{table*}

The spectral index of radio relic emission across frequencies from tens of MHz to several GHz can be used to estimate the Mach number of the underlying shock:
\begin{equation}
\mathcal{M} = v_{shock} / c_{s} = \sqrt{\frac{2\alpha_{inj}+3}{2\alpha_{inj}-1}}
\end{equation}
where $c_{s}$ is the sound speed in the upstream medium and $\alpha_{inj}$ is the injection spectral index. Mach numbers are frequently measured to be $\mathcal{M}\sim2-4$ assuming diffusive shock acceleration \citep[DSA:][]{DSA}. However, the correspondence between the Mach number and the spectral index is complicated by pre-accelerated particles, inclination angle of the relic, non-uniform magnetic fields or complex shock structures. 

The radio data presented here by no means exhausts the utility of radio analyses of galaxy clusters in terms of constraining the dynamics. As mentioned in \S\ref{sec:intro}, the polarization may be used to constrain the viewing angle of the merger axis, and the spectral steepening is a clear demonstration of the motion of the subclusters and shocks through the ICM. Radio observations of galaxy cluster mergers may well result in observational understanding of cosmic ray generation, a deeper understanding of shock propagation, turbulence, microphysics within the Mpc scale ICM in terms of plasma instabilities, as well as a better understanding of the timescale of radio plasma from AGN and the interplay between these galaxies and the variety of observed radio relic morphologies. 

A second form of diffuse radio emission often present in merging clusters are radio halos \citep{Giovannini:1993, Burns:1995,Feretti:1997, Giovannini:1999}. These tend to be well-aligned by the X-ray emission of a merging cluster and are thought to be associated with regions of turbulence in the ICM. Like radio relics, radio halos have no apparent associated galaxy. Radio halos tend to be large ($\sim$ 1 Mpc), have low surface brightness, and steep radio spectra. Several of the clusters studied in this paper have radio halos, and while they are an additional factor in determining that those systems are merging galaxy clusters, we have not factored these features into our modeling explicitly. We do include studies of these features in our literature reviews of individual clusters and make note of the features where present. 

\subsection{X-ray Data} 

The X-ray images listed in Table \ref{tab:radio} represent the best publicly available data for each cluster. Chandra images were downloaded from the archive and reduced with the {\tt chav} package following the process described in \citet{Vikhlinin2005} using CALDB 4.6.5. The calibration includes the application of gain maps to calibrate photon energies, filtering of counts with ASCA grade 1, 5, or 7 and from bad pixels, and a correction for the position-dependent charge transfer inefficiency. Periods with count rates a factor of 1.2 above the mean count rate in the 6--12 keV band were also removed. Standard blank sky files were used for background subtraction. The final exposure corrected image was made in the 0.5--2.0~keV band using a pixel binning of a factor of four.

For the XMM images, the data were reduced using version 16.0 of the Extended Source Analysis Software (ESAS) pipeline following the example reduction of Abell 1795 in the ESAS cookbook\footnote{\url{https://heasarc.gsfc.nasa.gov/docs/xmm/esas/cookbook/xmm-esas.html}}.  In particular, raw event files were created from the Observation Data Files (ODF) using the {\tt emchain} and {\tt epchain} routines. The data were filtered of flaring from soft protons using the {\tt mos-filter} and {\tt pn-filter} routines. These apply a 1.5$\sigma$ cutoff from a Gaussian fitted to the light curve. Point sources above a flux threshold of $3\times10^{-15}$ erg cm$^{-2}$s$^{-1}$ were masked out using the {\tt cheese} routine. Spectra and images were extracted using {\tt mos-spectra} and {\tt pn-spectra}. The data were then background subtracted using the XMM-ESAS CalDB files\footnote{\url{ftp://xmm.esac.esa.int/pub/ccf/constituents/extras/esas caldb/}} and the {\tt mos-back} and  {\tt pn-back} routines. CCD $\#$6 became unoperational in 2005, and is automatically excluded by XMM-ESAS for relevant observations.  

For ZwCl 1447, no XMM-Newton or Chandra data are available, so the cluster's emission detected by the ROSAT All-sky Survey \citep{RASS} is presented. ZwCl 1447 appeared in exposure 931236. The presented image is background subtracted and exposure corrected; however, the number of counts is much lower for this image compared with the rest of the sample. In this paper, these X-ray images are presented as X-ray surface brightness maps, which help in the inference of the merger scenarios, and in a few cases clear up which of several subclusters have merged and which have not. We discuss these more in \S\ref{subsec:xray} and \S\ref{sec:results_individual}.

X-ray data analysis is similarly rich as radio data analysis. In the context of merging clusters, bulk motion of the ICM such as cold fronts and cool-core survival is studied. Furthermore, in sufficiently deep Chandra exposures, the data are spatially sensitive enough to directly measure the emission profile near radio relics and measure shock properties.  XMM-Newton has less spatial resolution; however, its spectral capabilities allow for the detection of shocks in the form of temperature and density jumps between extracted regions on either side of a proposed shock. Shocks have been detected in several clusters \citep[e.g.,][]{Markevitch:2001b,Russell:2010}. Original interpretations of the shock speed from the Mach number suggested that the Bullet Cluster perhaps merged too fast for $\Lambda$CDM \citep[e.g.,][among others]{Markevitch:2001b,Hayashi:2006}. This controversy was resolved when it was noted that the inferred shock speed includes the relative motion of infalling gas \citep{springel2007}. 

However, shocks still spawn controversy when comparing inferred Mach numbers from X-ray and radio data. Several examples of discrepancy have been uncovered. For example, in the famous ``toothbrush'' relic in 1RXSJ0603, the radio analysis indicates a Mach number of $2.8^{+0.5}_{-0.3}$ \citep{vanWeeren:2016} versus $\mathcal{M}\lesssim1.5$ from the X-rays \citep{Ogrean:2013c}. If the X-ray measured Mach numbers are taken at face value, DSA seems unable to explain the observed radio relic brightnesses. Presently, the leading explanation in the literature suggests that DSA can explain radio relics only if the passing shocks are creating bright emission by reaccelerating a pre-accelerated semi-relativistic population of plasma. One such source of pre-accelerated plasma could very well be radio galaxies in the cluster. \citet{vanWeeren:2017} found direct evidence for such a connection in Abell 3411; however, detections of specific connections remain rare. Further evidence of this picture is emerging from the Low-Frequency Array (LOFAR), which has been used to show several low-frequency diffuse sources in A2034 \citep{Shimwell:2016}. The observation of a radio relic and AGN in A3411 presents a clear example of this phenomenon. Future radio surveys of merging clusters seem poised to uncover more examples. 


\section{Analysis Tools}\label{sec:analysis}

In this section, we will describe the steps of analysis starting with photometric and spectroscopic galaxy catalogs from our optical and multi-object spectroscopy observations discussed in \citetalias{Golovich:2017b}. The analysis has been tailored for our sample analysis but is largely based on the the analyses presented in \citet{Dawson:2015, Golovich:2016, Golovich:2017, vanWeeren:2017, Benson:2017}, which study five clusters from this sample: CIZAJ2242, MACSJ1149, ZwCl0008, A3411, and ZwCl2341, respectively. The following subsections describe individual analysis steps using 1RXSJ0603 or RXCJ1053 as informative examples for the figures that follow. The details for individual clusters vary and are discussed in the corresponding subsections of \S\ref{sec:results_individual}.

\subsection{Galaxy Density Maps}

\begin{figure}
\centering
\includegraphics[width=\columnwidth]{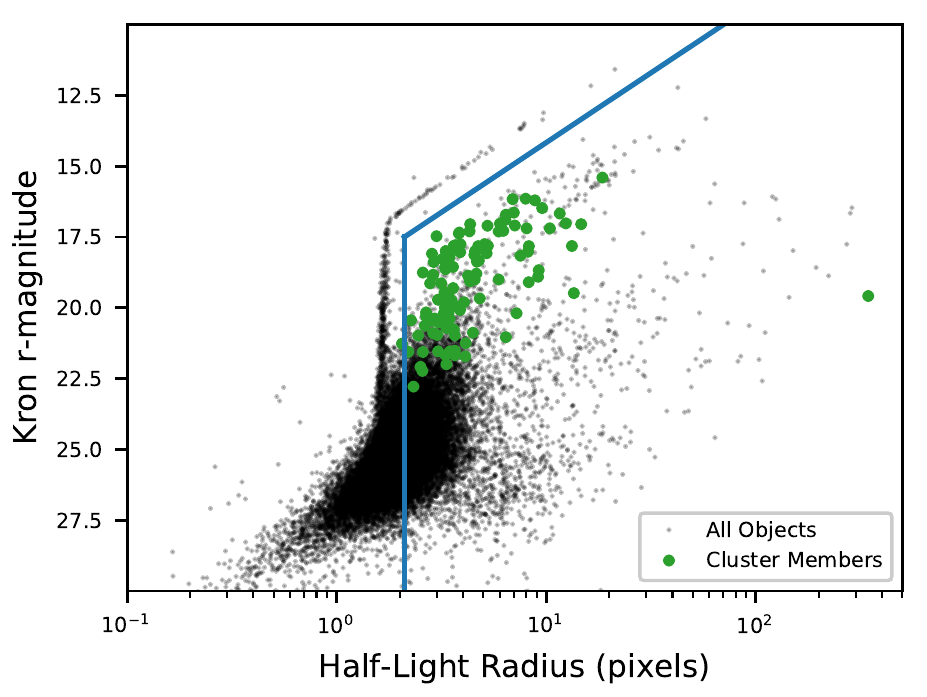}
\caption{Size--magnitude diagram for the photometric catalog of RXCJ1053 with overlaid spectroscopic matches. The stellar trace is easily eliminated to the left and above the delineating cut (blue line). Also eliminated are objects which are much fainter than the spectroscopic cluster members. Stars are defined as objects with half light radii less than 2.1 pixels up to $r\sim17.5$. Brighter stars saturate surrounding pixels, but the stellar trace is clearly separated by our cut.}
\label{fig:sep}

\end{figure}

We match objects from the spectroscopic and photometric catalogs using the Topcat \citep{Topcat} software utilizing the sky coordinate match with 1$\arcsec$\ tolerance. We define the color of the objects by selecting the two filters that most evenly straddle the 4000\,\AA\, break. Objects with large photometric errors ($>$0.5 magnitudes) are eliminated from the catalog. To separate stars and galaxies, we employ a size--magnitude selection. Stars follow a clear relation and are easily eliminated from galaxies. An example size--magnitude diagram is presented in Figure \ref{fig:sep}. 

Our spectroscopic survey is incomplete, so to select likely cluster members from the photometric catalog, we employ a red sequence selection by overlaying the spectroscopically confirmed cluster members in a color--magnitude diagram. The width of the red sequence depends on several factors including redshift, dynamical activity, early/late type galaxy population, etc., but to regularize the analysis, we restrict the red sequence width to 0.5 magnitudes along the color axis, and length to six magnitudes greater than the BCG, which we find sufficient to capture nearly all targeted red sequence galaxies. We eliminated contaminant stars and non-cluster member galaxies from the red sequence catalogs by cross checking the photometric catalog with the Subaru/SuprimeCam images and our spectroscopic catalog. In a few cases SExtractor objects were artificially bright due to proximity to stellar bleeds and diffraction spikes. These objects were eliminated from the red sequence catalogs as well. 

\begin{figure}
\centering
\includegraphics[width=\columnwidth]{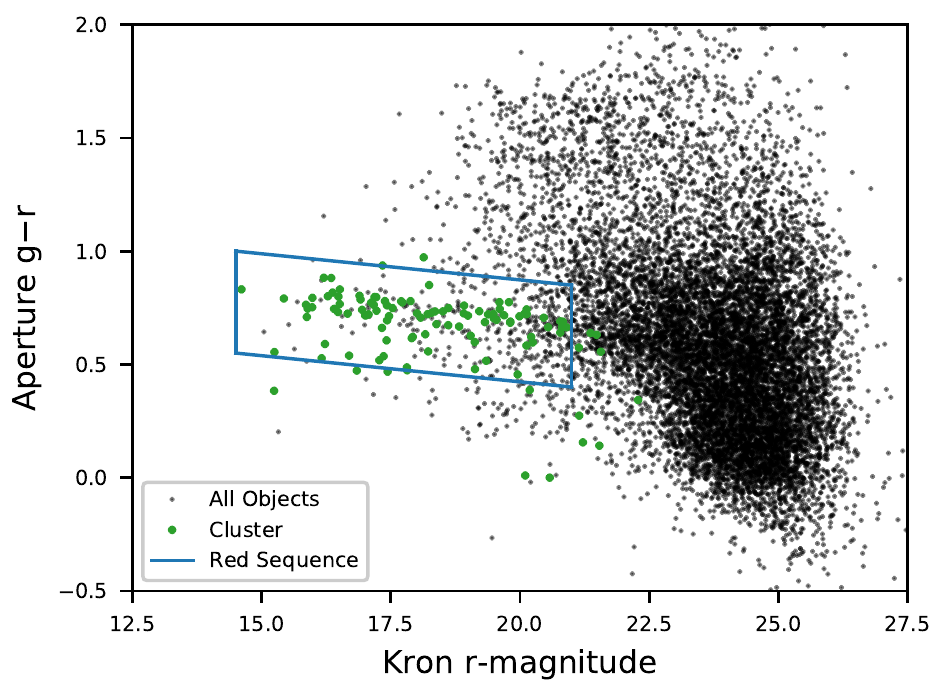}
\caption{Color--magnitude diagram for the photometric catalog of RXCJ1053 with overlaid spectroscopic matches. The red sequence selection box is shown in blue. Magnitudes are corrected for dust extinction, and all objects that passed the size--magnitude cut are plotted.}
\label{fig:redsequence}

\end{figure}

A clear and tight red sequence is visible in color--magnitude diagrams, such as the one for RXCJ1053 which is presented in Figure \ref{fig:redsequence}. For galaxies that were targeted for spectroscopic observations using a red sequence method, the spectroscopic points will largely fall within the red sequence region; whereas, for galaxies targeted using SDSS photometric redshifts, the blue cloud galaxies will be more completely sampled. In the case of RXCJ1053, a photometric redshift selection was implemented based on SDSS imaging and SDSS photometric redshifts, which is evident from the green points below the clear red sequence.

Typical red sequence populations range between 500 and 2500 galaxies depending on the richness, redshift, and depth of the optical image. We display the information via smoothed density maps weighted by luminosity in the reddest filter (typically r or i). We choose the smoothing bandwidth via a take one out, maximum likelihood, cross-validation, kernel density estimation (KDE) analysis. The KDE bandwidths range from 25$\arcsec$\ to 100$\arcsec$\ roughly proportional to the redshift. The optimal KDE bandwidths for each cluster are displayed on the DS-test plots described in \S \ref{subsec:DS}.

\subsection{Dressler-Schectman Test}\label{subsec:DS}

The Dressler-Schectman (DS)-test is performed by computing a statistic for local velocity information as compared to the global values \citep{Dressler:1988}. The statistic is given by:
\begin{equation}\label{eq:delta}
\delta^{2} = \frac{N_{local}}{\sigma^{2}}\big[\left(\bar{v}_{local}-\bar{v}\right)^{2} + \left(\sigma_{local} - \sigma\right)^{2}\big]
\end{equation}
where $N_{local}$ is the number of nearest neighbors (self-inclusive) to include when calculating $\bar{v}_{local}$, the local-average LOS velocity, and $\sigma_{local}$, the local velocity dispersion. We let $N_{local} \equiv \lceil\sqrt{N_{total}} \rceil$, where $N_{total}$ is the number of galaxies in the full spectroscopic catalog. This follows the best practice identified by \cite{Pinkney:1996}. Galaxies with larger $\delta$ values are highly correlated with their neighbors and different from the global parameters thus identifying local structure.

\begin{figure}
\centering
\includegraphics[width=\columnwidth]{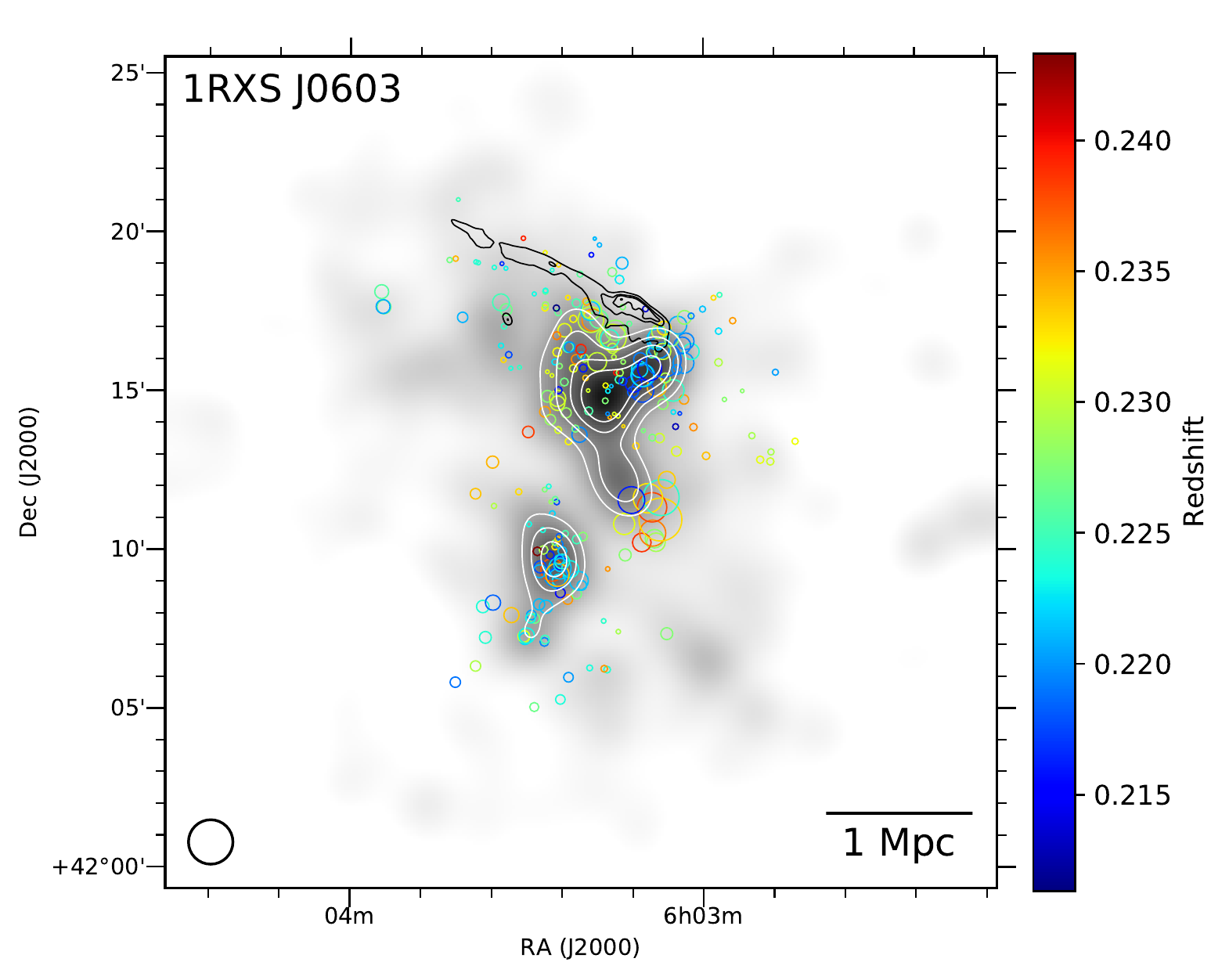}
\caption{DS-test for 1RXSJ0603. The circles each correspond to spectroscopically confirmed cluster members and are colored according to their redshift. The size of the circles is given by 10$^{\delta}$. The gray scale map corresponds to the red sequence luminosity density smoothed by the optimal KDE bandwidth, and the corresponding white contours start at 3$\sigma$ above the median pixel luminosity value and increase in units of $\sigma$, where $\sigma$ is the standard deviation of the pixel values within $R_{200}$. The smoothing scale is chosen with an optical KDE bandwidth estimation, and is displayed in the lower left. The black contours are linearly spaced and show the so-called ``Toothbrush'' radio relic based upon the 610 MHz WSRT image of \citet{vanWeeren:2012}. The field is $2R_{200}\times2R_{200}$ based on the global velocity dispersion and scaling relations \citep{Evrard:2008}.}
\label{fig:dstest}

\end{figure}

In Figure \ref{fig:dstest}, we plot the projected location of the galaxies in the spectroscopic catalog as circles with radii proportional to $10^{\delta}$ over the red sequence luminosity density distribution for 1RXSJ0603. The remaining 28 DS plots are presented in Figure A.1 in the Appendix. We singled out 1RXSJ0603 as an informative example to enlarge and show the details. We color-coded the circles in Figures \ref{fig:dstest} and A.1 by redshift. This helps us infer the source of inflated $\delta$; groups of large circles with a narrow band of redshift indicate an interloping substructure with a line of sight velocity difference, while clustered groups with a large velocity range indicate either massive subclusters or chance alignments of high and low velocity cluster members and thus an inflated local velocity dispersion.

Plotting the DS-test over the red sequence luminosity density distribution mitigates the main fail mode of the DS-test, which is insensitivity to equal mass subclusters distributed in projected space with similar redshifts. In this scenario, the test is effectively reduced to a two dimensional test, which is a strength of the red sequence luminosity for detecting substructure. This technique also provides a check against false subcluster detections in the spectroscopic and photometric data. The spectroscopic selection with DEIMOS causes some regions to be over-sampled relative to the cluster core since DEIMOS slits were placed on as many galaxies as possible. However, since the slits may not overlap along the dispersion axis, the cluster center can't be fully sampled. Thus, small yet correlated groups of galaxies may appear as a group of very large circles in the DS-test; meanwhile, bright substructure may be under-sampled spectroscopically. For this reason, we will only assign potential subclusters to areas that are high in photometric luminosity and were covered with the spectroscopic survey. However, the photometric red sequence selection means there is contamination, so false positives are possible in the red sequence luminosity map as well. Together, high red sequence luminosity density with corresponding high $\delta$ groups are constitute strong evidence for an associated subcluster, and this combination of evidence provides the basis for our subcluster priors. As a final step, placing the radio relic contours on these diagnostic plots immediately reveals the likely merger scenarios.

\subsection{Gaussian Mixture Models}\label{subsec:gmm} 

We seek to fit subcluster properties quantitatively by a clustering algorithm that assigns specific spectroscopically confirmed cluster members to the subcluster that most likely hosts them. Here we describe the subclustering method that we will implement on each cluster's spectroscopic data. We will utilize a three dimensional Gaussian mixture model (GMM) analysis. The basic aim is to simultaneously fit multiple Gaussians to the projected location and redshift distribution of a population of galaxies. Standard packages have been developed for GMM methods, but these codes often fail to fit physically motivated subclusters. It is important to consider the data we are attempting to fit. Galaxy spectroscopy within a cluster environment is a discrete set of data that serve as tracer particles for the dynamics of the cluster. Yet, galaxies make up a small fraction of the total mass, and thus, informative priors can be developed based on other observations that probe the DM halo and ICM of the clusters. Several of our clusters have detailed lensing studies in the literature, and all have dedicated X-ray observations that have mapped the X-ray surface brightness of the ICM. Additionally, each cluster has at least one radio relic that traces the underlying shocks in the ICM. With all this information, we developed a scheme to include informative priors on the location and scale of the component Gaussians. For these reasons, we have completed a detailed literature review for X-ray, radio, galaxy, and lensing analyses for each cluster. Priors are based upon results from the literature coupled with the DS-test and red sequence luminosity density as in Figure \ref{fig:dstest}.

When determining potential subclusters, we look for areas that are at least 3$\sigma$ above the median pixel value in the red sequence luminosity maps and also have spectroscopic coverage. We then run a Markov-Chain Monte Carlo (MCMC) code that fits the data to Gaussian mixture models that contain all possible combinations of the potential subclusters. We utilize the python package \emph{emcee} \citep{emcee} for the MCMC sampling. We do not run a model for each subcluster by itself instead running a single one halo model using the global Gaussian for the redshift information and an uninformative prior on the right ascension (RA) and declination (DEC) components. Thus if there are two potential subclusters we run two models,  (1) a single halo model with the redshift Gaussian from the global velocity information as a prior and (2) a two halo model with RA and DEC priors drawn from the red sequence luminosity density map and a conservative redshift prior based on the color of the DS circles in that potential subcluster. For a system with three potential subclusters we run five GMM models, and for a system with four potential subclusters, we run twelve. In general for a system with n potential subclusters we run N total models given by Equation \ref{eq:models}.

\begin{equation}\label{eq:models}
N=1+\sum_{k=2}^{n} \binom{n}{k}
\end{equation}

We summarize possible subclusters for all 29 systems in Table \ref{tab:subclusters}, and we detail each model run on the data for all 29 clusters in Table \ref{tab:mcmcgmm}. These tables will serve as a master list that will be referred to in each subsection of \S \ref{sec:results_individual}. The number of models for each cluster is determined by the number of potential subclusters listed for a given system in Table \ref{tab:subclusters} following the prescription of Equation \ref{eq:models}. 

\begin{table*}
\centering
\caption[MCMC-GMM Subcluster Priors]{Priors for potential subclusters as deemed such by the routine discussed in \S\ref{subsec:gmm} for 29 merging clusters. The first two systems are presented here. The remaining 27 systems can be found online in a machine readable table.}
\begin{tabular}{lllll}
Cluster	& Halo & X range (Mpc) & Y range (Mpc) & z range (1000 km s$^{-1}$) \\
\hline
1RXSJ0603 	& a  	& [$-0.58$,$-0.36$] 	& [$+0.32$,$+0.57$]	& [$-2.5$,$-0.50$]\\
		     	& b	& [$-0.24$,$-0.02$] 	& [$+0.10$,$+0.37$]	& [$-0.50$,$+0.50$]\\
			& c	& [$-0.67$,$-0.33$] 	& [$-0.78$,$-Ä0.14$]	& [$+0.75$,$+2.5$]\\
			& d	& [$+0.11$,$+0.33$] & [$-1.1$,$-0.75$]	& [$-2.0$,$-1.0$]\\
\hline
A115			& a	& [$-0.45$,$-0.23$] 	& [$+0.06$,$+0.38$]	& [$+0.05$,$+1.3$]\\
			& b	& [$-0.10$,$+0.13$] 	& [$-0.69$,$-Ä0.42$]	& [$-0.90$,$+0.30$]\\
			\\
\end{tabular}
\label{tab:subclusters}

\end{table*}

\begin{table*}
\centering
\caption[MCMC-GMM model results]{MCMC-GMM models run for each system in our ensemble. The first two systems are displayed for demonstration of the form of the table. The remaining 27 systems can be found online in a machine readable table.}
\begin{tabular}{llllll}
Cluster & Model & Halos included & N$_{back}$ & Acceptance & BIC-BIC$_{min}$ \\
\hline
1RXSJ0603	&	1	& -		& 18		& 0.160 	& 147\\
			&	2	& ab		& 38 		& 0.093 	&178\\
			&	3	& ac		& 66		& 0.078 	& 250\\
			&	4	& ad		& 87		& 0.121 	& 165\\
			&	5	& bc		& 28		& 0.086 	& 199\\
			&	6	& bd 		& 0		& 0.105 	& 8.73\\
			&	7	& cd		& 75		& 0.070 	& 219\\
			&	8	& abc	& 0 		& 0.072 	& 242\\
			&	9	& abd	& 0		& 0.084 	& 4.36\\
			&	10	& acd	& 8		& 0.083 	& 62.7\\
			&	11	& bcd	& 1		& 0.052	& 444\\
			&	12	& abcd	&14		& 0.074	& 0\\
\hline
A115			&	1	& - 	 	&  0		& 0.170	& 33.3 \\
			&	2	& ab		& 26		& 0.093	& 0 \\ 
			\\
			\\
			\\
			\\
\end{tabular}
\label{tab:mcmcgmm}

\end{table*}

The 3D Gaussian mixture models are defined by the means and covariance matrices. For the $i^{th}$ Gaussian in the model, these are given by: \begin{equation}
\bm{\mu}_i =  \begin{pmatrix}
    \mu_{\alpha_{i}} \\
    \mu_{\delta_{i}} \\
    \mu_{z_{i}}
\end{pmatrix}
\end{equation}
\begin{equation}
\bm{\Sigma}_i = \begin{pmatrix}
    \sigma_{\alpha_{i}}^{2} & \sigma_{\alpha_i{\text -}\delta_i}  & 0\\
    \sigma_{\alpha_i{\text -}\delta_i} & \sigma_{\alpha_{i}}^{2}  & 0\\
    0 & 0 &  \sigma_{z_{i}}^{2} 
\end{pmatrix}
\end{equation} where $\mu_\alpha$, $\mu_\delta$, and $\mu_z$ are the means of the RA, DEC, and z axes, respectively. $ \sigma_{\alpha_{i}}^{2}$, $\sigma_{\delta{i}}^{2}$, and $\sigma_{z_{i}}^{2}$ are the variances of the RA, DEC, and z axes, and $\sigma_{\alpha_i{\text -}\delta_i}$ is the covariance between the RA and DEC coordinates, which allows the projected Gaussians to be aligned at any angle on the sky. The $\sigma_{\alpha_i{\text -}z_i}$ and $\sigma_{\delta_i{\text -}z_i}$ covariance terms are zero because we do not expect appreciable rotation (compared to the velocity dispersion) within galaxy clusters \citep{Hwang:2007}.

Additionally, for each model, we include a background group, to account for the possibility of field galaxies. This background is modeled as a diagonal multivariate Gaussian with uninformative priors. 

We assume uniform priors on all other parameters. Specifically, we enforce the ratio of semi-minor to semi-major axes to be between 0.4 and 1 to avoid highly elliptical projections \citep{Schneider:2012}. We enforce the semi-major axis, as defined by the $\sigma$ of the projected Gaussian, to be between 0.2 and 1 Mpc to avoid overfitting or generating nonphysically large subclusters. Finally, we restrict the velocity dispersion of subclusters to be less than the global velocity dispersion of the spectroscopic population for each cluster.

\begin{figure}
\centering
\includegraphics[width=\columnwidth]{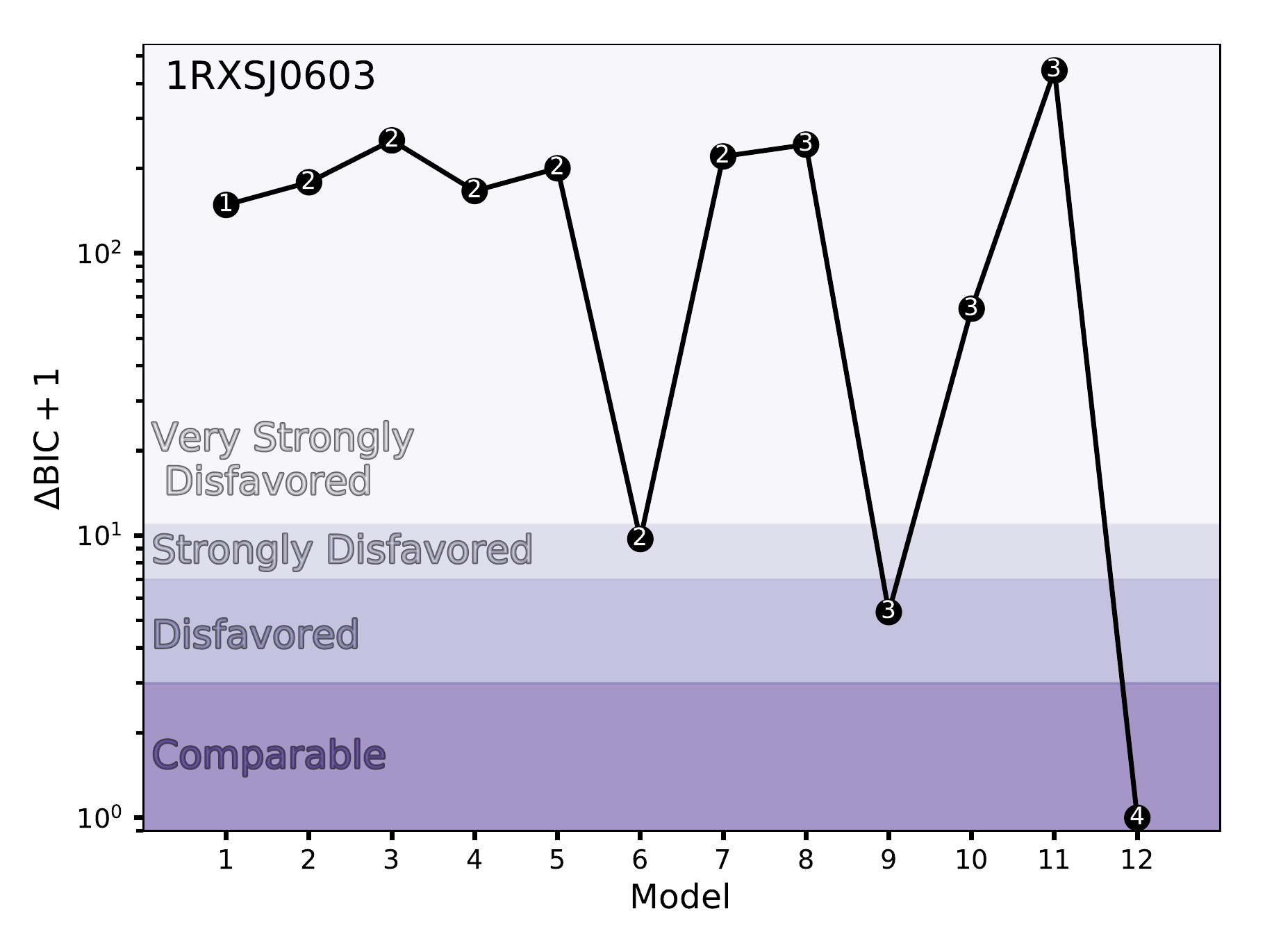}
\caption[$\Delta$BIC model selection for 1RXS J0603.3+4214]{Bayesian Information Criterion scores relative to the most economical model for twelve 3D (RA, DEC, z) Gaussian Mixture Models fit to the 242 spectroscopic cluster member galaxies of 1RXSJ0603. The numbers in the black points indicate the number of subclusters included in a given model. The shaded regions roughly denote how a given model compares with the lowest BIC score \citep{Kass:1995}. The most economical fit is a four component model (see Table \ref{tab:subclusters} and \ref{tab:mcmcgmm}). An analogous plot is not presented for each cluster; however, the $\Delta$BIC scores are presented in Table \ref{tab:mcmcgmm}.}
\label{fig:delta_bic}

\end{figure}

We run the MCMC with 400 walkers each taking 10,000 steps burning the first 5,000 steps to allow for convergence before posteriors are used to infer parameters. We checked that the number of steps is more than ten times the autocorrelation time to ensure convergence. For each model, we select the realization that results in the maximum likelihood fit. We then select the best model using the Bayesian Information Criterion (BIC):\begin{equation}
BIC=-2 \log \hat{L} + k \left(\log n - \log 2\pi\right)
\end{equation} where $\log \hat{L}$ is the log-likelihood function, $k=8i+7$ is the number of free parameters to be estimated for $i$ subclusters, and $n$ is the number of datapoints (galaxies in our case). The BIC value measures the likelihood of each model while penalizing the models with more subclusters to avoid rewarding over-fitting. A lower BIC score indicates the data are more economically described by the model. We summarize the results of this analysis for 1RXSJ0603 in Figure \ref{fig:delta_bic} in which we plot the difference in BIC scores relative to the lowest BIC score among the models as an informative example. The shaded regions indicate the relative model success given the BIC scores \citep{Kass:1995}. There are three models with starkly lower BIC scores than the other nine. Each of these contain the two most massive subclusters \citep{Jee:2015}; leaving these subclusters out results in poor models. The BIC scores for all models for each system are displayed Table \ref{tab:mcmcgmm}, organized by cluster.

We choose the BIC to quantify model success based on the results of Appendix A of \citet{Benson:2017}, which demonstrates that the BIC is less likely to overestimate the number of halos and less likely to artificially split halos on a suite of simulated clusters as compared to the corrected Akaike information criterion (AIC$_c$). The AIC$_c$ is essentially similar to the BIC with a slightly lesser penalty for extra model parameters. An important conclusion is that the BIC virtually never underestimates the number of halos indicating that our models will place hard lower limits on the number of substructures in these clusters. For more details, we refer readers to \citet{Benson:2017}. 

\begin{figure}
\centering
\includegraphics[width=\columnwidth]{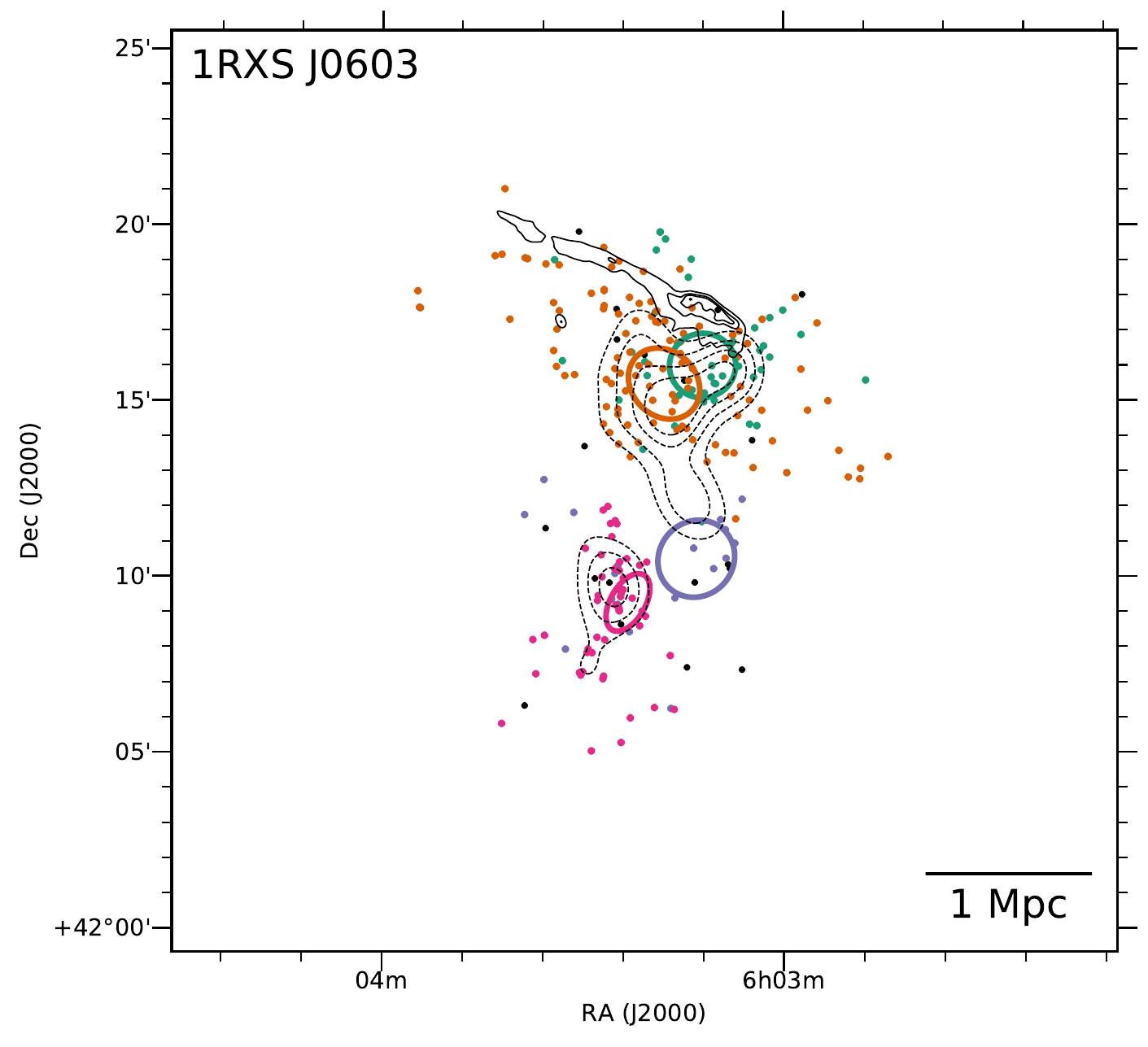}
\caption[Preferred 1RXS J0603.3+4214 subcluster membership]{Subcluster membership for the lowest BIC model (see Table \ref{tab:mcmcgmm}) for 1RXSJ0603. The dashed contours show the red sequence luminosity distribution starting at 3$\sigma$ and increasing in units of $\sigma$ where $\sigma$ is the standard deviation of the pixel value within $R_{200}$. The solid black contours show the confirmed radio relic in the 610 MHz GMRT radio image presented in \citet{vanWeeren:2012}. The colored ellipses show the most likely projected position and size of the corresponding Gaussians from the GMM. The field is $2R_{200}\times2R_{200}$ based on the global velocity dispersion and corresponding scaling relations \citep{Evrard:2008}.}
\label{fig:scatter}

\end{figure}

In Figure \ref{fig:scatter}, we plot a projected-space (RA and DEC) scatter plot for the preferred model of 1RXSJ0603. The the color-coded membership selection for each spectroscopic galaxy is assigned to the subcluster with the highest likelihood of hosting each galaxy based on the three dimensional Gaussian for each subcluster. The subcluster Gaussians are presented as projected ellipses centered on the most likely value with widths corresponding to the marginal 1-$\sigma$ confidence in the RA and DEC positions. The same galaxy luminosity contours and radio relic contours from Figure \ref{fig:dstest} are displayed showing the alignment of galaxy substructure and the radio relic. We display analogous figures for the remaining 28 systems in the appendix organized by cluster. 

\subsection{Redshift and Velocity Dispersion Estimation}\label{subsec:redshift_analysis}

With the galaxies separated into subclusters by our MCMC-GMM analysis, we implement the biweight statistic based on bootstrap samples of each subcluster's member galaxies and calculate the bias-corrected 68\% confidence limits for the redshift and velocity dispersion from the bootstrap sample (similar to the global redshift analysis in \citetalias{Golovich:2017b}). The redshift distributions of the subclusters for the lowest BIC model of 1RXSJ0603 are shown in Figure \ref{fig:subclusthist}. We present the analogous figures for the remaining 28 clusters in the appendix.

\begin{figure}
\centering
\includegraphics[width=\columnwidth]{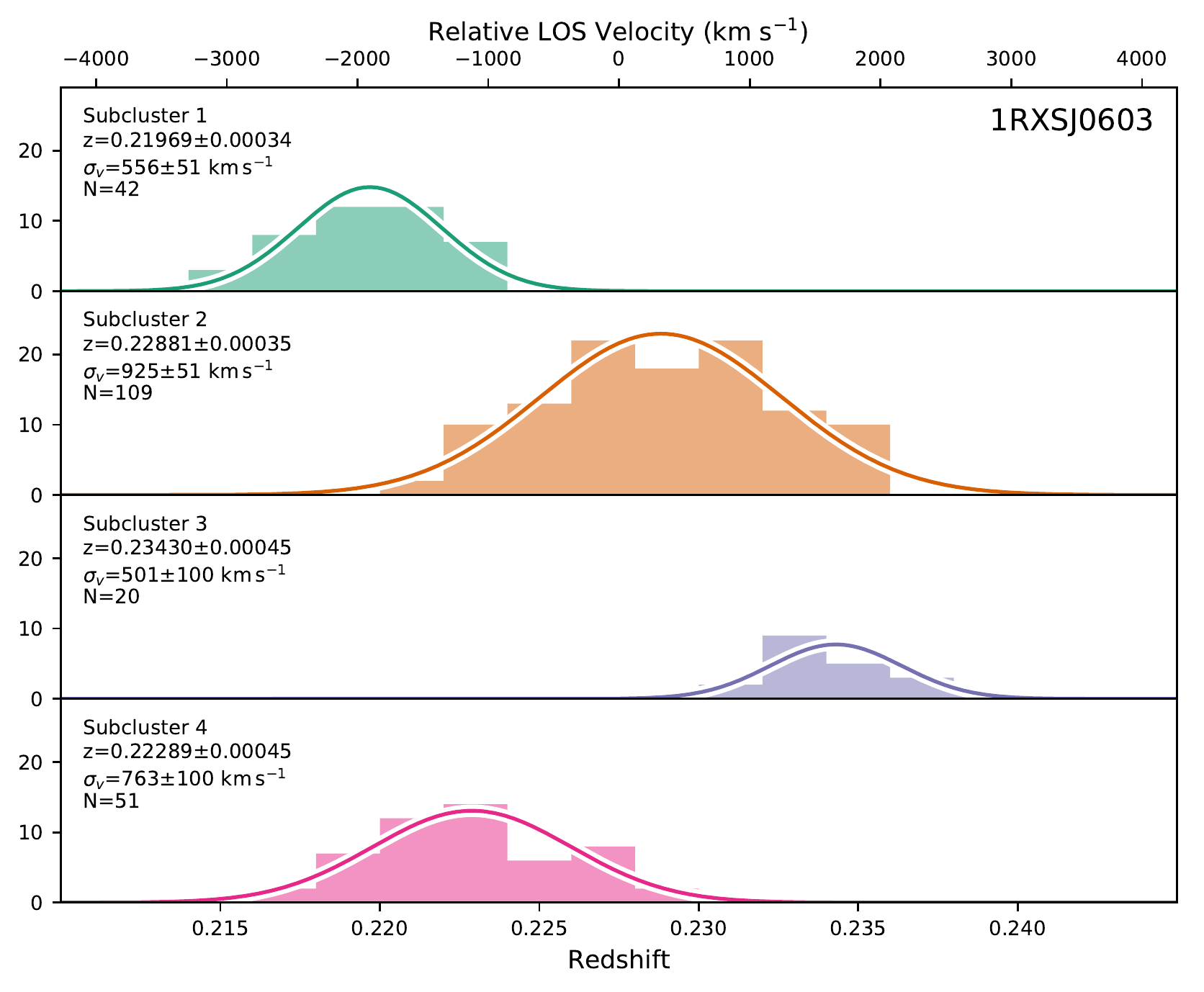}
\caption[1RXS J0603.3+4214 subcluster redshift distributions]{Subcluster redshift distributions for the lowest BIC model of 1RXSJ0603. The four subclusters are color coded according to the scatter points in Figure \ref{fig:scatter}. The overlaid Gaussians are centered on the biweight redshift and have widths from the bias-corrected 68\% confidence limits \citep{Beers1990}. The velocity scale at the top of the figure is in the cluster frame and is centered on $z=0.2275$.}
\label{fig:subclusthist}

\end{figure}

We use the biweight analysis to mitigate outliers from affecting the estimates of the subcluster redshifts and velocity dispersions, but this does not protect against the well known fact that mergers tend to bias high the velocity dispersion based mass estimates \citep{Takizawa:2010, Saro:2013}. The results of \citet{Dawson:2015, Ng:2015, Golovich:2016, Golovich:2017} suggest that many of the mergers in the sample are observed at least several hundred Myr after pericenter.

\subsection{X-ray Surface Brightness Maps}\label{subsec:xray}

\begin{figure}
\centering
\includegraphics[width=\columnwidth]{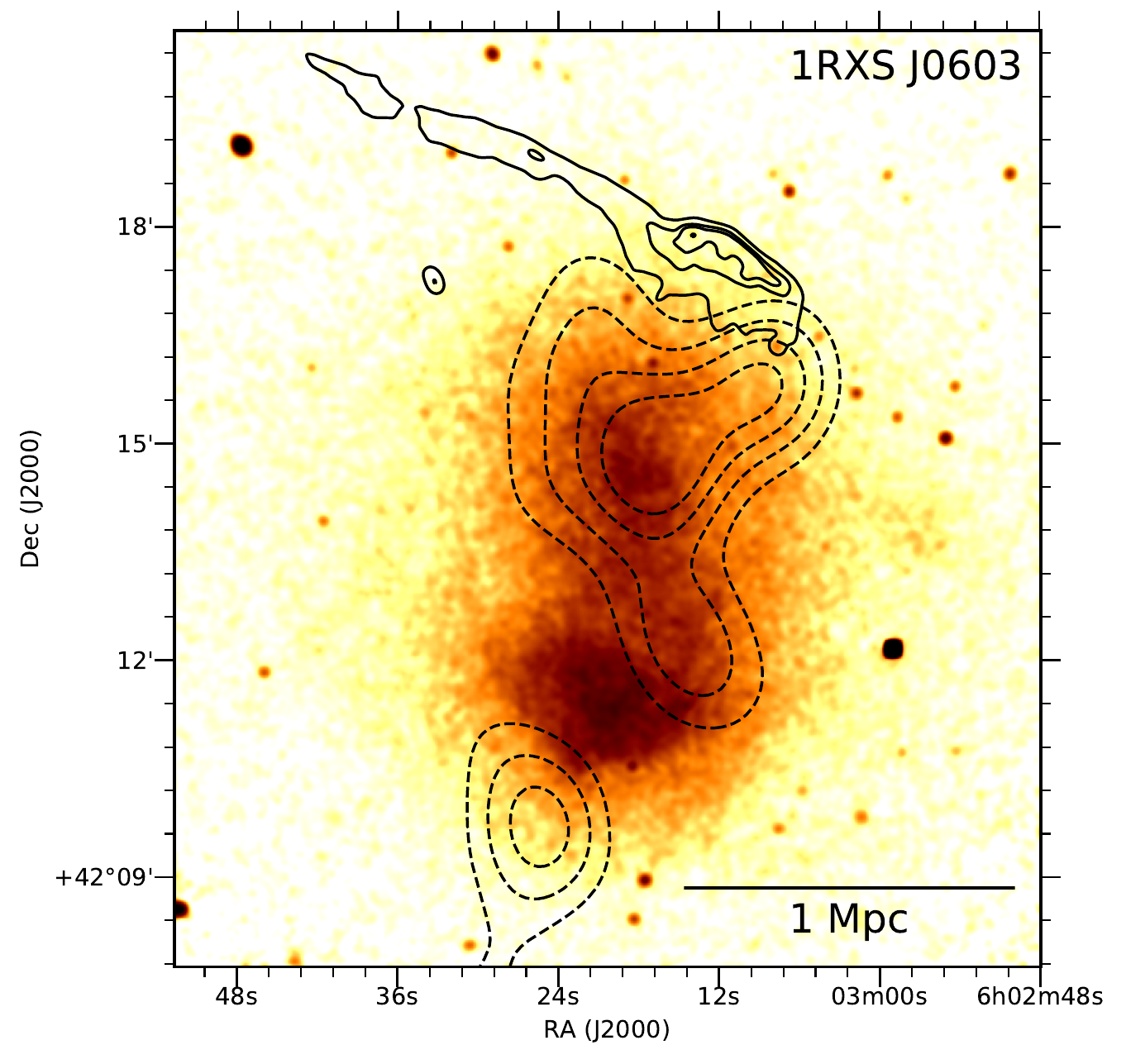}
\caption[1RXS J0603.3+4214 Chandra X-ray surface brightness profile]{X-ray surface brightness map for 1RXSJ0603 based on a 250 ks of Chandra ACIS exposure. The map is created in the 0.5-2.0 keV band and is exposure corrected. GMRT 610 MHz solid black contours and red sequence luminosity dashed black contours are presented.}
\label{fig:xray}
\end{figure}

X-ray surface brightness maps are invaluable for determining the merger scenario. The gas is strongly affected by the merger due to ram pressure, and thus, the merger geometry is imprinted in the gas in ways invisible to optical and lensing studies. In Figure \ref{fig:xray} shows the X-ray, radio, and optical features of 1RXSJ0603, and we present the remaining 28 systems in the appendix.

\subsection{BCG Thumbnails and Merger Scenarios}

\begin{figure}
\centering
\includegraphics[width=\columnwidth]{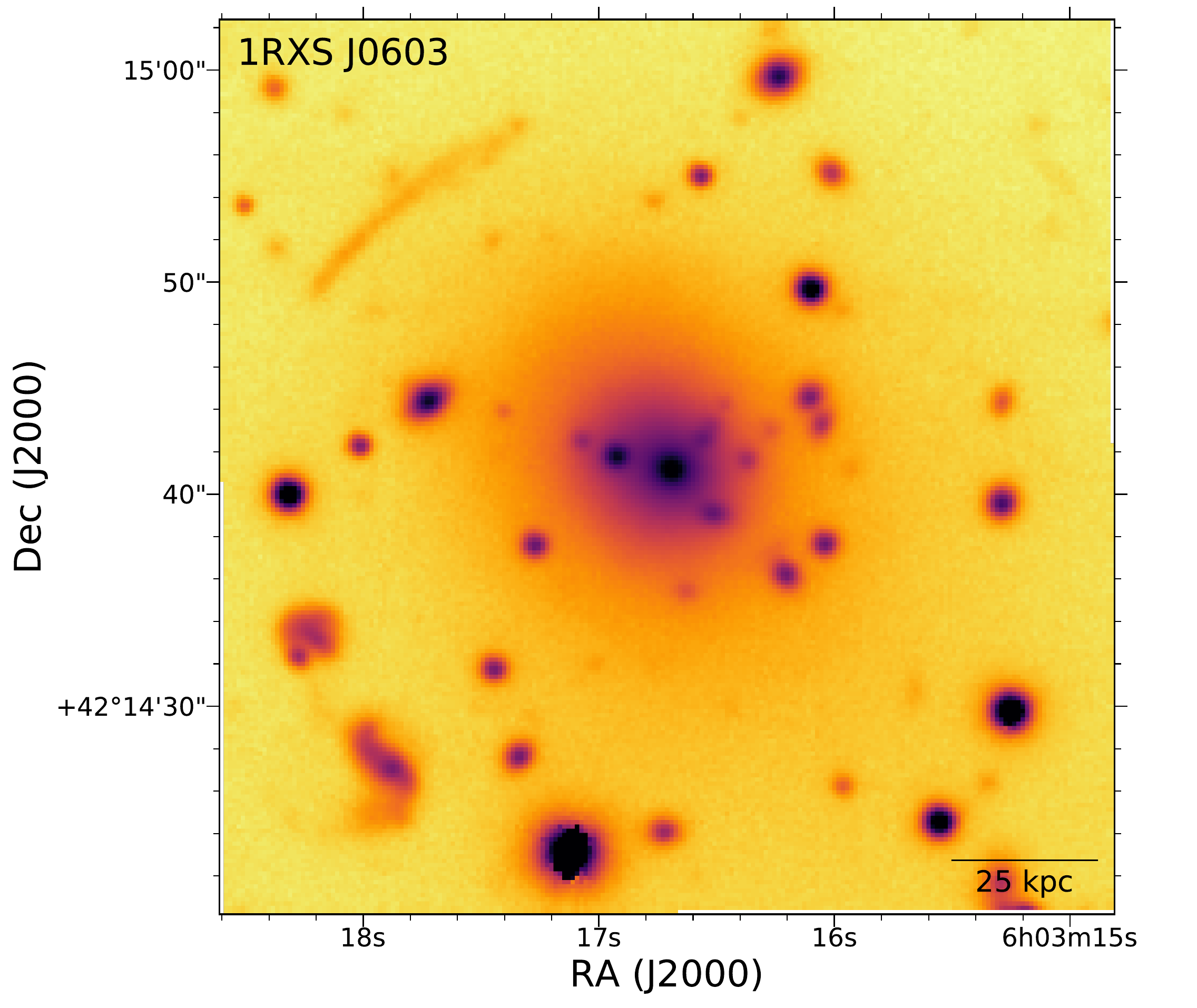}
\caption{Subaru/SuprimeCam r-band cutout of the inner 150$\times$150 kpc of the BCG of 1RXSJ0603. A strongly lensed arc is apparent along with several light peaks within the inner region of the overall light profile. This figure, along with the 28 analogous figures in the respective following subsections will be analyzed in detail in a forthcoming paper.}
\label{fig:bcg}
\end{figure}

\begin{figure}
\includegraphics[width=\columnwidth]{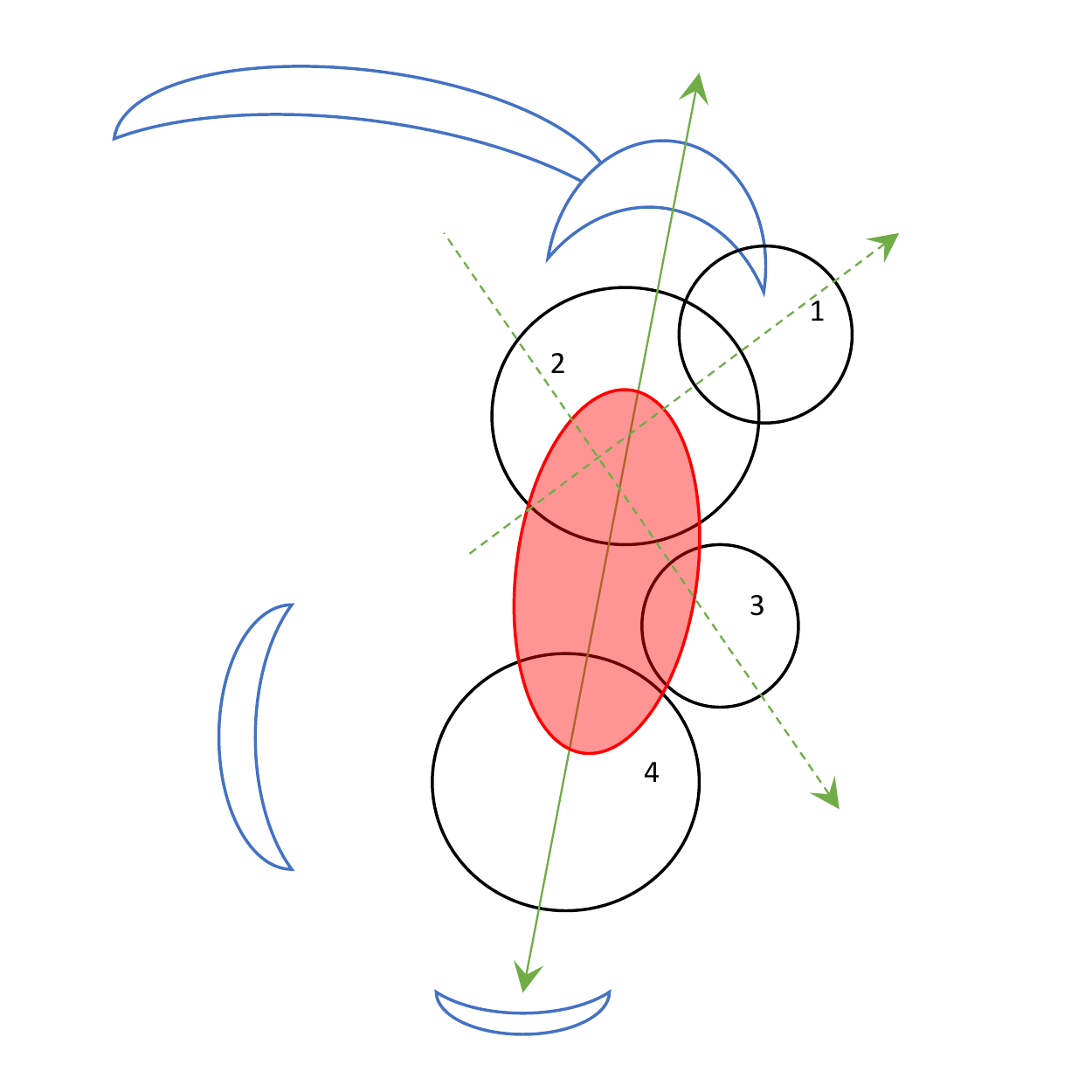}
\caption{Preferred merger scenario schematic for 1RXSJ0603. Subclusters are listed with numbers from our GMM analysis and match naming conventions listed in Figure \ref{fig:subclusthist}. The red shading is the approximate ICM location as interpreted by the X-ray surface brightness distribution. The blue objects represent radio relics, and solid lines represent merger axes that launched radio relics. Dash lines are proposed trajectories that are proposed with lower confidence. Analogous figures are presented for each system in Figure \ref{fig:A1} through Figure \ref{fig:A28}.}
\label{fig:cartoon}
\end{figure}

For each cluster, we present the inner 150$\times$150 kpc centered on the BCG. Figure \ref{fig:bcg} shows the central image for 1RXSJ0603. Finally, for each cluster we illustrate the proposed merger scenario; Figure \ref{fig:cartoon} contains the example for 1RXSJ0603. 

\section{Results: Individual Clusters}\label{sec:results_individual}

In this section we focus on each cluster individually. First, we present a detailed literature review and discuss the history of understanding each cluster as (1) a merging galaxy cluster and (2) and radio relic cluster. Next, we discuss results of our analysis for each cluster and discuss these results in the context of the literature review. We reserve the ensemble analysis for \S\ref{sec:results_sample}. In each subsection, we will frequently refer readers to Tables \ref{tab:subclusters} and \ref{tab:mcmcgmm} of the previous section while discussing the subcluster analyses. The potential halos discussed in the text and listed in Table \ref{tab:subclusters} are arranged north to south as halos a, b, c, etc. Subclusters of the lowest BIC model are listed north to south as well, but given a numbered convention instead. The color sequence for presentation of the numbered subclusters is as follows. The northern most subcluster of each lowest BIC model is presented with green colors and labeled ``Subcluster 1.'' The color sequence is then, orange, purple, and pink for subclusters 2, 3, and 4, respectively. No system is found to have greater than four subclusters; however, as mentioned in \S\ref{subsec:gmm}, our models represent a lower limit for the true number of subclusters.

For each cluster we collect the analogs of Figures \ref{fig:dstest}, \ref{fig:scatter}, \ref{fig:subclusthist}, \ref{fig:xray}, \ref{fig:bcg}, and \ref{fig:cartoon} into full page figures. These comprise Figuers 10-37 for the 28 clusters after 1RXSJ0603.

\subsection{1RXS J0603.3+4212}
\label{subsec:1RXSJ0603}
\input{1RXSJ0603.tex}

\subsection{Abell 115}
\label{subsec:A115}
\input{A115.tex}

\begin{figure*}
\centering
\includegraphics[height=8.25in]{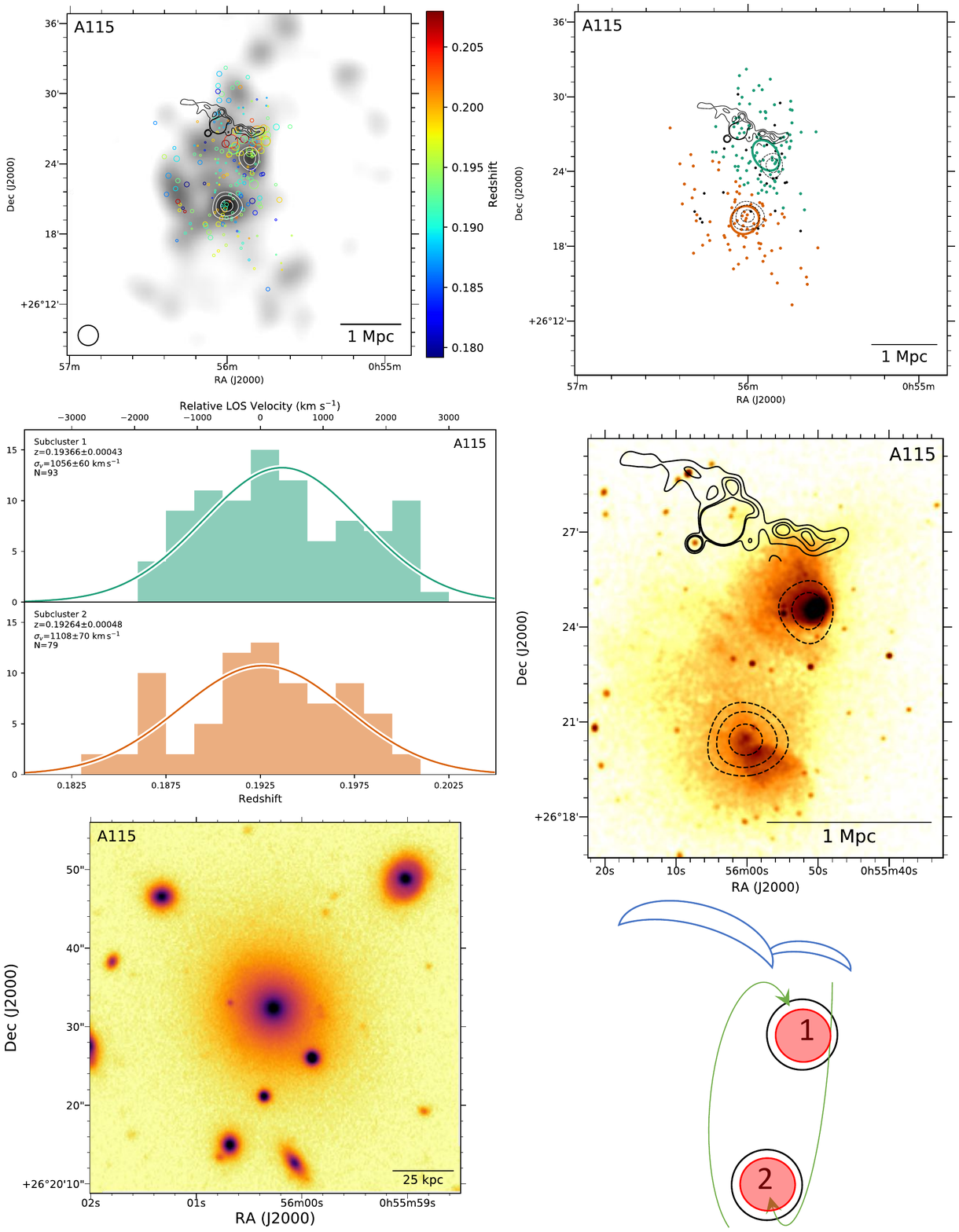}
\caption{\emph{Top left:} DS-test over red sequence i-band luminosity distribution with linearly space white contours. The black contours show the 1.4 GHz VLA contours \citep{Govoni:2001b}. \emph{Top right:} Subcluster membership of spectroscopic cluster members assigned with our GMM analysis. The same VLA contours (solid) and red sequence luminosity contours (dashed) are depicted. \emph{Middle left:} Subcluster redshift histograms with colors matching the image at the top right. \emph{Middle right:} 376 ks Chandra image with the same VLA radio contours (solid) and red sequence luminosity contours (dashed). The glancing merger scenario is clear from the swirling stream of X-ray emission. \emph{Bottom left:} 150 kpc cutout of the Subaru/SuprimeCam image centered on the BCG of subcluster 2. \emph{Bottom right:} Schematic of preferred merger scenario showing a large impact parameter merger and spiraling motion in the observed state.}
\label{fig:A1}
\end{figure*}

\subsection{Abell 521}
\label{subsec:A521}
\input{A521.tex}

\begin{figure*}
\centering
\includegraphics[height=8.25in]{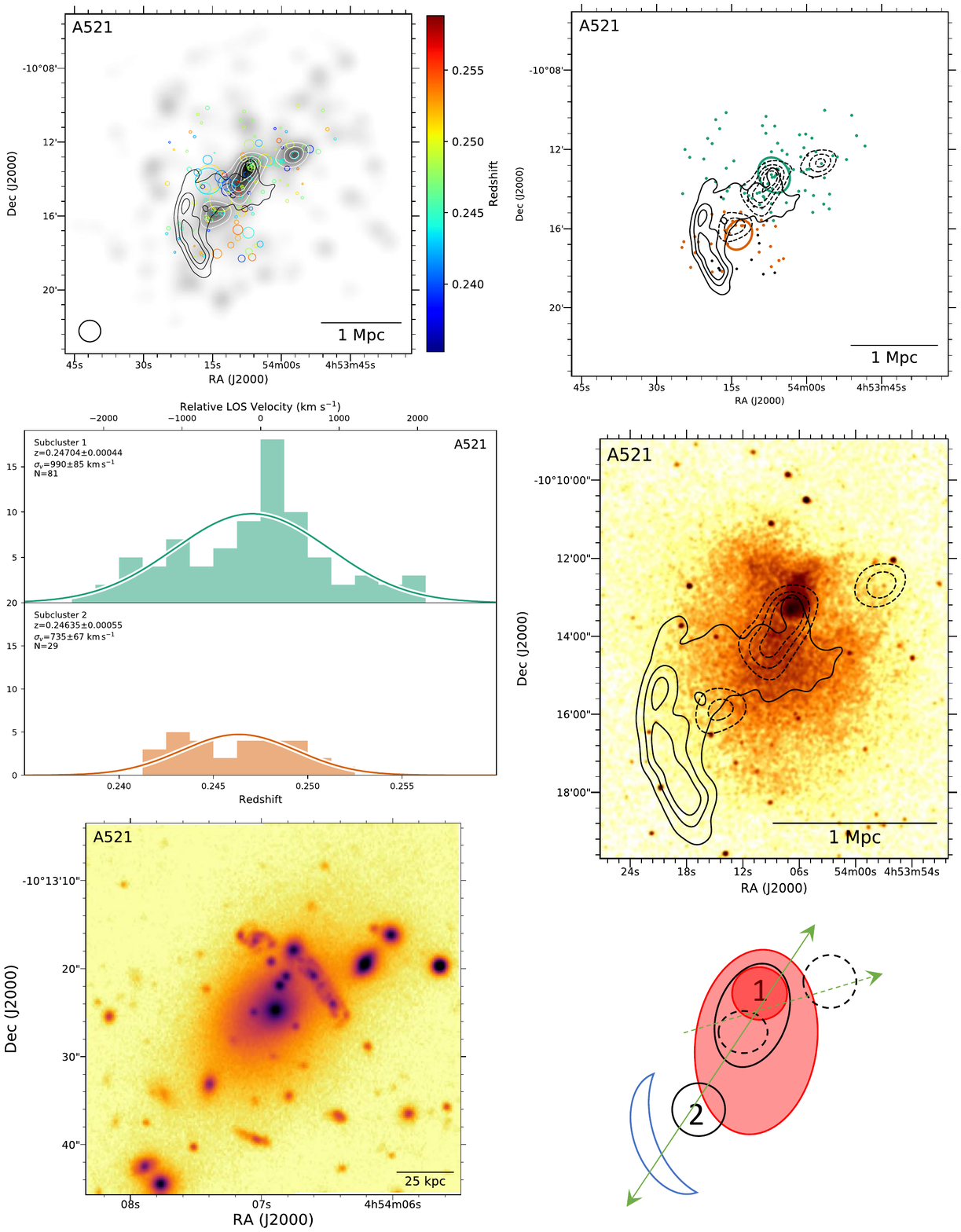}
\caption{\emph{Top left:} DS-test over red sequence i-band luminosity distribution with linearly space white contours. The black contours show the 240 MHz GMRT contours \citep{Venturi:2013}. \emph{Top right:} Subcluster membership of spectroscopic cluster members assigned with our GMM analysis. The same GMRT contours (solid) and red sequence luminosity contours (dashed) are depicted. \emph{Middle left:} Subcluster redshift histograms with colors matching the image at the top right. \emph{Middle right:} 170 ks Chandra image with the same GMRT radio contours (solid) and red sequence luminosity contours (dashed). A complex merger scenario is evident. \emph{Bottom left:} 150 kpc cutout of the Subaru/SuprimeCam i-band image centered on the BCG of subcluster 1. The multiple images of the face on spiral are evident. \emph{Bottom right:} Proposed merger scenario. Dashed black circles correspond to galaxy substructure not identified by the GMM analysis. The dark red region corresponds to the dense ICM core coincident with the BCG.} 
\label{fig:A2}
\end{figure*}

\subsection{Abell 523}
\label{subsec:A523}
\input{A523.tex}

\begin{figure*}
\centering
\includegraphics[height=8.25in]{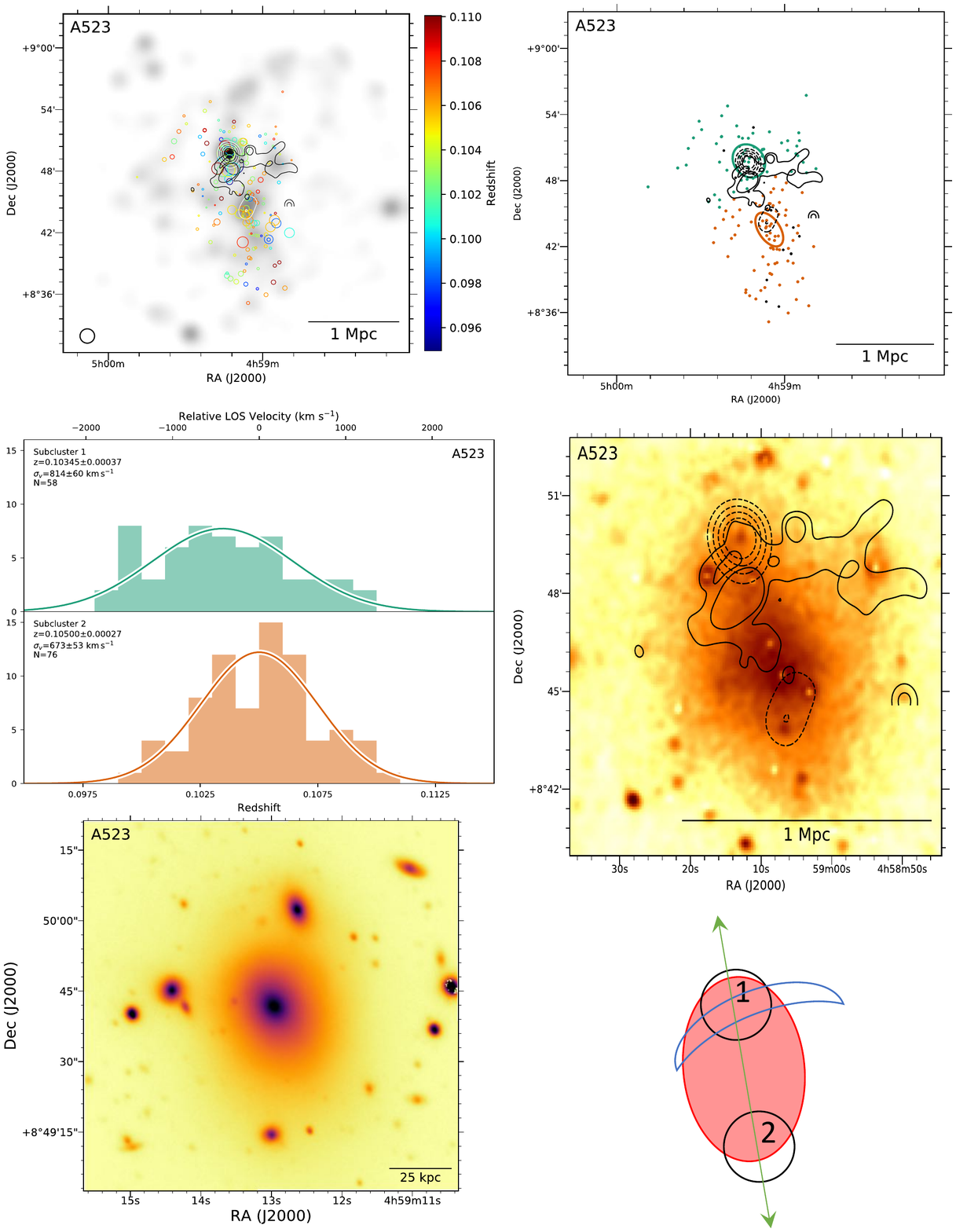}
\caption{\emph{Top left:} DS-test over red sequence r-band luminosity distribution with linearly space white contours. The black contours show the 1.4 GHz VLA contours \citep{vanWeeren:2011}. \emph{Top right:} Subcluster membership of spectroscopic cluster members assigned with our GMM analysis. The same VLA contours (solid) and red sequence luminosity contours (dashed) are depicted. \emph{Middle left:} Subcluster redshift histograms with colors matching the image at the top right. \emph{Middle right:} 30 ks Chandra image with the same VLA radio contours (solid) and red sequence luminosity contours (dashed). A bimodal merger scenario is evident, but the radio emission is not clearly tracing a potential merger induced shock between subclusters 1 and 2. \emph{Bottom left:} 150 kpc cutout of the Subaru/SuprimeCam r-band image centered on the BCG of subcluster 1. \emph{Bottom right:} Proposed merger scenario. A523 is a clean bimodal merger with a difficult to interpret radio feature that has been debated in the literature.}
\label{fig:A3}
\end{figure*}

\subsection{Abell 746}
\label{subsec:A746}
\input{A746.tex}
\begin{figure*}
\centering
\includegraphics[height=8.25in]{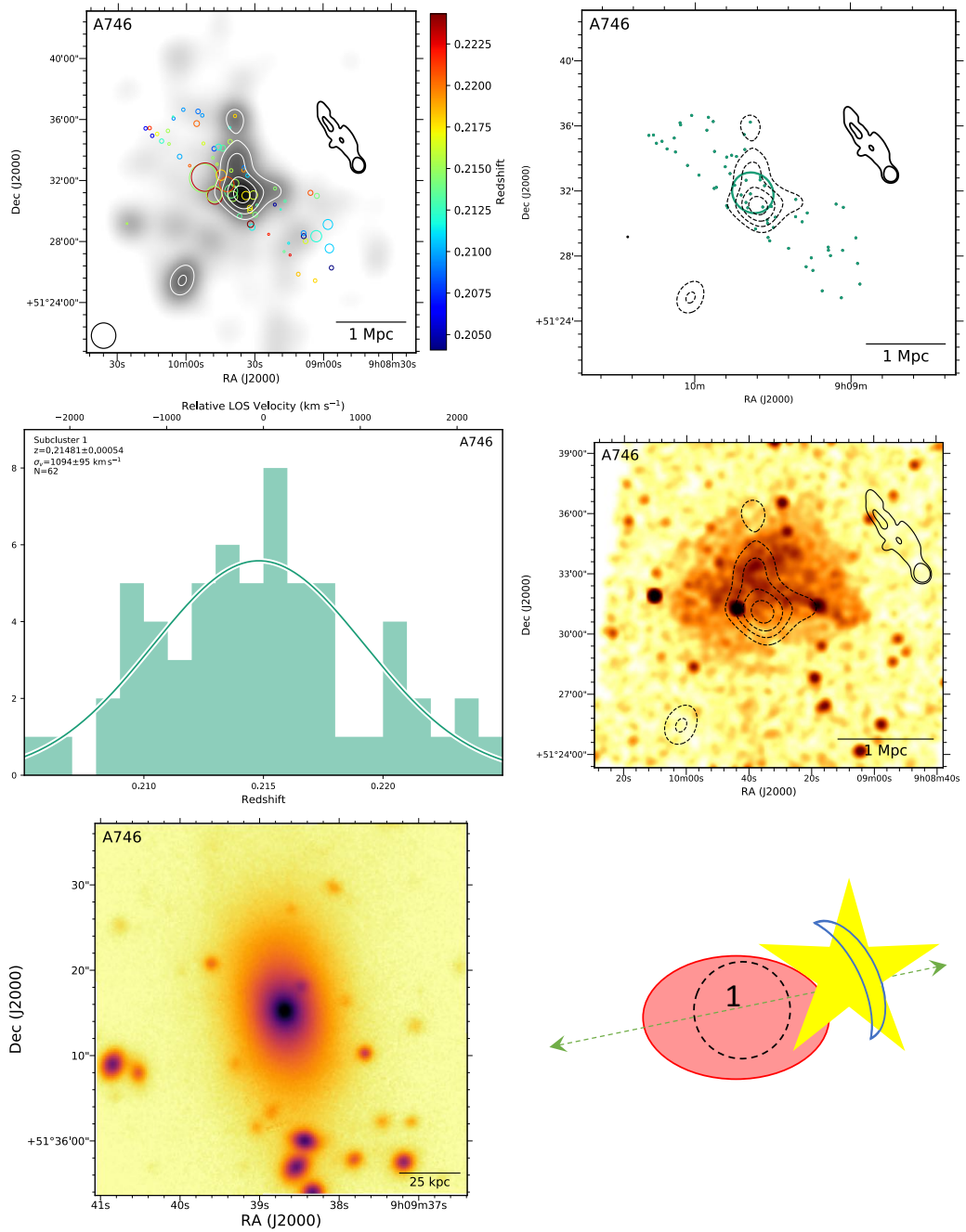}
\caption{\emph{Top left:} DS-test over red sequence r-band luminosity distribution with linearly space white contours. The absence of red sequence light in northwest is an artifact of a bright foreground star. The black contours show the 1.4 GHz WSRT contours \citep{vanWeeren:2011}. \emph{Top right:} Subcluster membership of spectroscopic cluster members assigned with our GMM analysis. We did not attempt to fit multiple Gaussians to the spectroscopic data due to the high contamination from the bright star. The same WSRT contours (solid) and red sequence luminosity contours (dashed) are depicted. \emph{Middle left:} Subcluster redshift histograms with colors matching the image at the top right. \emph{Middle right:} 30 ks Chandra image with the same WSRT radio contours (solid) and red sequence luminosity contours (dashed). While the optical analysis is hindered by the star, there appears to be and east to west elongation of the X-ray surface brightness, which is suggestive of the radio relic location. \emph{Bottom left:} 150 kpc cutout of the Subaru/SuprimeCam r-band image centered on the BCG. \emph{Bottom right:} Schematic of observed orientation. The bright star makes it impossible to interpret merger scenario beyond a likely east--west axis based on the X-ray and radio morphology. }
\label{fig:A4}
\end{figure*}

\subsection{Abell 781}
\label{subsec:A781}
\input{A781.tex}
\begin{figure*}
\centering
\includegraphics[height=8.25in]{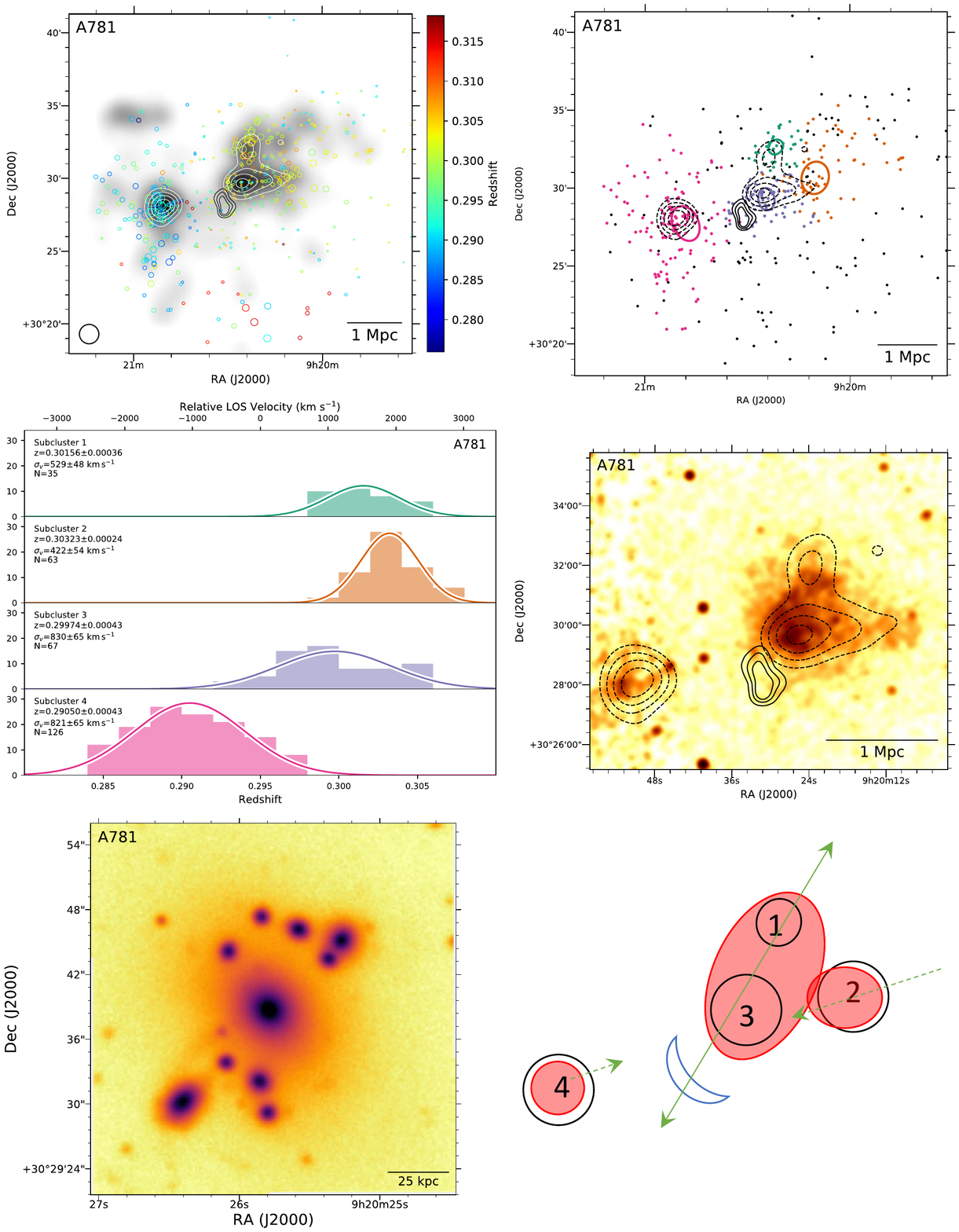}
\caption{\emph{Top left:} DS-test over red sequence i-band luminosity distribution with linearly space white contours. The black contours show the 325 MHz GMRT contours \citep{Venturi:2013}. \emph{Top right:} Subcluster membership of spectroscopic cluster members assigned with our GMM analysis. The same GMRT contours (solid) and red sequence luminosity contours (dashed) are depicted. \emph{Middle left:} Subcluster redshift histograms with colors matching the image at the top right. \emph{Middle right:} 48 ks Chandra image with the same WSRT radio contours (solid) and red sequence luminosity contours (dashed). The relic and disturbed ICM are suggestive of a merger between subclusters 1 and 3. \emph{Bottom left:} 150 kpc cutout of the Subaru/SuprimeCam i-band image centered on the cluster BCG, which is associated with subcluster 3. \emph{Bottom right:} Proposed merger scenario. The ICM is most disrupted between subclusters 1 and 3 with subcluster 2 having a less disturbed ICM coincident with its galactic light. Subcluster 4 has yet to interact, but it will coalesce with the other subclusters in the future.}
\label{fig:A5}
\end{figure*}

\subsection{Abell 1240}
\label{subsec:A1240}
\input{A1240.tex}
\begin{figure*}
\centering
\includegraphics[height=8.25in]{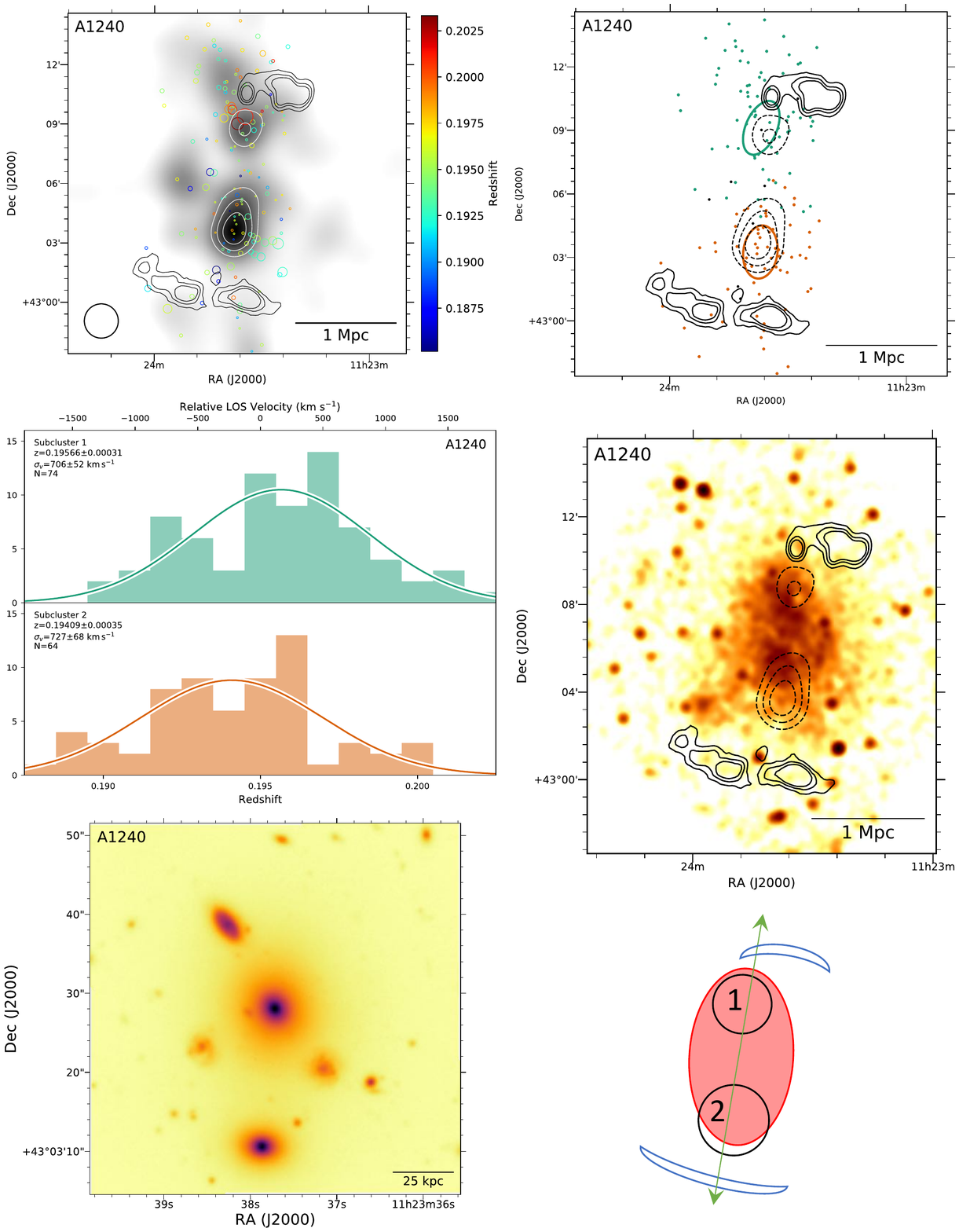}
\caption{\emph{Top left:} DS-test over red sequence r-band luminosity distribution with linearly space white contours. The black contours show the 1.4 GHz VLA contours \citep{Bonafede:2009}. \emph{Top right:} Subcluster membership of spectroscopic cluster members assigned with our GMM analysis. The same VLA contours (solid) and red sequence luminosity contours (dashed) are depicted. \emph{Middle left:} Subcluster redshift histograms with colors matching the image at the top right. \emph{Middle right:} 52 ks Chandra image with the same VLA radio contours (solid) and red sequence luminosity contours (dashed). The relics and disturbed ICM are suggestive of a merger between subclusters 1 and 2. \emph{Bottom left:} 150 kpc cutout of the Subaru/SuprimeCam r-band image centered on the southern BCG, which is associated with subcluster 2. \emph{Bottom right:} Proposed merger scenario. A1240 is a clean bimodal merger with two radio relics colinear with the two subclusters.}
\label{fig:A6}
\end{figure*}
f
\subsection{Abell 1300}
\label{subsec:A1300}
\input{A1300.tex}
\begin{figure*}
\centering
\includegraphics[height=8in]{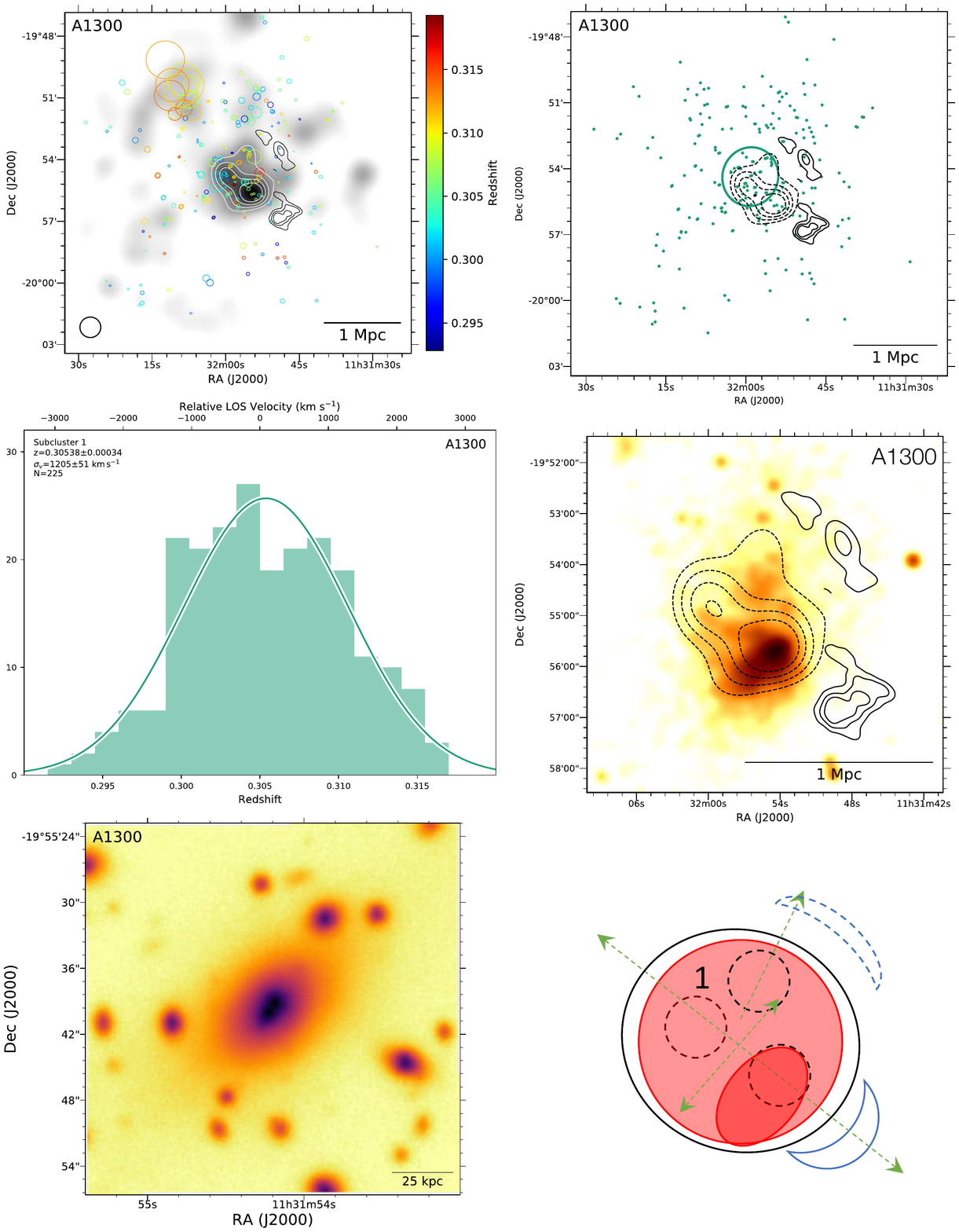}
\caption{\emph{Top left:} DS-test over red sequence r-band luminosity distribution with linearly space white contours. The black contours show the 325 MHz GMRT contours \citep{Venturi:2013}. \emph{Top right:} Subcluster membership of spectroscopic cluster members assigned with our GMM analysis. The same GMRT contours (solid) and red sequence luminosity contours (dashed) are depicted. The small projected separation and low spectroscopic survey density likely contributes to a single halo model being preferred by the BIC analysis. \emph{Middle left:} Subcluster redshift histograms with colors matching the image at the top right. \emph{Middle right:} 100 ks Chandra image with the same GMRT radio contours (solid) and red sequence luminosity contours (dashed). The relics and disturbed ICM are suggestive of a complex merger between at least three subclusters. \emph{Bottom left:} 150 kpc cutout of the Subaru/SuprimeCam r-band image centered on the cluster BCG. \emph{Bottom right:} Proposed merger scenario. A1300 is a difficult to interpret merger with likely at least three subclusters. The X-ray surface brightness morphology is suggestive of of an additional merger axis associated with the an additional merger in the southwest to northeast direction. The BCG is elongated along the same axis. Inspection of the higher resolution HST RELICS images at \url{https://www.spacetelescope.org/images/potw1745a/} show the BCG is actually composed of two luminosity peaks along this axis.}
\label{fig:A7}
\end{figure*}

\subsection{Abell 1612}
\label{subsec:A1612}
\input{A1612.tex}
\begin{figure*}
\centering
\includegraphics[height=8.25in]{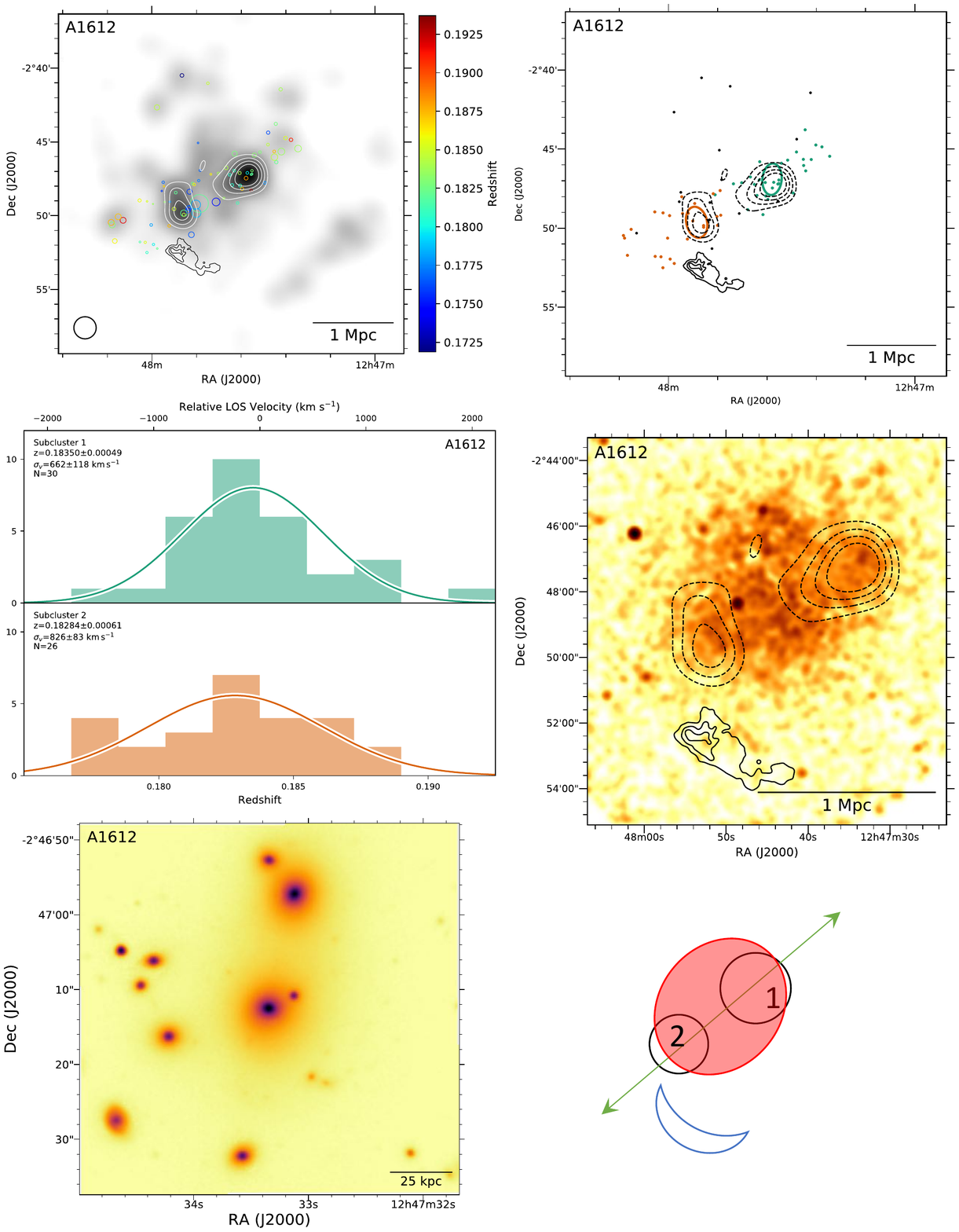}
\caption{\emph{Top left:} DS-test over red sequence r-band luminosity distribution with linearly space white contours. The black contours show the 325 MHz GMRT contours \citep{vanWeeren:2011}. \emph{Top right:} Subcluster membership of spectroscopic cluster members assigned with our GMM analysis. The same GMRT contours (solid) and red sequence luminosity contours (dashed) are depicted. \emph{Middle left:} Subcluster redshift histograms with colors matching the image at the top right. \emph{Middle right:} 100 ks Chandra image with the same GMRT radio contours (solid) and red sequence luminosity contours (dashed). The relics and disturbed ICM are suggestive of a complex merger between at least three subclusters. \emph{Bottom left:} 150 kpc cutout of the Subaru/SuprimeCam r-band image centered on the cluster BCG. \emph{Bottom right:} Proposed merger scenario. The radio relic location perhaps suggests an off axis merger; however, inspection of the 610 MHz GMRT in Figure 2 of \citet{vanWeeren:2011} suggests the relic is closely linked to the AGN of the BCG suggesting the shock may be larger than the radio relic, but the weak shock is only accelerating the electrons where they were pre-accelerated by the AGN \citep[e.g.,][]{vanWeeren:2017}. \emph{Bottom right:} Proposed merger scenario.}
\label{fig:A8}
\end{figure*}

\subsection{Abell 2034}
\label{subsec:A2034}
\input{A2034.tex}
\begin{figure*}
\centering
\includegraphics[height=8.25in]{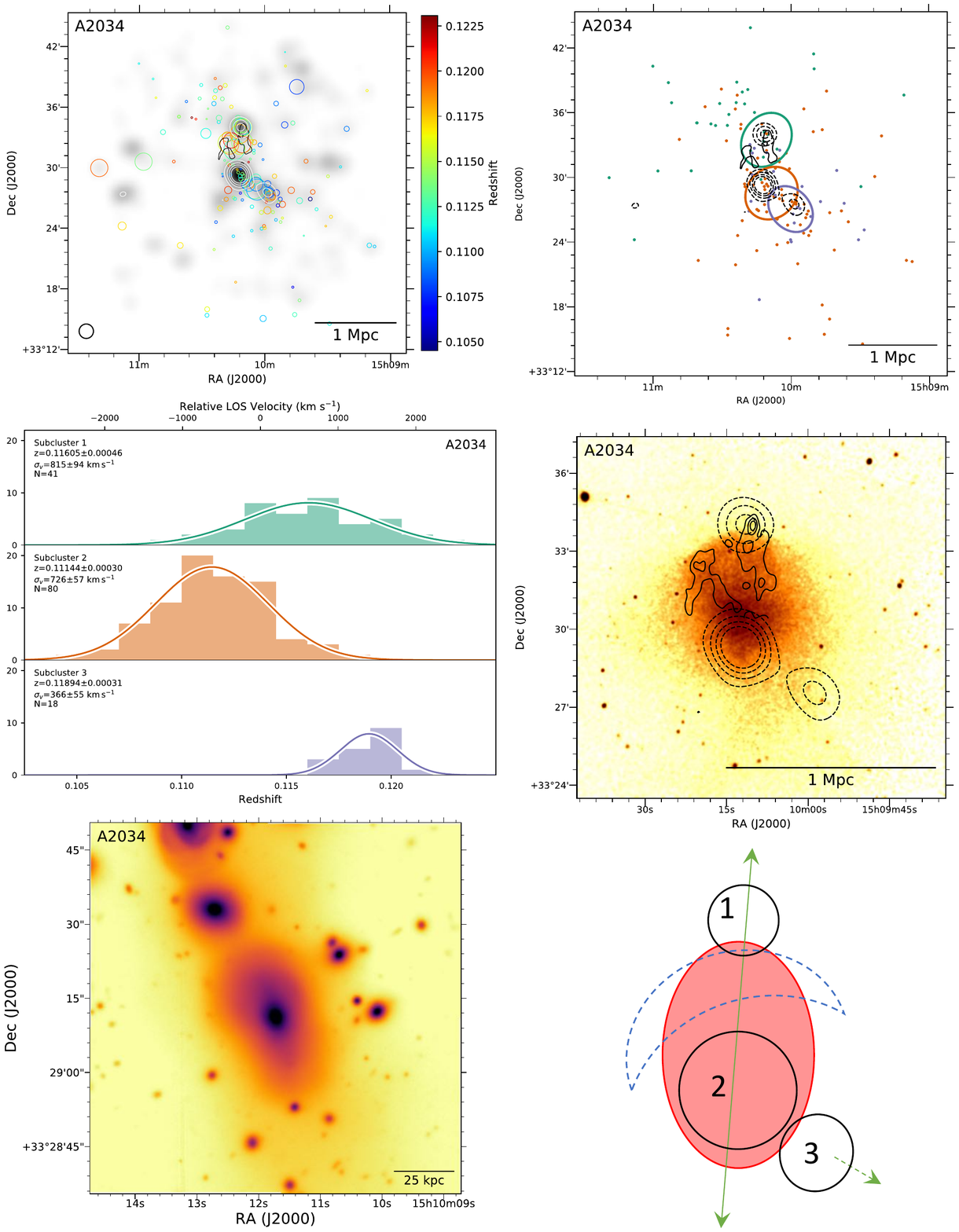}
\caption{\emph{Top left:} DS-test over red sequence R-band luminosity distribution with linearly space white contours. The black contours show the 1.4 GHz WSRT contours \citep{vanWeeren:2011}. \emph{Top right:} Subcluster membership of spectroscopic cluster members assigned with our GMM analysis. The same WSRT contours (solid) and red sequence luminosity contours (dashed) are depicted. \emph{Middle left:} Subcluster redshift histograms with colors matching the image at the top right. \emph{Middle right:} 261 ks Chandra image with the same WSRT radio contours (solid) and red sequence luminosity contours (dashed). The relics and disturbed ICM are suggestive of a merger between at least two subclusters; however, the lack of X-ray emission with subcluster 3 suggests it was also involved. \emph{Bottom left:} 150 kpc cutout of the Subaru/SuprimeCam R-band image centered on the cluster BCG in subcluster 2. \emph{Bottom right:} Proposed merger scenario. A2034 is primarily a north to south merger with two complications. First, the shock has been classified as a cold front in the literature, which would better explain subcluster 1's location; however, the relic candidate is coincident with this edge in the ICM. Second, subcluster 3 appears to have exited toward the southwest having its ICM stripped.}
\label{fig:A9}
\end{figure*}

\subsection{Abell 2061}
\label{subsec:A2061}
\input{A2061.tex}
\begin{figure*}
\centering
\includegraphics[height=8.25in]{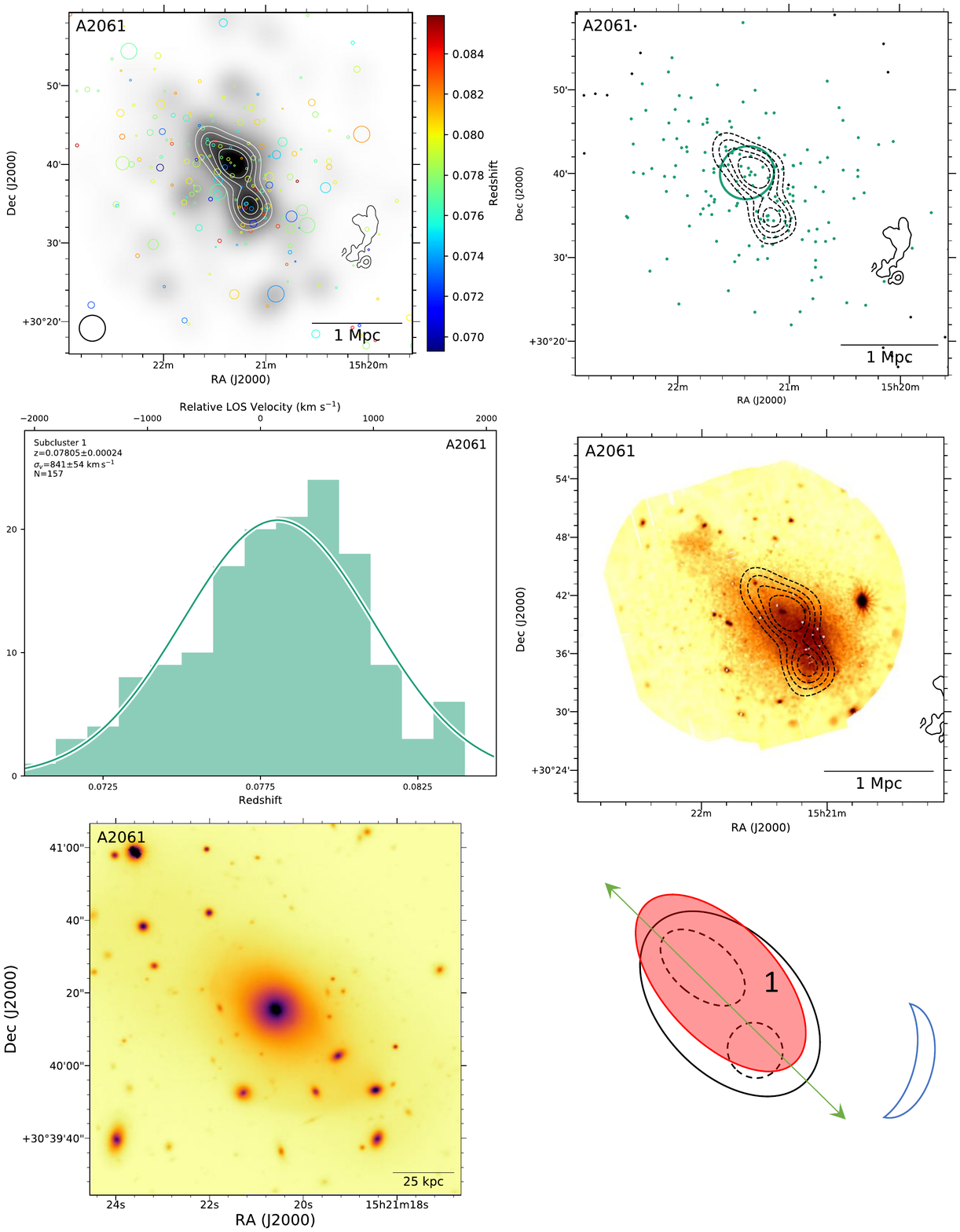}
\caption{\emph{Top left:} DS-test over red sequence r-band luminosity distribution with linearly space white contours. The black contours show the 1.4 GHz WSRT contours \citep{vanWeeren:2011}. \emph{Top right:} Subcluster membership of spectroscopic cluster members assigned with our GMM analysis. The same WSRT contours (solid) and red sequence luminosity contours (dashed) are depicted. The small projected separation and low spectroscopic survey density likely contributes to a single halo model being preferred by the BIC analysis. \emph{Middle left:} Subcluster redshift histograms with colors matching the image at the top right. \emph{Middle right:} 50 ks XMM image with the same WSRT radio contours (solid) and red sequence luminosity contours (dashed). The relic at the edge of the frame in the southwest and the disturbed ICM are suggestive of a merger between at least two subclusters. The extended stream of X-ray emission to the northeast toward A2067 is unlikely due to a merger between the two. A2067 sits 2.7 Mpc away. \emph{Bottom left:} 150 kpc cutout of the Subaru/SuprimeCam r-band image centered on the cluster BCG in the southwest. \emph{Bottom right:} Proposed merger scenario. Two subclusters are apparent in multiple wavelengths, but the GMM in unable to separate them due to low spectroscopic sampling and a small projected separation.}
\label{fig:A10}
\end{figure*}

\subsection{Abell 2163}
\label{subsec:A2163}
\input{A2163.tex}
\begin{figure*}
\centering
\includegraphics[height=8in]{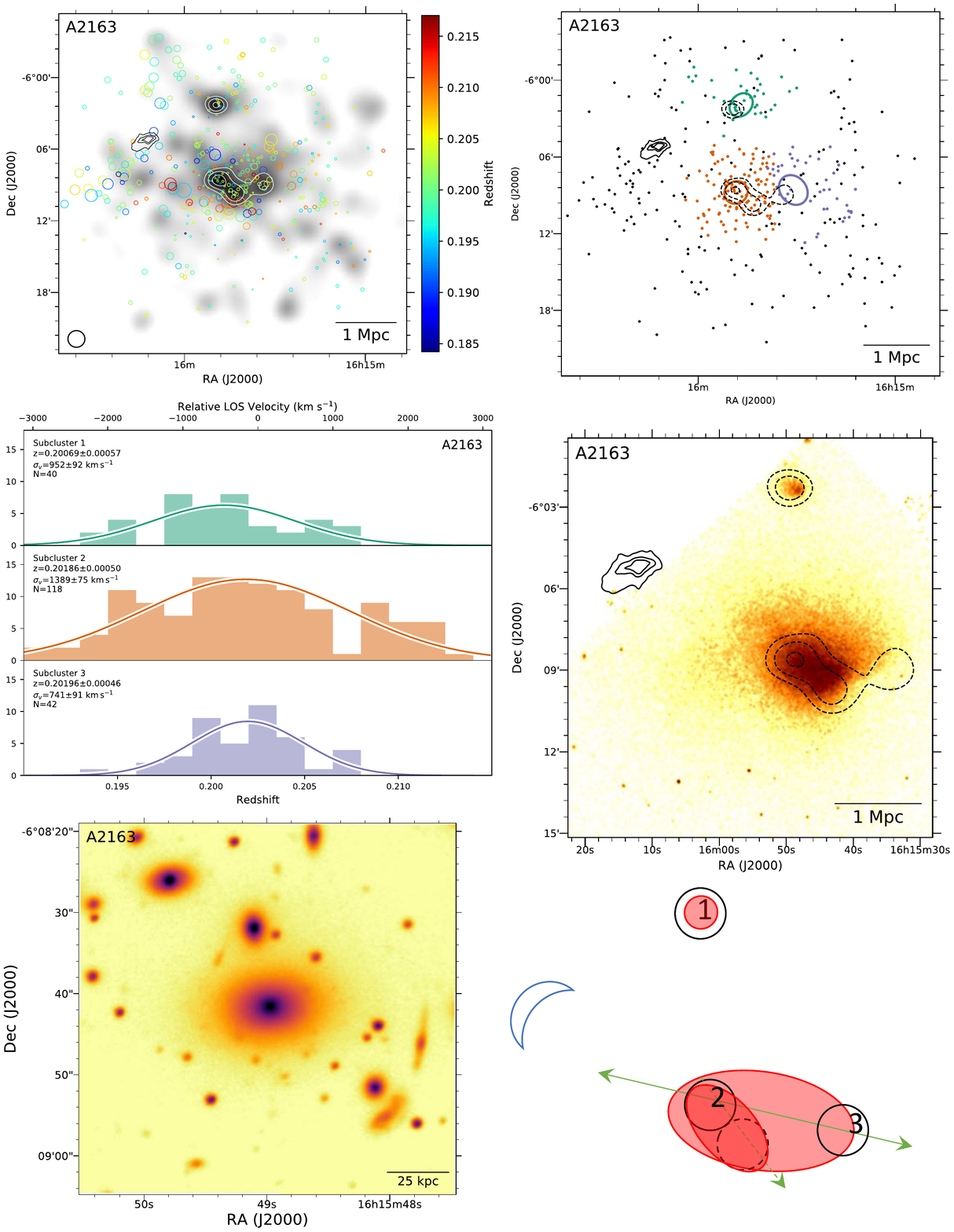}
\caption{\emph{Top left:} DS-test over red sequence R-band luminosity distribution with linearly space white contours. The black contours show the 1.4 GHz VLA contours \citep{Feretti:2001}. \emph{Top right:} Subcluster membership of spectroscopic cluster members assigned with our GMM analysis. The same VLA contours (solid) and red sequence luminosity contours (dashed) are depicted. The small projected separation between the two apparent subclusters comprising subcluster 2 likely contributes to a single halo describing this region. \emph{Middle left:} Subcluster redshift histograms with colors matching the image at the top right. \emph{Middle right:} 90 ks Chandra image with the same VLA radio contours (solid) and red sequence luminosity contours (dashed). The relic at the northeast and the disturbed ICM are suggestive of a merger between at least three subclusters.  The northernmost subcluster 1 has yet to merge. \emph{Bottom left:} 150 kpc cutout of the Subaru/SuprimeCam R-band image centered on the cluster BCG in subcluster 2. \emph{Bottom right:} Preferred merger scenario. The large projected distance to the radio relic suggest an older merger. Subcluster 1 has yet to merge, and subcluster 3 has been stripped of its ICM suggesting it is post pericenter. A fourth subcluster is likely associated with the cool core remnant identified by \citet{Bourdin:2011}. This merger is very near core-crossing and likely contributing to the very high X-ray luminosity and SZ inferred mass.}
\label{fig:A11}
\end{figure*}

\subsection{Abell 2255}
\label{subsec:A2255}
\input{A2255.tex}
\begin{figure*}
\centering
\includegraphics[height=8.1in]{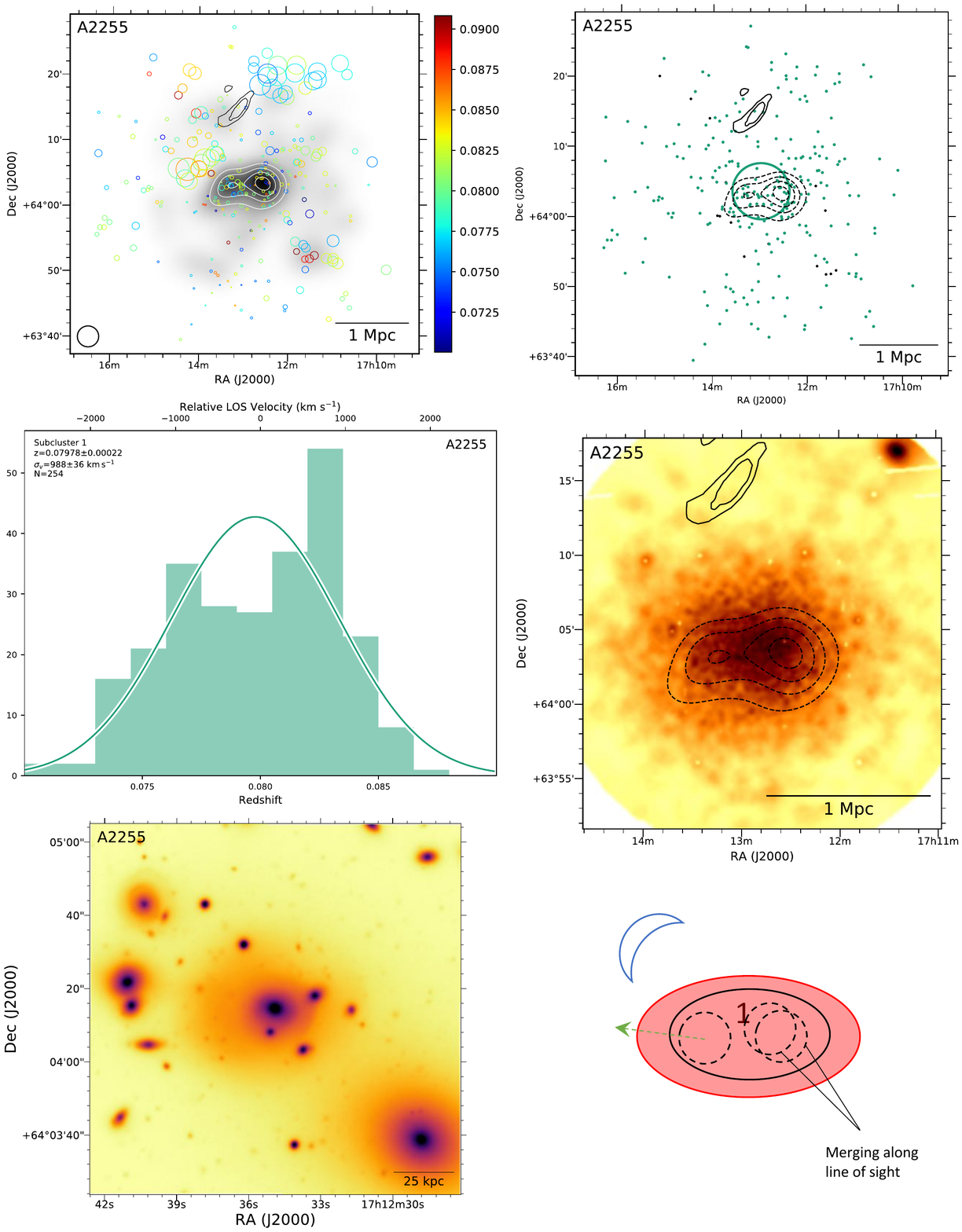}
\caption{\emph{Top left:} DS-test over red sequence R-band luminosity distribution with linearly space white contours. The black contours show the 350 MHz WSRT contours \citep{Pizzo:2009}. \emph{Top right:} Subcluster membership of spectroscopic cluster members assigned with our GMM analysis. The same WSRT contours (solid) and red sequence luminosity contours (dashed) are depicted. The small projected separation and low spectroscopic survey density likely contributes to a single halo model being preferred. \emph{Middle left:} Subcluster redshift histograms with colors matching the image at the top right. \emph{Middle right:} 42 ks XMM image with the same WSRT radio contours (solid) and red sequence luminosity contours (dashed). The relic at the northeast and the disturbed ICM are suggestive of a merger between at least three subclusters.  \emph{Bottom left:} 150 kpc cutout of the Subaru/SuprimeCam R-band image centered on the cluster BCG in the western half of the cluster. The second BCG apparent in the image has a $\sim$2000 km s$^{-1}$ line of sight velocity difference, which is suggestive of a merger along the line of sight. \emph{Bottom right:} Preferred merger scenario. The elongation of the X-ray surface brightness suggests a second plane of the sky merger component in addition to the line of sight merger. We do not make a prediction of which subclusters merged to launch the radio relic. We can not rule out further complications in this system.}
\label{fig:A12}
\end{figure*}

\subsection{Abell 2345}
\label{subsec:A2345}
\input{A2345.tex}
\begin{figure*}
\centering
\includegraphics[height=8.1in]{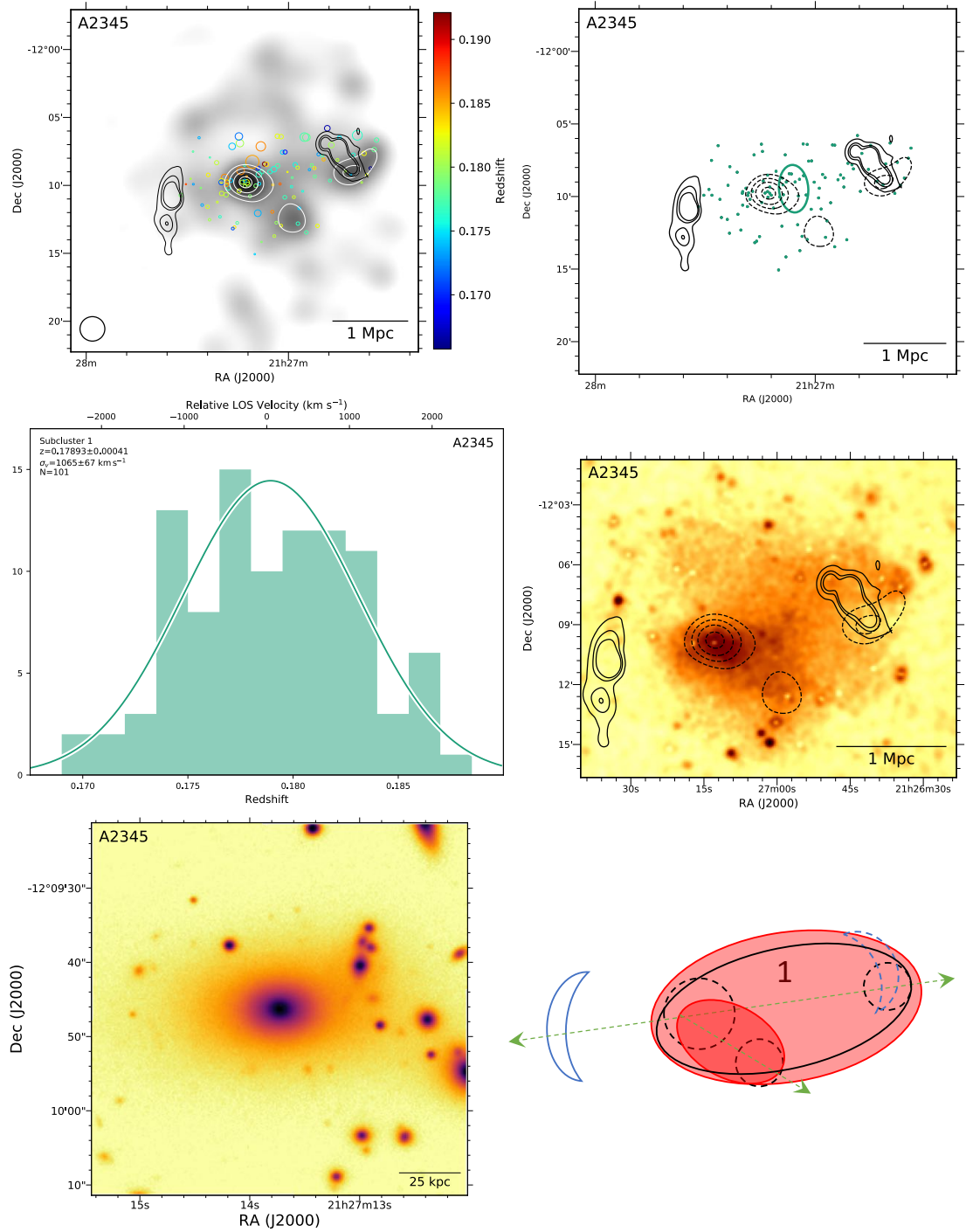}
\caption{\emph{Top left:} DS-test over red sequence R-band luminosity distribution with linearly space white contours. The black contours show the 1.4 GHz VLA contours \citep{Bonafede:2009}. \emph{Top right:} Subcluster membership of spectroscopic cluster members assigned with our GMM analysis. The same VLA contours (solid) and red sequence luminosity contours (dashed) are depicted. The low spectroscopic survey density  likely contributes to a single halo model being preferred. \emph{Middle left:} Subcluster redshift histogram with colors matching the image at the top right. \emph{Middle right:} 93 ks XMM image with the same VLA radio contours (solid) and red sequence luminosity contours (dashed). The relic at the northeast and the disturbed ICM are suggestive of a merger between at least three subclusters, perhaps with a large impact parameter for the northwest subcluster. \emph{Bottom left:} 150 kpc cutout of the Subaru/SuprimeCam R-band image centered on the cluster BCG in the northeastern subcluster. \emph{Bottom right:} Preferred merger scenario. The proximity of the western relic to the corresponding subcluster is difficult to explain in a scenario where the two relics were launched with the same merger.}
\label{fig:A13}
\end{figure*}

\subsection{Abell 2443}
\label{subsec:A2443}
\input{A2443.tex}
\begin{figure*}
\centering
\includegraphics[height=8.2in]{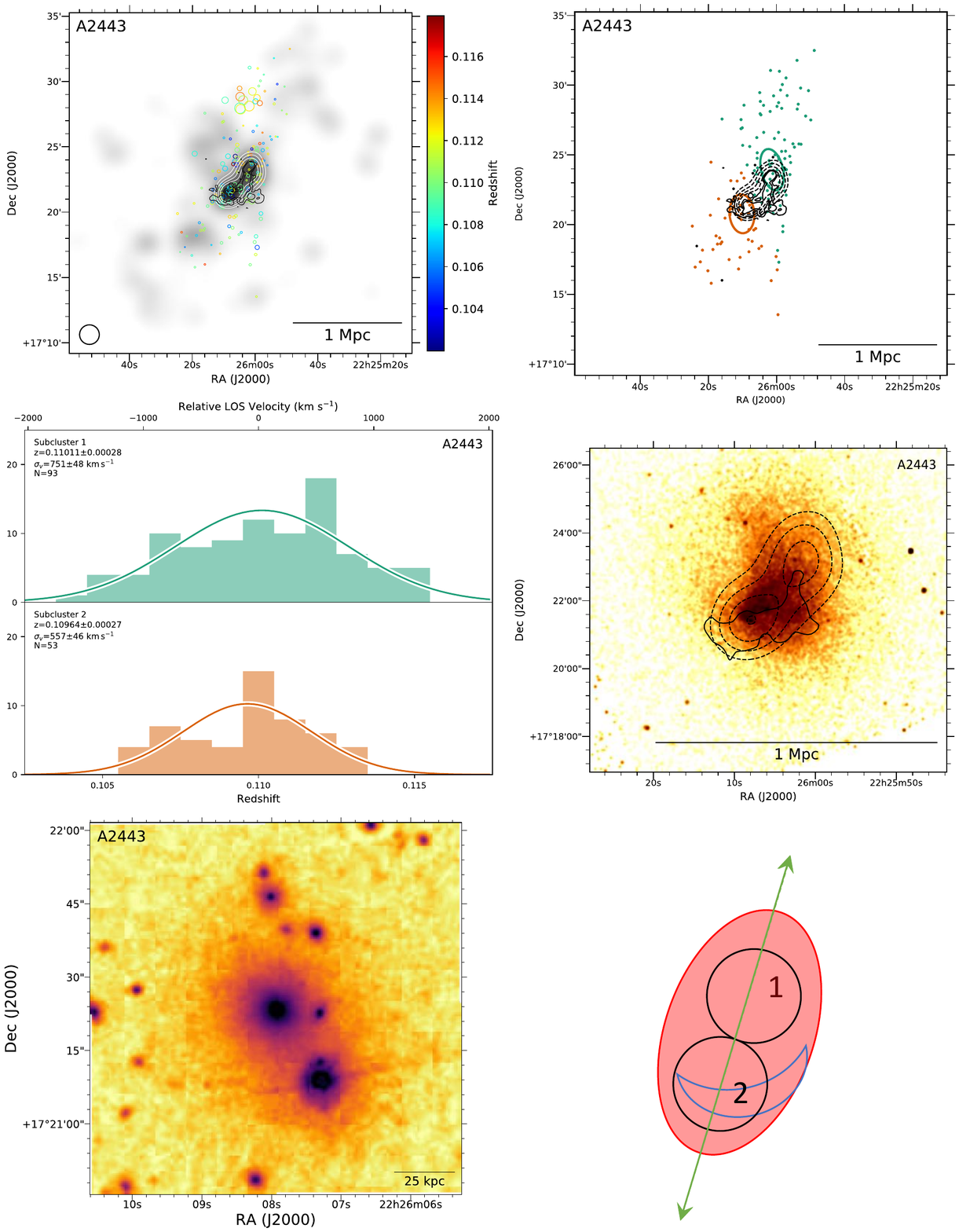}
\caption{\emph{Top left:} DS-test over red sequence SDSS r-band luminosity distribution with linearly space white contours. The black contours show the 325 MHz VLA contours \citep{Cohen:2011}. \emph{Top right:} Subcluster membership of spectroscopic cluster members assigned with our GMM analysis. The same VLA contours (solid) and red sequence luminosity contours (dashed) are depicted. \emph{Middle left:} Subcluster redshift histogram with colors matching the image at the top right. \emph{Middle right:} 116 ks Chandra image with the same VLA radio contours (solid) and red sequence luminosity contours (dashed). \emph{Bottom left:} 150 kpc cutout of the SDSS r-band image centered on the cluster BCG in subcluster 2. \emph{Bottom right:} Preferred merger scenario. The small projected separation between the two subclusters and radio phoenix type relic close to the galaxies suggests a young merger.}
\label{fig:A14}
\end{figure*}

\subsection{Abell 2744}
\label{subsec:A2744}
\input{A2744.tex}
\begin{figure*}
\centering
\includegraphics[height=8.2in]{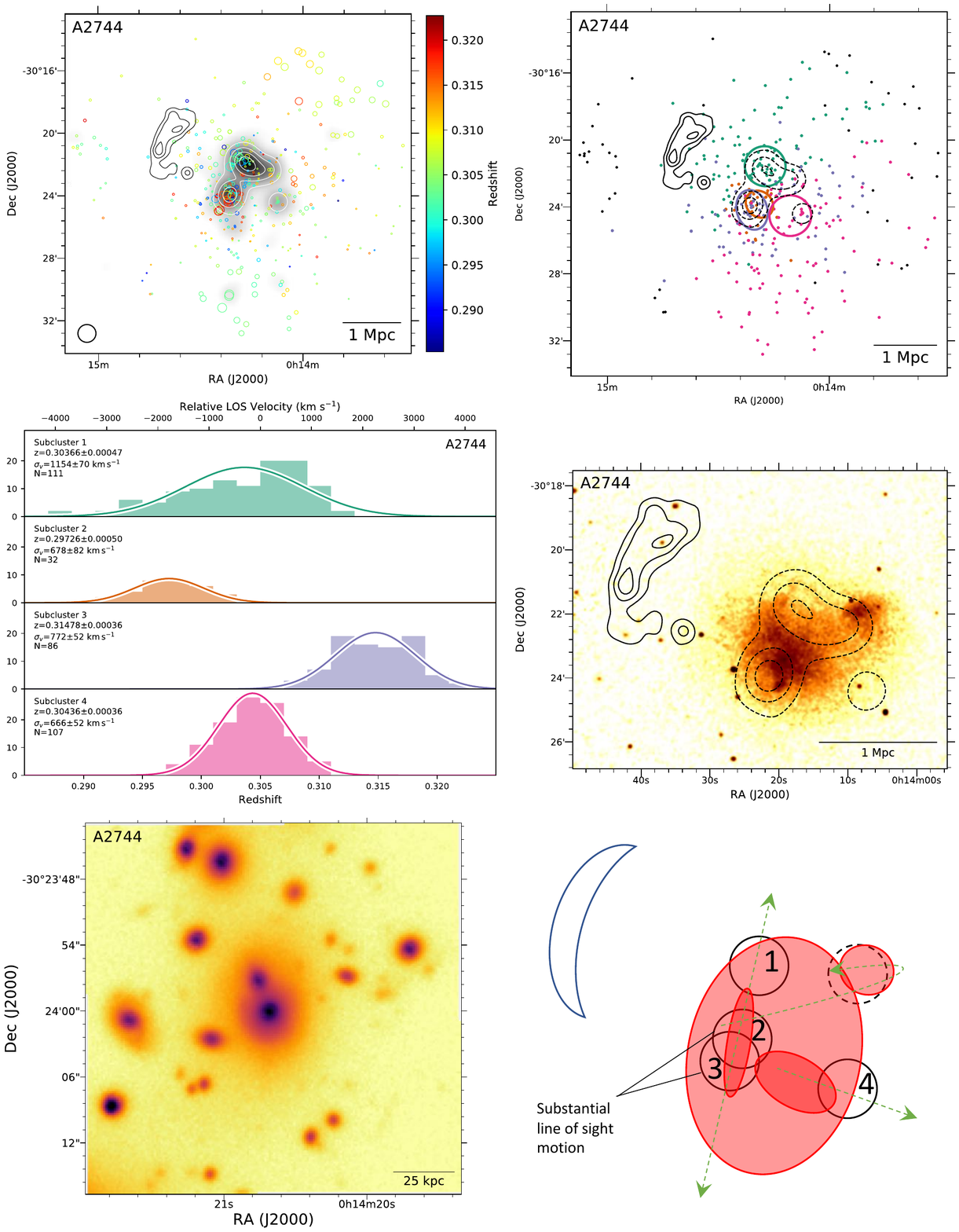}
\caption{\emph{Top left:} DS-test over red sequence Subaru/SuprimeCam R-band luminosity distribution with linearly space white contours. The black contours show the 325 MHz GMRT contours \citep{Venturi:2013}. \emph{Top right:} Subcluster membership of spectroscopic cluster members assigned with our GMM analysis. The same GMRT contours (solid) and red sequence luminosity contours (dashed) are depicted. \emph{Middle left:} Subcluster redshift histogram with colors matching the image at the top right. \emph{Middle right:} 132 ks Chandra image with the same GMRT radio contours (solid) and red sequence luminosity contours (dashed). \emph{Bottom left:} 150 kpc cutout of the Subaru/SuprimeCam R-band image centered on the cluster BCG in subcluster 2. \emph{Bottom right:} Preferred merger scenario. A2744 is the most complex merger in our sample. The core of the cluster (subclusters 2 and 3) has been studied in detail as a HST Frontier Field and has a substantial line of sight motion. Subclusters 1 and 4 have been stripped of their ICM suggesting post merger scenarios. Finally, a fifth subcluster with an associated ICM further from the cluster than the galaxies suggests a ram-pressure slingshot type scenario.}
\label{fig:A15}
\end{figure*}

\subsection{Abell 3365}
\label{subsec:A3365}
\input{A3365.tex}
\begin{figure*}
\centering
\includegraphics[height=8.15in]{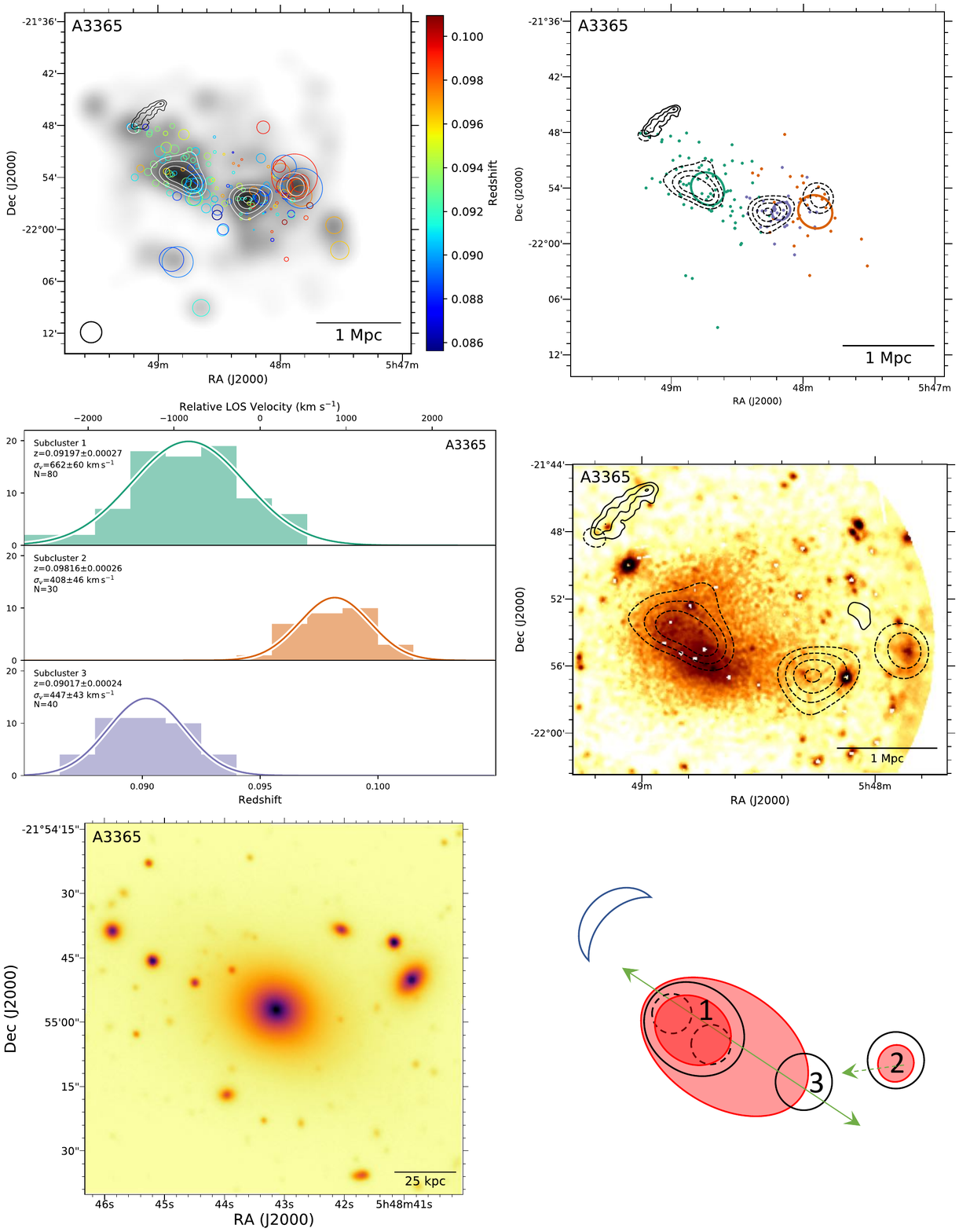}
\caption{\emph{Top left:} DS-test over red sequence Subaru/SuprimeCam r-band luminosity distribution with linearly space white contours. The black contours show the 1.4 GHz VLA contours \citep{vanWeeren:2011}. \emph{Top right:} Subcluster membership of spectroscopic cluster members assigned with our GMM analysis. The same VLA contours (solid) and red sequence luminosity contours (dashed) are depicted. \emph{Middle left:} Subcluster redshift histogram with colors matching the image at the top right. \emph{Middle right:} 161 ks XMM image with the same VLA radio contours (solid) and red sequence luminosity contours (dashed). \emph{Bottom left:} 150 kpc cutout of the Subaru r-band image centered on the cluster BCG in subcluster 2. \emph{Bottom right:} Preferred merger scenario. A3365 is composed of at least three subclusters with a merger occurring between subclusters 1 and 3 giving rise to the disturbed ICM and northeast relic. Subcluster 2 has a large line of sight velocity difference and undisturbed ICM suggesting it is still infalling with some component along the line of sight from the foreground, since it is redshifted relative to the rest of the cluster. There is evidence of composite structure in subcluster 1 which has two bright galaxies at either end of its associated red sequence luminosity peak as well as an elongated ridge of X-ray emission.}
\label{fig:A16}
\end{figure*}

\subsection{Abell 3411}
\label{subsec:A3411}
\input{A3411.tex}
\begin{figure*}
\centering
\includegraphics[height=8.25in]{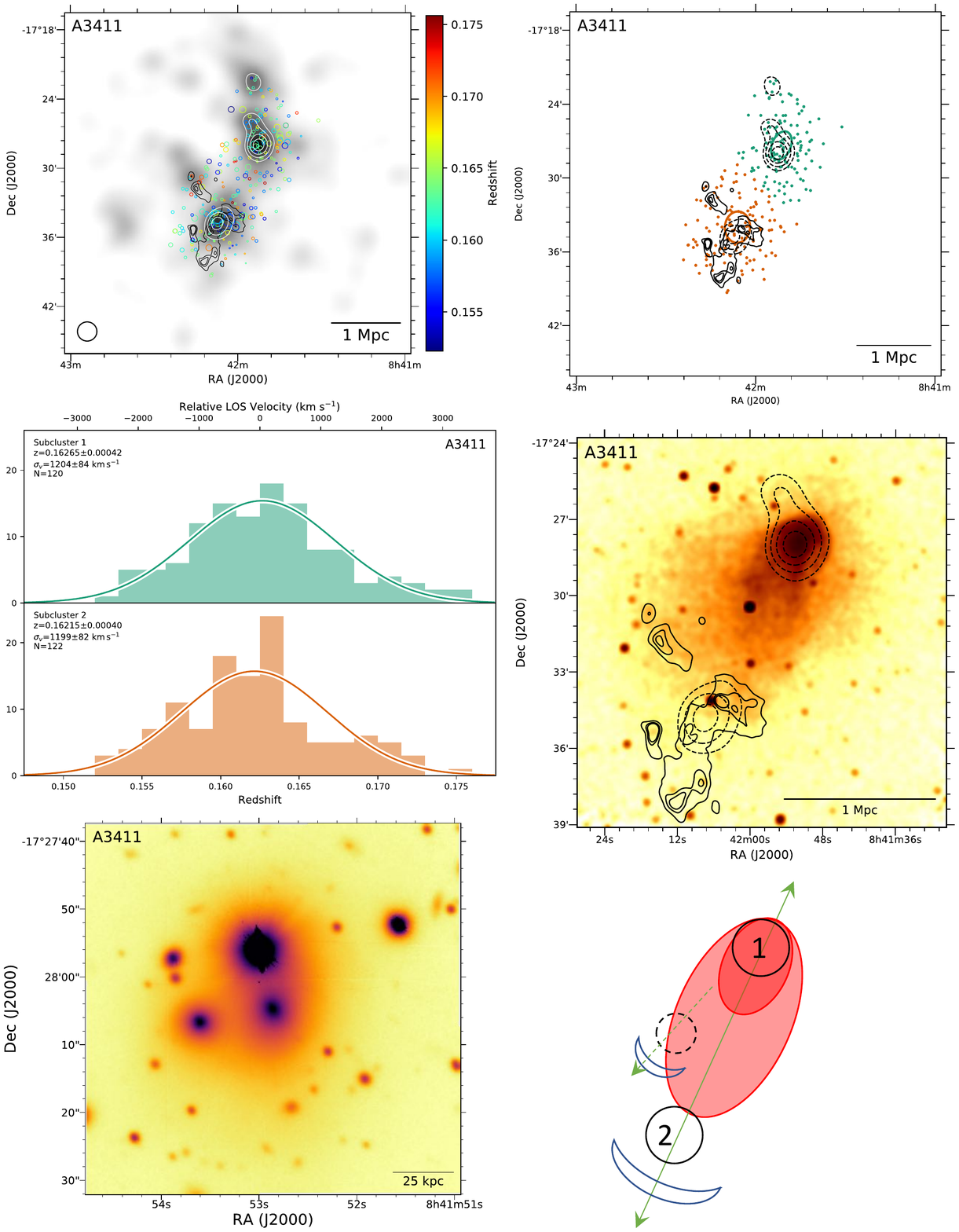}
\caption{\emph{Top left:} DS-test over red sequence Subaru/SuprimeCam r-band luminosity distribution with linearly space white contours. The black contours show GMRT 325 MHz contours \citep{vanWeeren:2017}. \emph{Top right:} Subcluster membership of spectroscopic cluster members assigned with our GMM analysis. The same GMRT contours (solid) and red sequence luminosity contours (dashed) are depicted. \emph{Middle left:} Subcluster redshift histogram with colors matching the image at the top right. \emph{Middle right:} 215 ks Chandra image with the same GMRT radio contours (solid) and red sequence luminosity contours (dashed). \emph{Bottom left:} 150 kpc cutout of the Subaru r-band image centered on the cluster BCG in subcluster 1. \emph{Bottom right:} Preferred merger scenario. A3411 is composed of at least two subclusters, where the northern subcluster 1 is A3411 and the southern 2 is A3412. Subcluster 1 is associated with a cool-core remnant and wake feature indicating the ICM is traveling northward. The two subclusters have very similar velocity dispersions and redshifts suggesting a 1:1 mass ratio merger in the plane of the sky. There is evidence of composite structure in subcluster 2 as presented in \citet{vanWeeren:2017}, where a small subcluster has created a shock that reaccelerated particles associated with an AGN.}
\label{fig:A17}
\end{figure*}

\subsection{CIZA J2242.8+5301}
\label{subsec:CIZAJ2242}
\input{CIZAJ2242.tex}
\begin{figure*}
\centering
\includegraphics[height=8.25in]{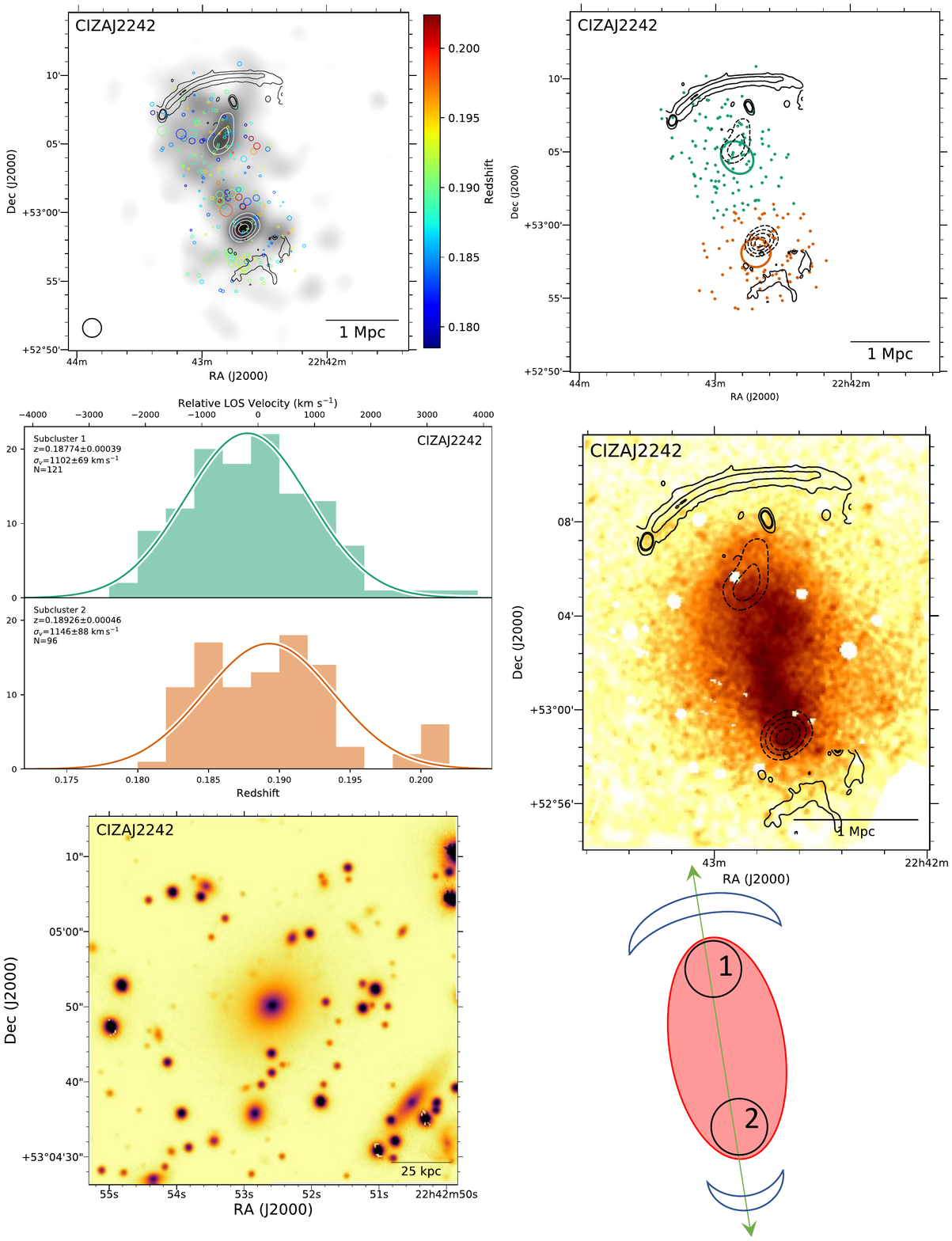}
\caption{\emph{Top left:} DS-test over red sequence Subaru/SuprimeCam i-band luminosity distribution with linearly space white contours. The black contours show WSRT 1382 MHz contours \citep{vanWeeren:2010}. \emph{Top right:} Subcluster membership of spectroscopic cluster members assigned with our GMM analysis. The same WSRT contours (solid) and red sequence luminosity contours (dashed) are depicted. \emph{Middle left:} Subcluster redshift histogram with colors matching the image at the top right. \emph{Middle right:} 206 ks Chandra image with the same WSRT radio contours (solid) and red sequence luminosity contours (dashed). \emph{Bottom left:} 150 kpc cutout of the Subaru i-band image centered on the cluster BCG in subcluster 1. \emph{Bottom right:} Preferred merger scenario. CIZAJ2242 is composed of at least two subclusters with similar velocity dispersions and redshifts suggesting a 1:1 merger in the plane of the sky. The ICM has been extremely disrupted and is situated between the two subclusters. The famous `Sausage' radio relic in the north has been studied across a wide range of frequencies, and dark matter-galaxy offsets have been detected in this cluster \citep{Jee:2016}. The high mass, simple geometry, and extreme shock features have motivated a number of simulation studies as well.}
\label{fig:A18}
\end{figure*}

\subsection{MACS J1149.5+2223}
\label{subsec:MACSJ1149}
\input{MACSJ1149.tex}

\begin{figure*}
\centering
\includegraphics[height=8.25in]{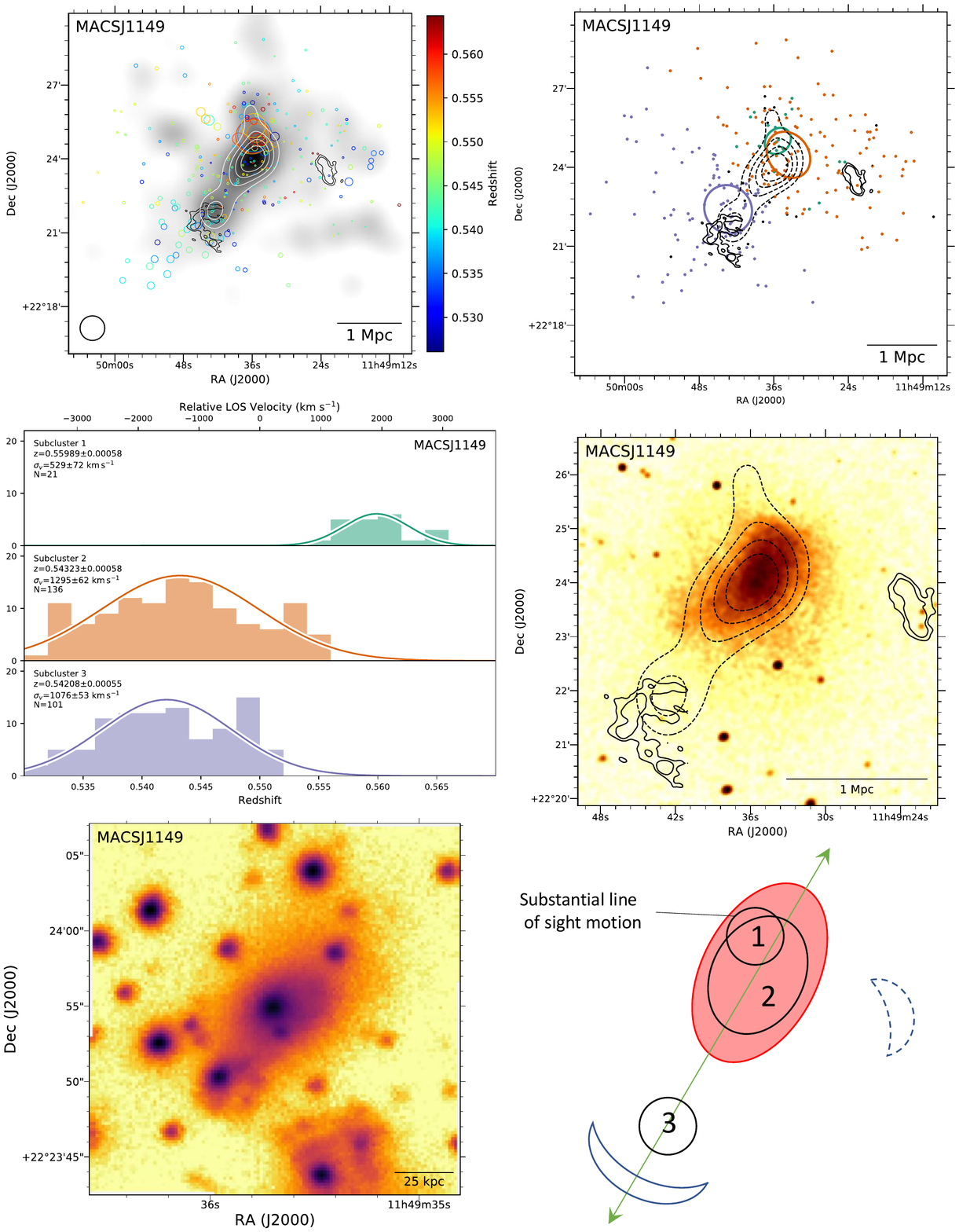}
\caption{\emph{Top left:} DS-test over red sequence Subaru/SuprimeCam R-band luminosity distribution with linearly space white contours. The black contours show GMRT 323 MHz contours \citep{Bonafede:2012}. \emph{Top right:} Subcluster membership of spectroscopic cluster members assigned with our GMM analysis. The same GMRT contours (solid) and red sequence luminosity contours (dashed) are depicted. \emph{Middle left:} Subcluster redshift histogram with colors matching the image at the top right. \emph{Middle right:} 372 ks Chandra image with the same GMRT radio contours (solid) and red sequence luminosity contours (dashed). \emph{Bottom left:} 150 kpc cutout of the Subaru R-band image centered on the cluster BCG in subcluster 2. \emph{Bottom right:} Preferred merger scenario. A detailed dynamical study of MACSJ1149 was presented in \citet{Golovich:2016}.}
\label{fig:A19}
\end{figure*}

\subsection{MACS J1752.0+4440}
\label{subsec:MACSJ1752}
\input{MACSJ1752.tex}

\begin{figure*}
\centering
\includegraphics[height=8.25in]{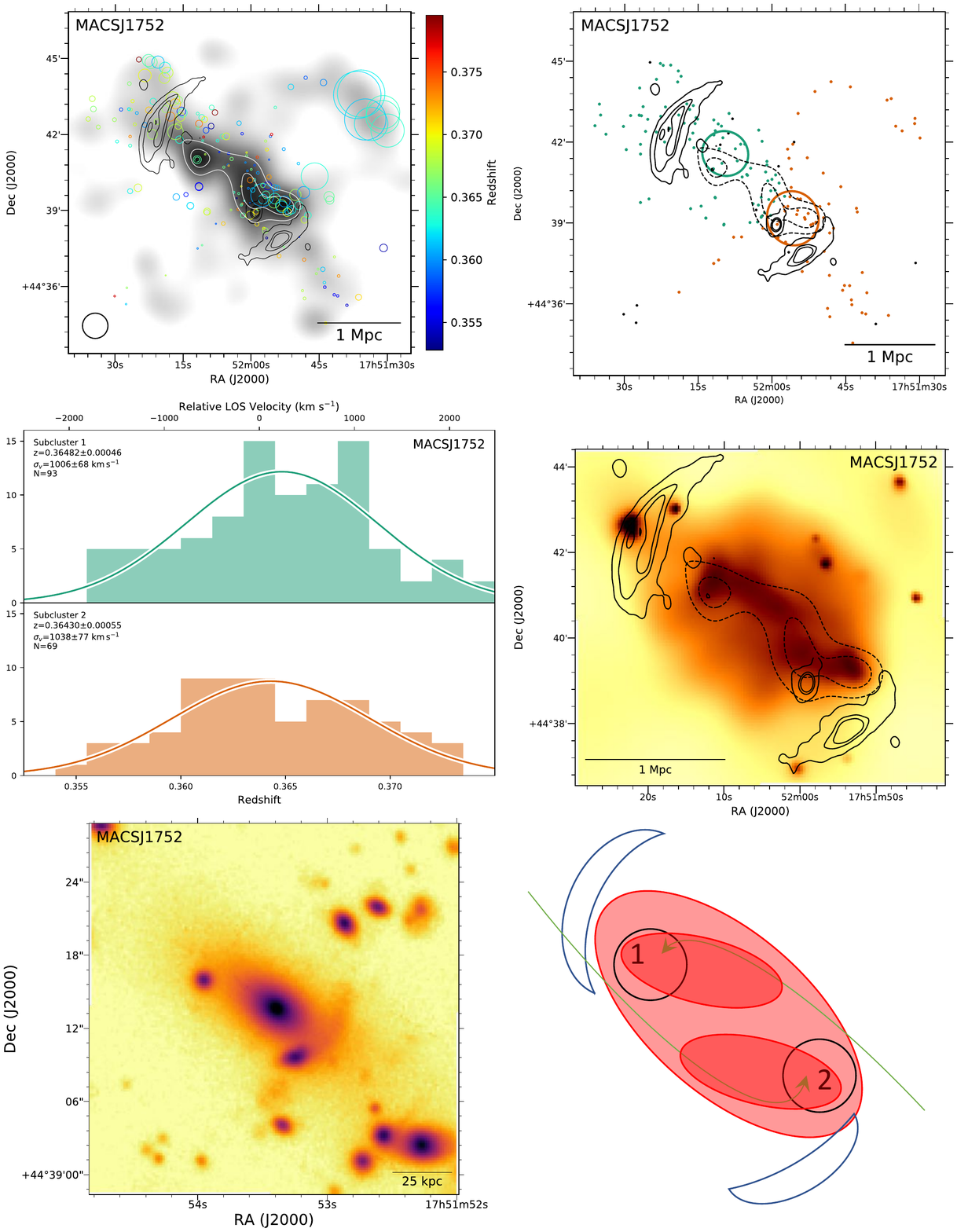}
\caption{\emph{Top left:} DS-test over red sequence Subaru/SuprimeCam i-band luminosity distribution with linearly space white contours. The black contours show WSRT 1.4 GHz contours \citep{vanWeeren:2012b}. \emph{Top right:} Subcluster membership of spectroscopic cluster members assigned with our GMM analysis. The same WSRT contours (solid) and red sequence luminosity contours (dashed) are depicted. \emph{Middle left:} Subcluster redshift histogram with colors matching the image at the top right. \emph{Middle right:} 13 ks XMM image with the same WSRT radio contours (solid) and red sequence luminosity contours (dashed). \emph{Bottom left:} 150 kpc cutout of the Subaru i-band image centered on the cluster BCG in subcluster 2. \emph{Bottom right:} Preferred merger scenario. The X-ray image shows evidence for two `bullets' that just missed and are beginning to orbit one another counter-clockwise. There two relics are bright and suggest a major merger despite not being head-on. The two subclusters have similar velocity dispersions and line of sight velocities suggesting the orbital plane is perpendicular to the line of sight.}
\label{fig:A20}
\end{figure*}

\subsection{PLCK G287.0+32.9}
\label{subsec:PLCKG287}
\input{PLCKG287.tex}

\begin{figure*}
\centering
\includegraphics[height=8.25in]{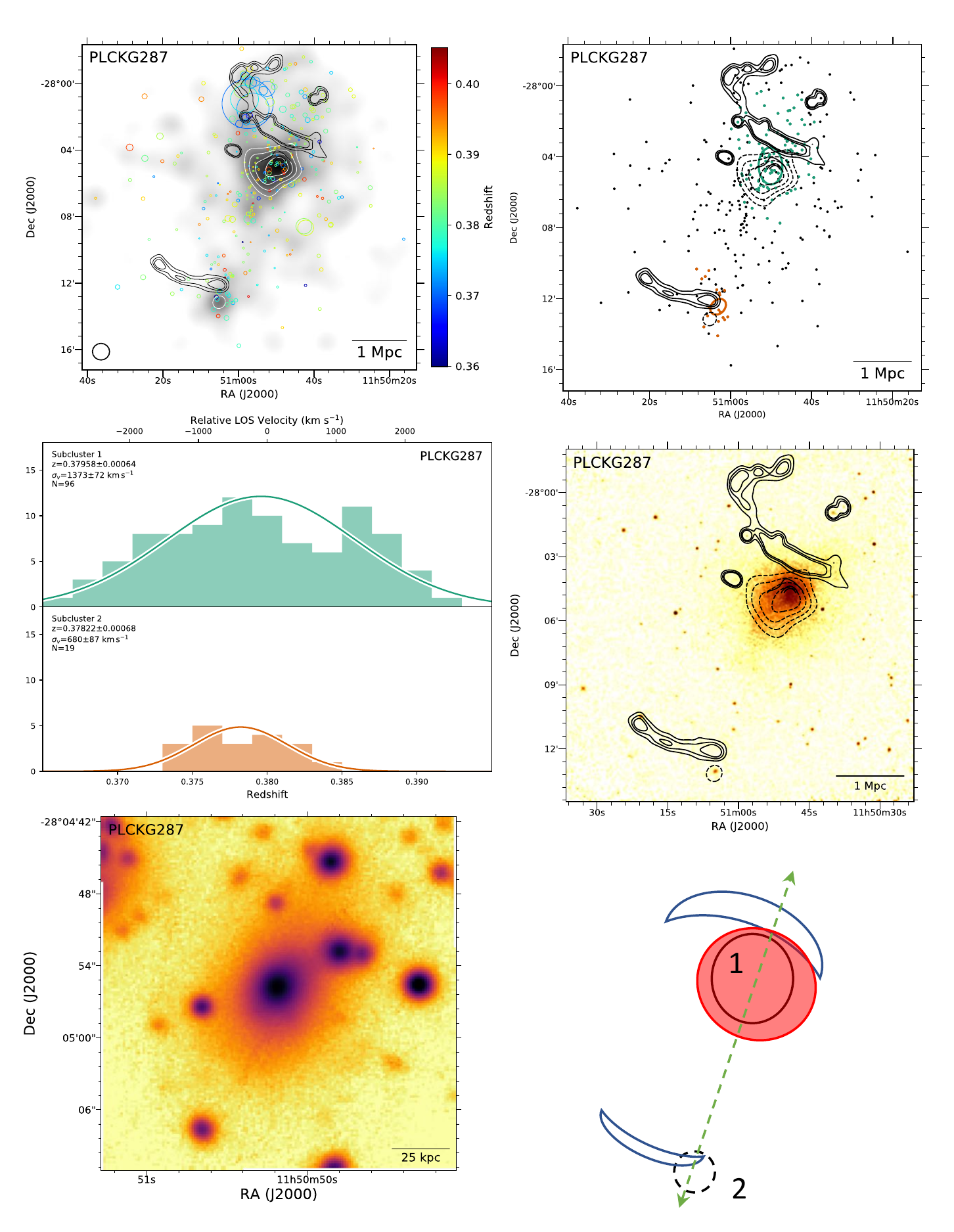}
\caption{\emph{Top left:} DS-test over red sequence Subaru/SuprimeCam r-band luminosity distribution with linearly space white contours. The black contours show GMRT 325 GHz contours \citep{Bonafede:2014}. \emph{Top right:} Subcluster membership of spectroscopic cluster members assigned with our GMM analysis. The same GMRT contours (solid) and red sequence luminosity contours (dashed) are depicted. \emph{Middle left:} Subcluster redshift histogram with colors matching the image at the top right. \emph{Middle right:} 200 ks Chandra image with the same GMRT radio contours (solid) and red sequence luminsity contours (dashed). \emph{Bottom left:} 150 kpc cutout of the Subaru i-band image centered on the cluster BCG in subcluster 1. \emph{Bottom right:} Preferred merger scenario. The merger scenario is unclear. There is evidence for a small subcluster in the south near the southern radio relic; however, it leads the relic. There is disturbed X-ray emission streaming to the south of subcluster 1, which could motivate this merger scenario. A detailed weak gravitational lensing analysis detected composite structure in subcluster 1 \citep{Finner:2017}, and more complex scenarios have been proposed \citep{Bonafede:2014}.}
\label{fig:A21}
\end{figure*}

\subsection{PSZ1 G108.18-11.53}
\label{subsec:PSZ1G108}
\input{PSZ1G108.tex}

\begin{figure*}
\centering
\includegraphics[height=8.25in]{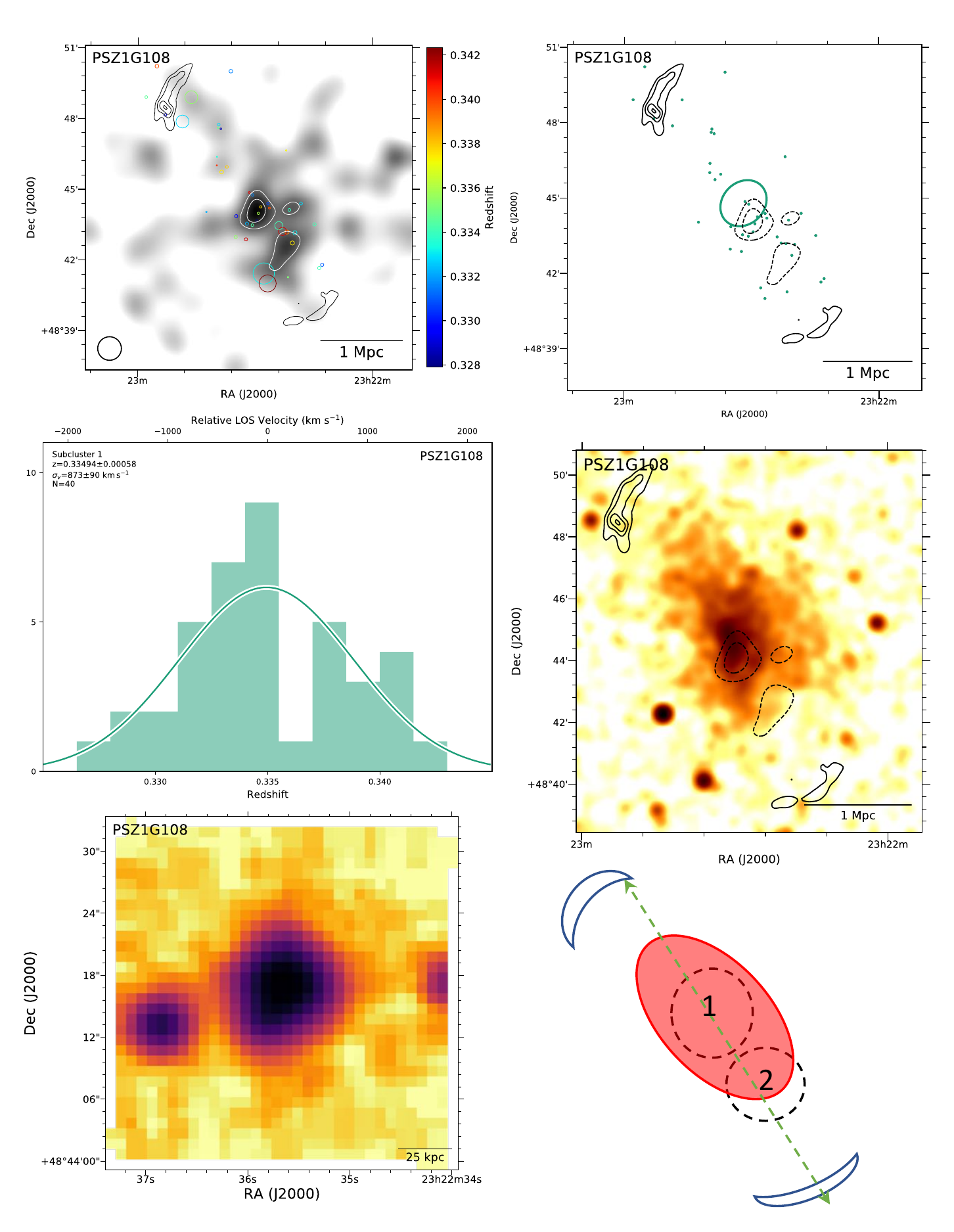}
\caption{\emph{Top left:} DS-test over red sequence DSS R-band luminosity distribution with linearly space white contours. The black contours show GMRT 323 GHz contours \citep{deGasperin:2015}. \emph{Top right:} Subcluster membership of spectroscopic cluster members assigned with our GMM analysis. The same GMRT contours (solid) and red sequence luminosity contours (dashed) are depicted. \emph{Middle left:} Subcluster redshift histogram with colors matching the image at the top right. \emph{Middle right:} 27 ks Chandra image with the same GMRT radio contours (solid) and red sequence luminsity contours (dashed). \emph{Bottom left:} 150 kpc cutout of the DSS R-band image centered on the cluster BCG. \emph{Bottom right:} Preferred merger scenario. We do not have enough spectra to resolve composite structure, but the red sequence luminosity is multi-modal and aligned with the two radio relics and extended X-ray emission. Since the X-ray, radio, and optical light is all aligned, we propose a simple bimodal merger along this axis.}
\label{fig:A22}
\end{figure*}

\subsection{RXC J1053.7+5452}
\label{subsec:RXCJ1053}
\input{RXCJ1053.tex}
\begin{figure*}
\centering
\includegraphics[height=8.25in]{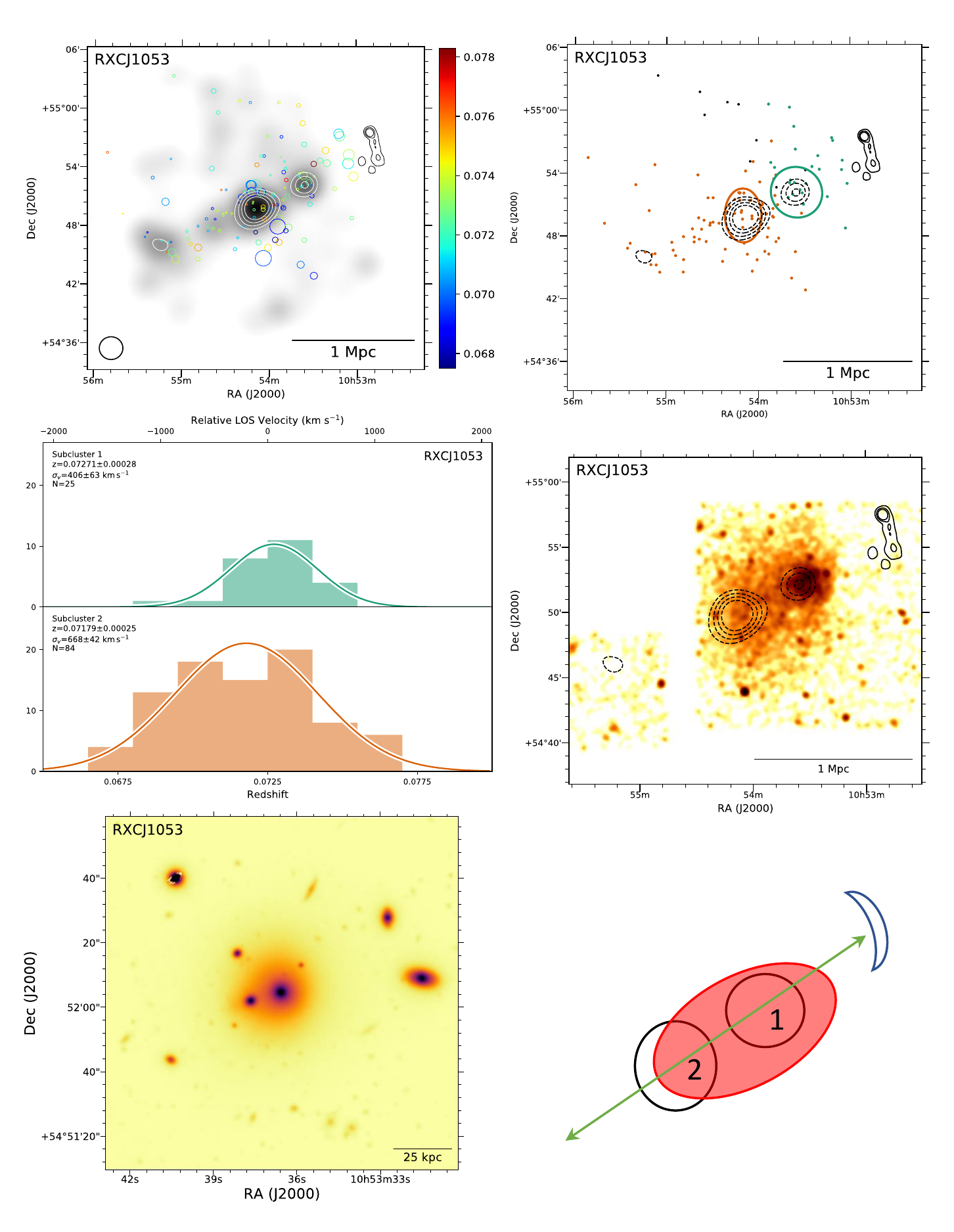}
\caption{\emph{Top left:} DS-test over red sequence Subaru/SuprimeCam r-band luminosity distribution with linearly space white contours. The black contours show WSRT 1382 MHz contours \citep{Bonafede:2014}. \emph{Top right:} Subcluster membership of spectroscopic cluster members assigned with our GMM analysis. The same WSRT contours (solid) and red sequence luminosity contours (dashed) are depicted. \emph{Middle left:} Subcluster redshift histogram with colors matching the image at the top right. \emph{Middle right:} 31 ks Chandra image with the same WSRT radio contours (solid) and red sequence luminosity contours (dashed). \emph{Bottom left:} 150 kpc cutout of the Subaru r-band image centered on the cluster BCG in subcluster 1. \emph{Bottom right:} Preferred merger scenario. The two subclusters and radio relic are aligned with the extended X-ray emission. Most of the mass appears to be associated with subcluster 2; however, the X-ray emission is associated with subcluster 1, which is perhaps a signature of a cool-core remnant. Deeper X-ray observations are necessary to reveal the nature of the ICM.}
\label{fig:A23}
\end{figure*}

\subsection{RXCJ1314.4-2515}
\label{subsec:RXCJ1314}
\input{RXCJ1314.tex}
\begin{figure*}
\centering
\includegraphics[height=8.25in]{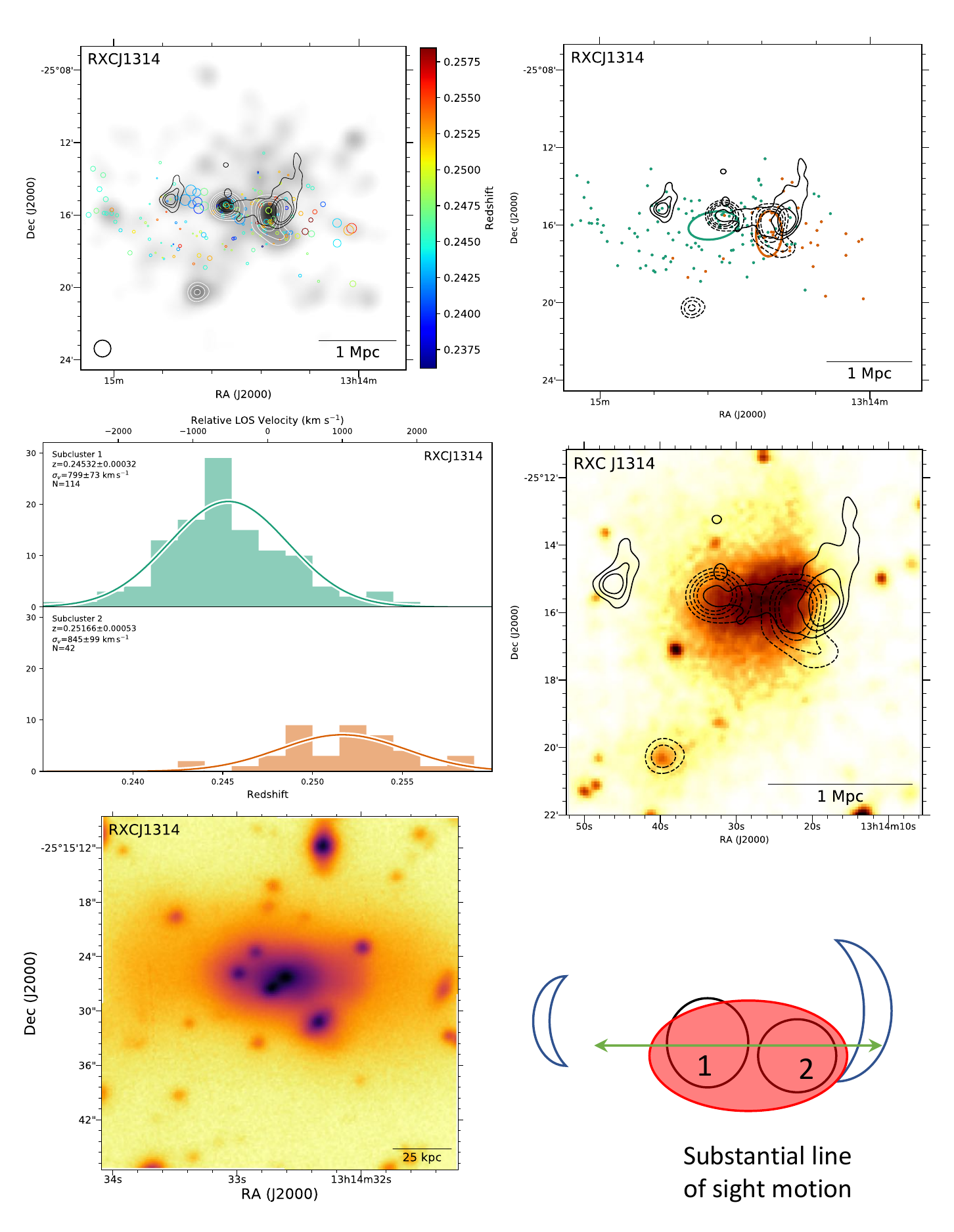}
\caption{\emph{Top left:} DS-test over red sequence Subaru/SuprimeCam r-band luminosity distribution with linearly space white contours. The black contours show VLA 2-4 GHz contours (P.I. Dawson). \emph{Top right:} Subcluster membership of spectroscopic cluster members assigned with our GMM analysis. The same VLA contours (solid) and red sequence luminosity contours (dashed) are depicted. \emph{Middle left:} Subcluster redshift histogram with colors matching the image at the top right. \emph{Middle right:} 110 ks XMM image with the same VLA radio contours (solid) and red sequence luminosity contours (dashed). \emph{Bottom left:} 150 kpc cutout of the Subaru r-band image centered on the cluster BCG in subcluster 1. \emph{Bottom right:} Preferred merger scenario. The two subclusters and radio relics are aligned with the extended X-ray emission. A large line of sight velocity difference between the two subclusters is discovered suggesting a substantial component of the merger axis is along the line of sight, which is an outlier among other bimodal, double relic systems. A third subcluster is detected to the southeast, in the red sequence luminosity and X-ray surface brightness. It may have interacted in a glancing collision given the stream of X-ray emission connecting it to the cluster center.}
\label{fig:A24}
\end{figure*}

\subsection{ZwCl 0008.8+5215}
\label{subsec:ZwCl0008}
\input{ZwCl0008.tex}
\begin{figure*}
\centering
\includegraphics[height=8.25in]{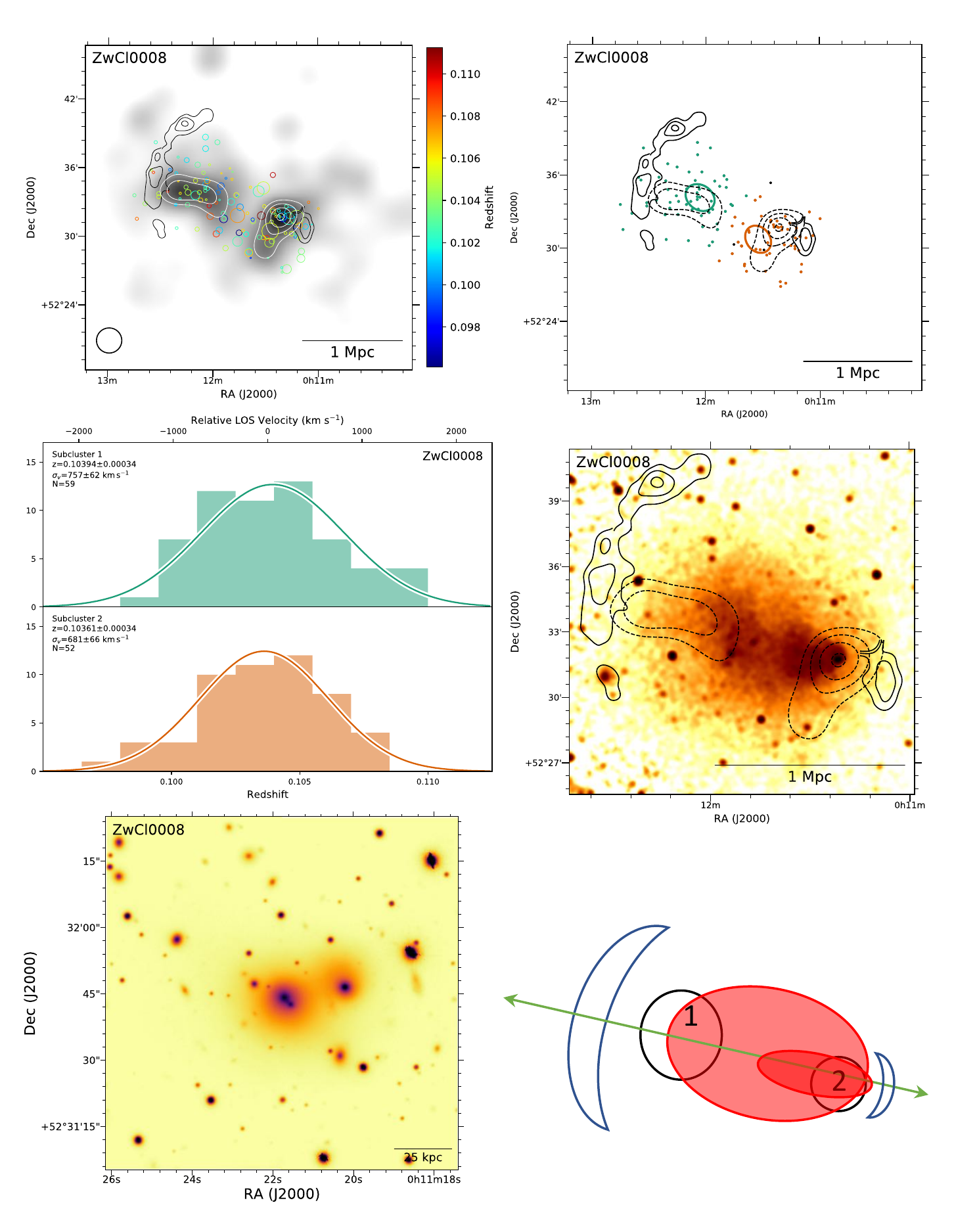}
\caption{\emph{Top left:} DS-test over red sequence Subaru/SuprimeCam r-band luminosity distribution with linearly space white contours. The black contours show WSRT 1382 MHz contours \citep{vanWeeren:2011c}. \emph{Top right:} Subcluster membership of spectroscopic cluster members assigned with our GMM analysis. The same WSRT contours (solid) and red sequence luminosity contours (dashed) are depicted. \emph{Middle left:} Subcluster redshift histogram with colors matching the image at the top right. \emph{Middle right:} 411 ks Chandra image with the same WSRT radio contours (solid) and red sequence luminosity contours (dashed). \emph{Bottom left:} 150 kpc cutout of the Subaru r-band image centered on the cluster BCG in subcluster 2. \emph{Bottom right:} Preferred merger scenario. The two subclusters and radio relics are aligned with the extended X-ray emission. A `bullet' like cool-core remnant and wake feature are associated with subcluster 2, which is traveling to the west. The cluster exhibits simple, bimodal geometry indicative of the Bullet Cluster.}
\label{fig:A25}
\end{figure*}

\subsection{ZwCl 1447+2619}
\label{subsec:ZwCl1447}
\input{ZwCl1447.tex}
\begin{figure*}
\centering
\includegraphics[height=8.25in]{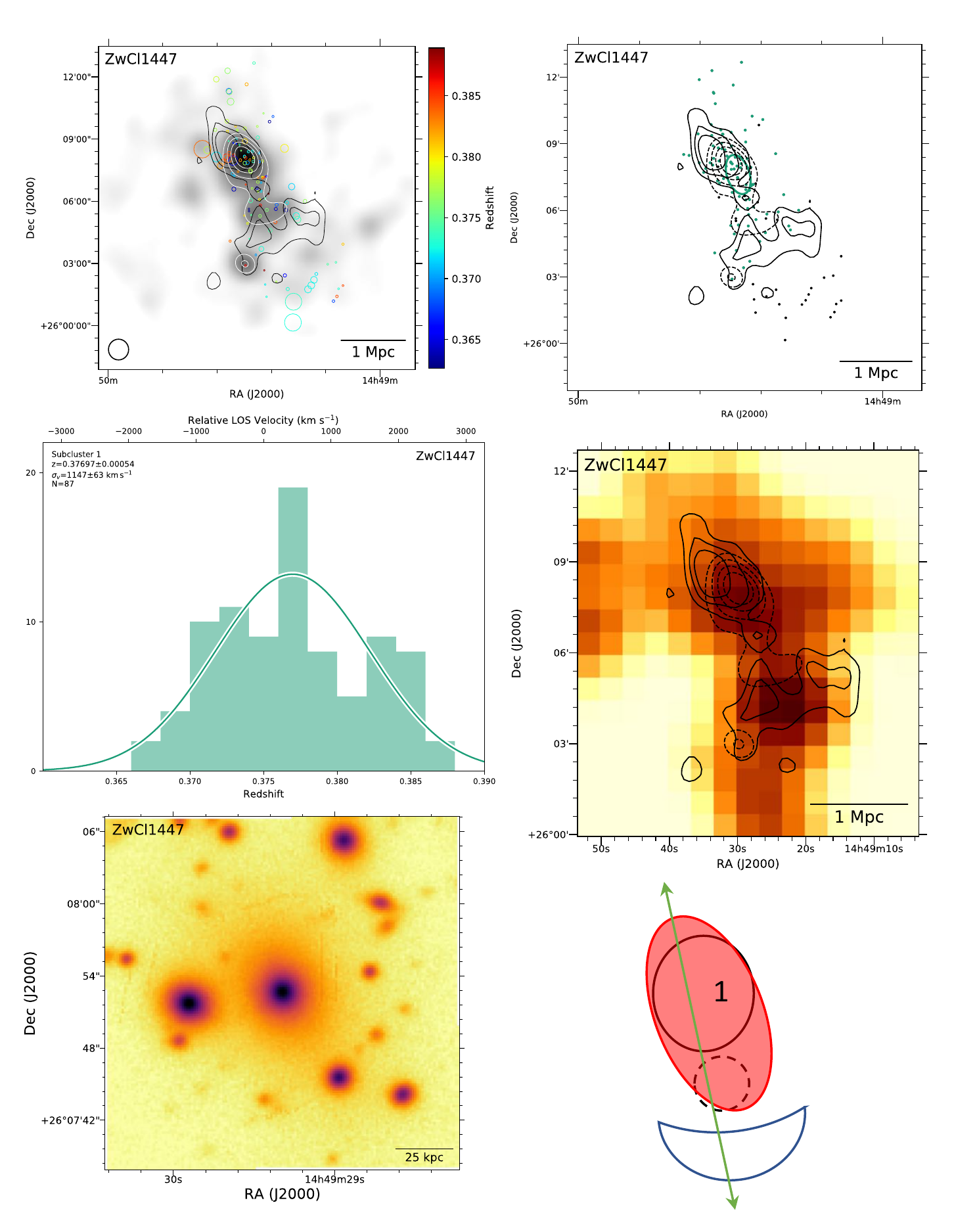}
\caption{\emph{Top left:} DS-test over red sequence Subaru/SuprimeCam r-band luminosity distribution with linearly space white contours. The black contours show VLA 1.4 GHz contours \citep{Govoni:2012}. \emph{Top right:} Subcluster membership of spectroscopic cluster members assigned with our GMM analysis. The same VLA contours (solid) and red sequence luminosity contours (dashed) are depicted. \emph{Middle left:} Subcluster redshift histogram with colors matching the image at the top right. \emph{Middle right:} 21 ks ROSAT image with the same VLA radio contours (solid) and red sequence luminosity contours (dashed). \emph{Bottom left:} 150 kpc cutout of the Subaru r-band image centered on the cluster BCG. \emph{Bottom right:} Preferred merger scenario. The GMM did not detect two subclusters; however, the red sequence luminosity is elongated along the same axis as the X-ray surface brightness and radio relics. We propose a north-south merger axis. Further spectroscopy in the south is necessary to better constrain the subcluster velocities.}
\label{fig:A26}
\end{figure*}

\begin{figure*}
\centering
\includegraphics[height=8.25in]{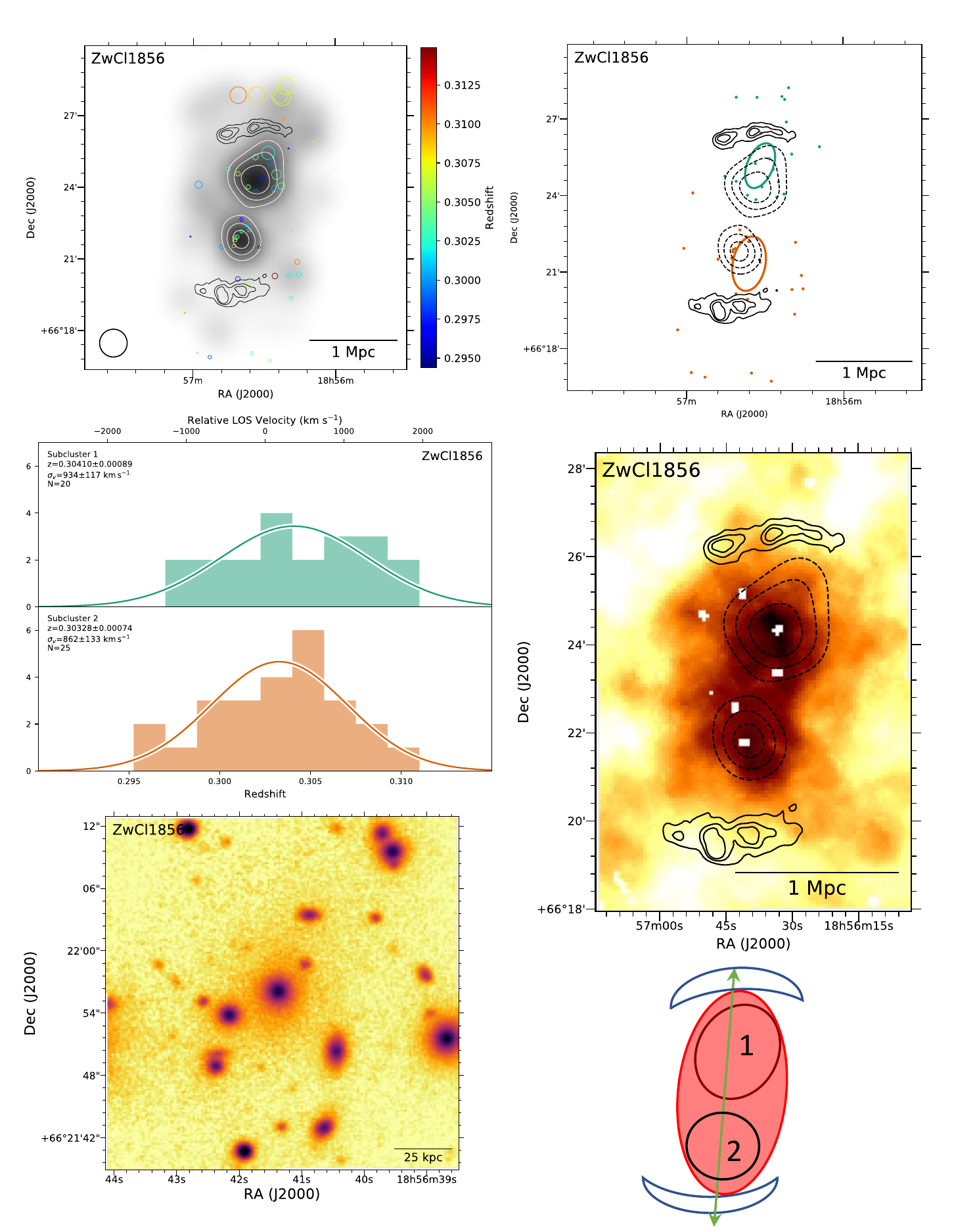}
\caption{\emph{Top left:} DS-test over red sequence Gemini/GMOS r-band luminosity distribution with linearly space white contours. The black contours show WSRT 1382 GHz contours \citep{deGasperin:2014}. \emph{Top right:} Subcluster membership of spectroscopic cluster members assigned with our GMM analysis. The same WSRT contours (solid) and red sequence luminosity contours (dashed) are depicted. \emph{Middle left:} Subcluster redshift histogram with colors matching the image at the top right. \emph{Middle right:} 12 ks XMM image with the same WSRT radio contours (solid) and red sequence luminosity contours (dashed). \emph{Bottom left:} 150 kpc cutout of the Gemini r-band image centered on the cluster BCG in subcluster 2. \emph{Bottom right:} Preferred merger scenario. We detect two subclusters aligned with the X-ray surface brightness elongation and two radio relics suggesting a simple bimodal merger along a north-south axis. The two subclusters have similar velocity dispersions and redshifts; although, these are based on few galaxies per subcluster.}
\label{fig:A27}
\end{figure*}

\begin{figure*}
\centering
\includegraphics[height=8.25in]{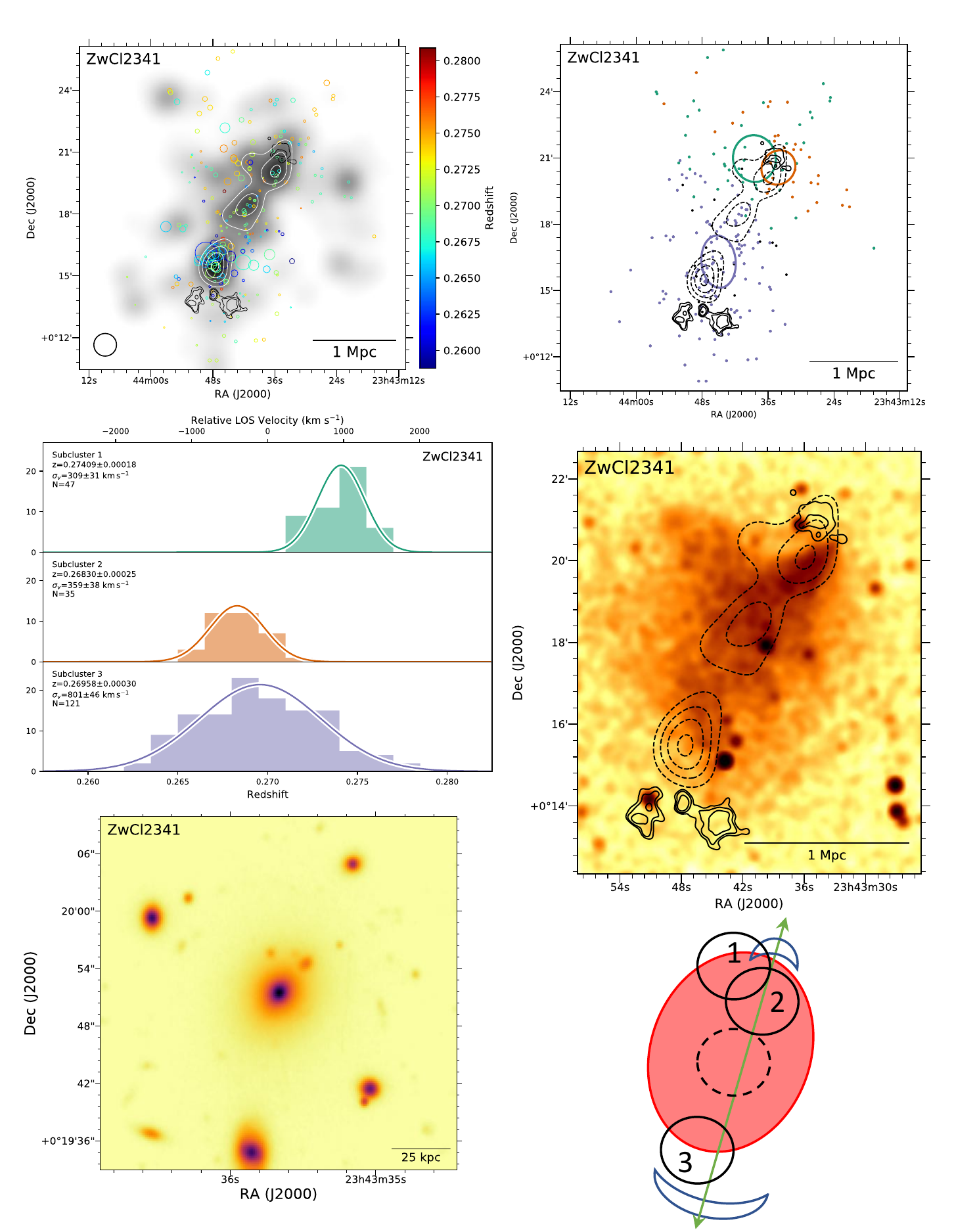}
\caption{\emph{Top left:} DS-test over red sequence Subaru/SuprimeCam r-band luminosity distribution with linearly space white contours. The black contours show GMRT 610 MHz contours \citep{vanWeeren:2009}. \emph{Top right:} Subcluster membership of spectroscopic cluster members assigned with our GMM analysis. The same WSRT contours (solid) and red sequence luminosity contours (dashed) are depicted. \emph{Middle left:} Subcluster redshift histogram with colors matching the image at the top right. \emph{Middle right:} 227 ks Chandra image with the same GMRT radio contours (solid) and red sequence luminosity contours (dashed). \emph{Bottom left:} 150 kpc cutout of the Subaru r-band image centered on the cluster BCG in subcluster 1. \emph{Bottom right:} Preferred merger scenario. We detect three subclusters with our GMM as well as a fourth that is evident in the red sequence luminosity distribution. Two subclusters (subclusters 2 and 3) and the candidate third situated in the middle of the cluster are aligned with the two radio relics. This suggests the dominant component of the merger is occurring along this axis with at least two subclusters partaking. Subcluster 1 is positioned to the east of subcluster 2, and is evident in both the X-ray surface brightness profile and red sequence luminosity. The disturbed gas suggests it has already merged, and the large line of sight velocity difference suggests line of sight motion.}
\label{fig:A28}
\end{figure*}

\subsection{ZwCl 1856+6616}
\label{subsec:ZwCl1856}
\input{ZwCl1856.tex}

\subsection{ZwCl 2341+0000}
\label{subsec:ZwCl2341}
\input{ZwCl2341.tex}


\section{Results: Sample}\label{sec:results_sample}

In the previous section, we discussed the results of the photometric and spectroscopic survey for 29 merging clusters. We used these results in conjunction with radio relic and X-ray surface brightness morphology to infer the most likely merger scenario that launched the shocks associated with the radio relics in each system. In this section, we compile results and study the ensemble characteristics of radio selection with respect to the velocity and geometry information. 

\subsection{Line of Sight Motion}

Our spectroscopic survey and GMM analysis on each system allows for direct measure of the line of sight velocity difference between merging subclusters. Simply put, if we assume the radio relics indicate a merger is occurring, a large line of sight velocity difference between subclusters indicates a merger axis along the line of sight while a small line of sight velocity difference indicates a merger occurring within the plane of the sky and/or near apocenter. 

To estimate the velocity difference, we draw random subcluster members (with replacement) from the subcluster histograms (see Figure \ref{fig:subclusthist} and the analogous figures in the appendix) and calculate the line of sight velocity difference for each bootstrap. For the $i$th bootstrap from each subcluster, \begin{equation}
v_{i_{1,2}}=\frac{(1+z_{i_{1,2}})^{2}-1}{(1+z_{i_{1,2}})^{2}+1} c,
\end{equation} where c is the speed of light. The line of sight velocity difference is then, 
\begin{equation}
\Delta v_{\text{LOS}} = \frac{|v_{i_{1}}-v_{i_{2}}|}{1-\frac{v_{i_{1}}v_{i_{2}}}{c^{2}}}.
\end{equation} We take the median of this distribution after 10,000 samples as the subcluster line of sight velocity difference and the 95\% confidence interval for the remaining analysis. 

Three systems appear to be bimodal based on all the evidence, yet are considered unimodal by the BIC test due to insufficient spectroscopic sampling. For these ``single halo'' clusters (A1300, A2061, ZwCl1447), we define the line of sight velocity difference by the BCG redshifts and assume a 200 km s$^{-1}$ uncertainty, which is in line with spectroscopic studies of BCG peculiar velocities \citep[e.g.,][]{Oegerle:2001}. For three and four subcluster systems, we look for a preferred merger axis in the X-ray surface brightness map, the red sequence luminosity distribution, GMM substructure, and the radio relic(s). In A2744, no pair of subclusters aligned with the radio relic; however, subcluster one, in the north, has an east to west elongated red sequence luminosity distribution, which has a bright galaxy at either end. We use these two galaxies to define the line of sight velocity difference and assume 200 km s$^{-1}$ uncertainty. 

Note that we eliminated several systems from this analysis: A746 due to the extreme stellar contamination, which makes it impossible to map the red sequence luminosity distribution; A2255 due to the ambiguous merger axis across all wavelengths; A2345 due to the complex merger where two separate mergers may have generated separate radio relics \citep{Boschin:2010}; finally, PSZ1G108 due to insufficient optical imaging quality to map the red sequence luminosity distribution. This leaves 25 systems with estimates of the line of sight velocity difference between merging subclusters. 

We then calculated the expected free-fall velocity from infinity ($v_{\text{max}}$) of two truncated NFW \citep{NFW} halos with mass determined using the $M_{200}$\,--\,$\sigma_{v}$ scaling relations of \citet{Evrard:2008} assuming the mass--concentration relation of \citet{Duffy}. Following \citet{Dawson:2012}, we calculate the maximum free-fall velocity using conservation of energy and the potential energy, $V$, of overlapping truncated (at $R_{200}$) NFW profiles, 
\begin{equation}
v_{\text{max}}=\sqrt{-\frac{2}{\mu}V(r=0)}.
\end{equation}

\begin{figure}
\centering
\includegraphics[width=\columnwidth]{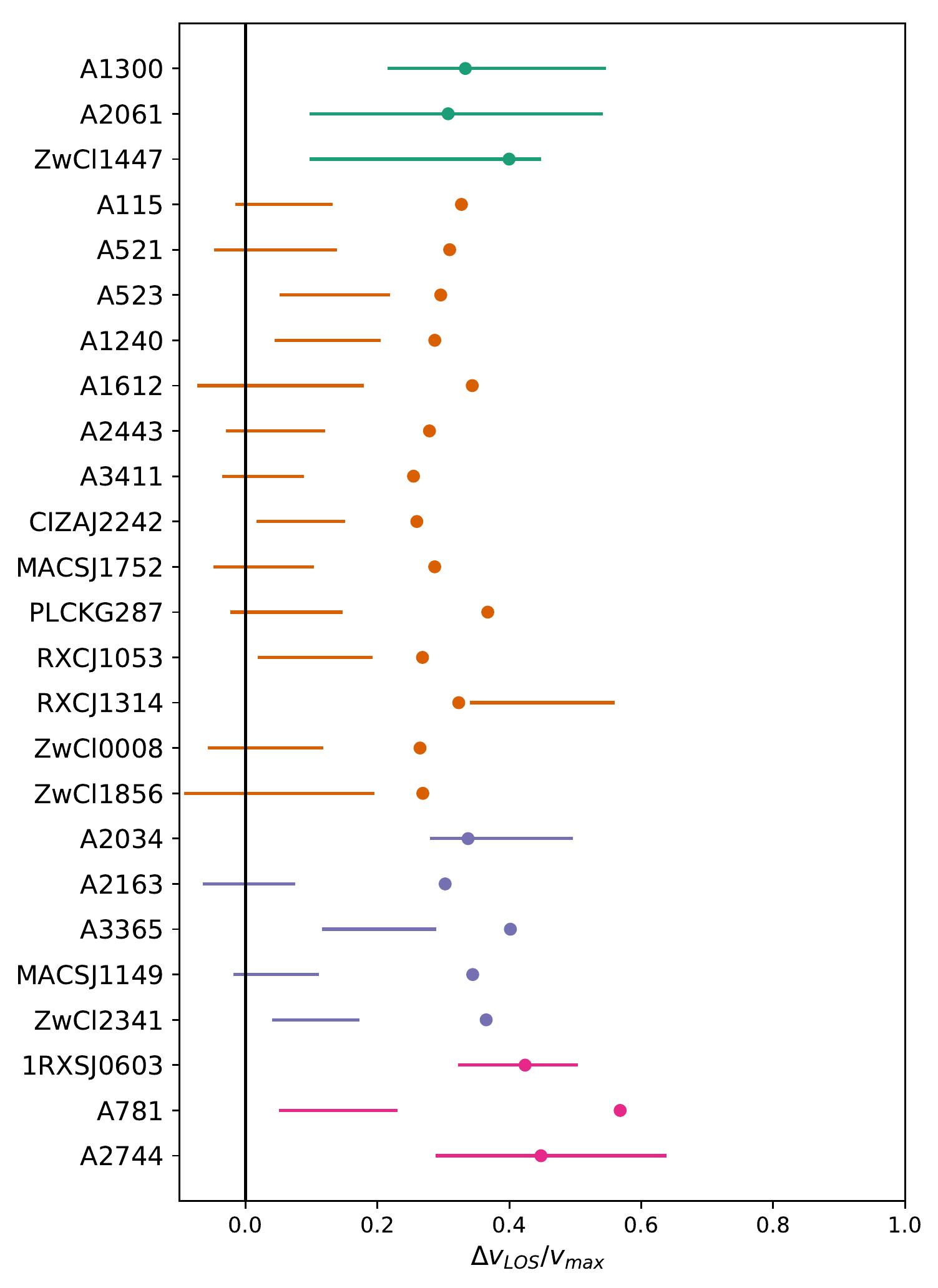}
\caption[Ensemble line of sight velocity difference estimates]{Line of sight velocity differences between merging subclusters associated with radio relics. The clusters are organized by the number of subclusters in the lowest BIC model (see Table \ref{tab:mcmcgmm}). Green, orange, purple, and pink refer to systems comprised of one, two, three, and four GMM subclusters, respectively. The bars represent the 95\% confidence interval for the ratio of line of sight velocity difference to maximum three dimensional infall velocity. The colored points are the ratio of the global velocity dispersion to the maximum free-fall velocity, for comparison.}
\label{fig:velocity}

\end{figure}

We plot the ratio $\Delta v_{\text{LOS}} / v_{\text{max}}$ for each system in Figure \ref{fig:velocity}. The 95\% confidence intervals are color-coded according to the number of subclusters identified by our GMM analysis (green, orange, purple, and pink for one, two, three, and four subclusters, respectively). With the exception of a few outliers we see low line of sight velocity differences compared with the free-fall velocity. In particular, with the exception of RXCJ1314, bimodal systems have line of sight velocity differences that are generally consistent with zero at 1\,--\,3$\sigma$. 

For comparison, we plotted the global velocity dispersion of each cluster as a point along the same axis, scaled to the free-fall velocity. This helps explain why the GMM was able to identify substructure in some systems and not in others. First, for the three one-halo systems at the top of Figure \ref{fig:velocity} (A1300, A2061, and ZwCl1447), the red sequence luminosity distribution clearly shows multiple peaks (see white contours in appendix figures analogous to Figure \ref{fig:dstest}), but the projected separation in these systems is small. Also, the spectroscopic completeness in the inner regions of these systems is insufficient to overcome the BIC penalty for the extra parameters necessary for bimodal models of these systems. The large line of sight velocity difference between the respective BCGs suggests that these mergers have a substantial fraction of their merger axis situated along the line of sight, which should help the GMM identify overlapping substructures; however, the velocity difference is not larger than the velocity dispersion of the cluster velocity distribution. In the case of RXCJ1314, which has a similar orientation to the three one-halo clusters at hand, the velocity difference is larger than the velocity dispersion, so the BIC penalty for extra model parameters does not outweigh the likelihood of having subclusters with a large line of sight velocity difference. The purple and pink points at the bottom of Figure \ref{fig:velocity} represent more complex systems. Relics may preferentially select for mergers occurring in the plane of the sky (as suggested by the results for the bimodal systems); however, this simple geometry could be washed out by more complex three and four body orbits by the time of our observations several hundred Myr after pericenter. Despite this, we find small line of sight velocity differences between the selected merging pairs in five of the eight most complex systems in our ensemble. However, the interpretation of these merger scenarios is difficult without detailed simulations. 

The results presented in Figure \ref{fig:velocity} also explain why the global velocity histograms are nearly universally well fit by a single Gaussian as demonstrated in \citetalias{Golovich:2017b}. The line of sight velocity difference between radio selected merging subclusters is generally smaller than the velocity dispersion of the global cluster velocity distribution. 

\subsection{Merger Geometry}

\begin{figure}
\centering
\includegraphics[width=\columnwidth]{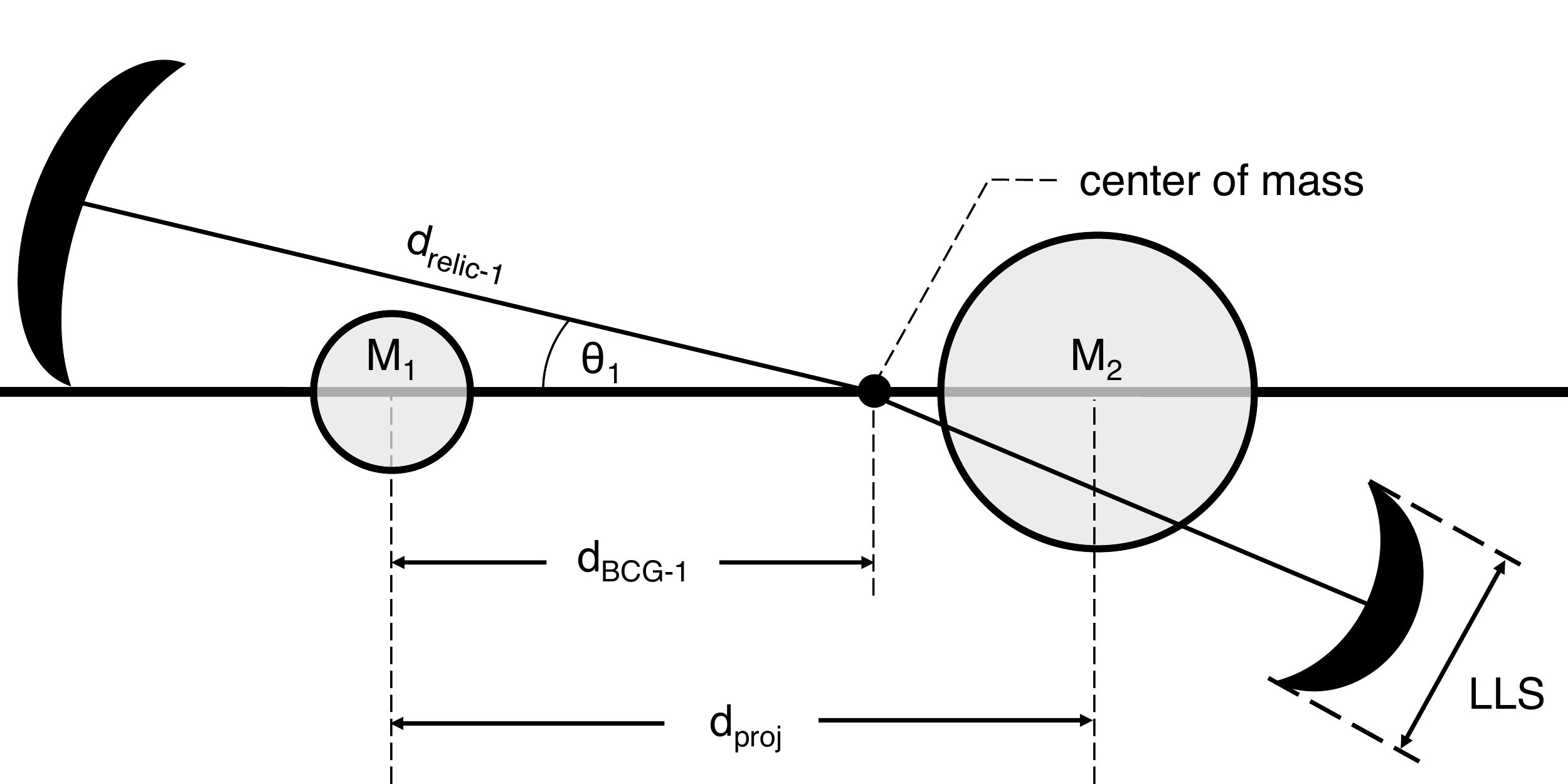}
\caption[Schematic representation of radio relic cluster merger geometry]{Schematic representation of the relic geometry for a double relic system (e.g. A1240). Each relic is located with a position angle and distance from the center of mass of the system determined from the velocity dispersions of subclusters aligned with the radio relics. Single relic systems follow the same conventions. LLS is an abbreviation for the largest linear scale of a radio relic, which is often quoted in radio astronomy literature and has been shown to correlate with the propagation distance of a shock.}
\label{fig:geometry}
\end{figure}

In 28 of 29 systems in our sample (all except A2255), there is evidence of merger activity aligned with the radio relic. For the majority of these, our GMM has identified two substructures within the spectroscopy that are apparently the merging subclusters. In clusters with insufficient spectroscopic coverage to prefer multimodality, the red sequence luminosity distribution generally is elongated along the axis that connects with the radio relic (e.g. A1300, A2061, A2345, A2744, PSZ1G108, and ZwCl1447). In this subsection, we compare the axis connecting the two BCGs of apparent merging substructures with the axis connecting the center of mass to the center of the radio relic. The basic geometry under examination is presented in Figure \ref{fig:geometry}. 

For clusters without GMM identification of substructure, we connect the brightest galaxies at the ends of the extension of the red sequence luminosity distribution. We withhold A2443 from this analysis because it is a known radio phoenix. This type of radio relic was shown to bias similar analyses \citep{vanWeeren:2009b, deGasperin:2014}. Also, we withhold A746 due to the inability to identify substructure due to the bright star in the northwest. We also remove PSZ1G108 due to the lack of good quality optical imaging for galaxy identification. Finally, we withhold A2255 and A2345 due to the complexity of the mergers and inability to define the merger geometries. 

\begin{figure*}
\centering
\includegraphics[width=\textwidth]{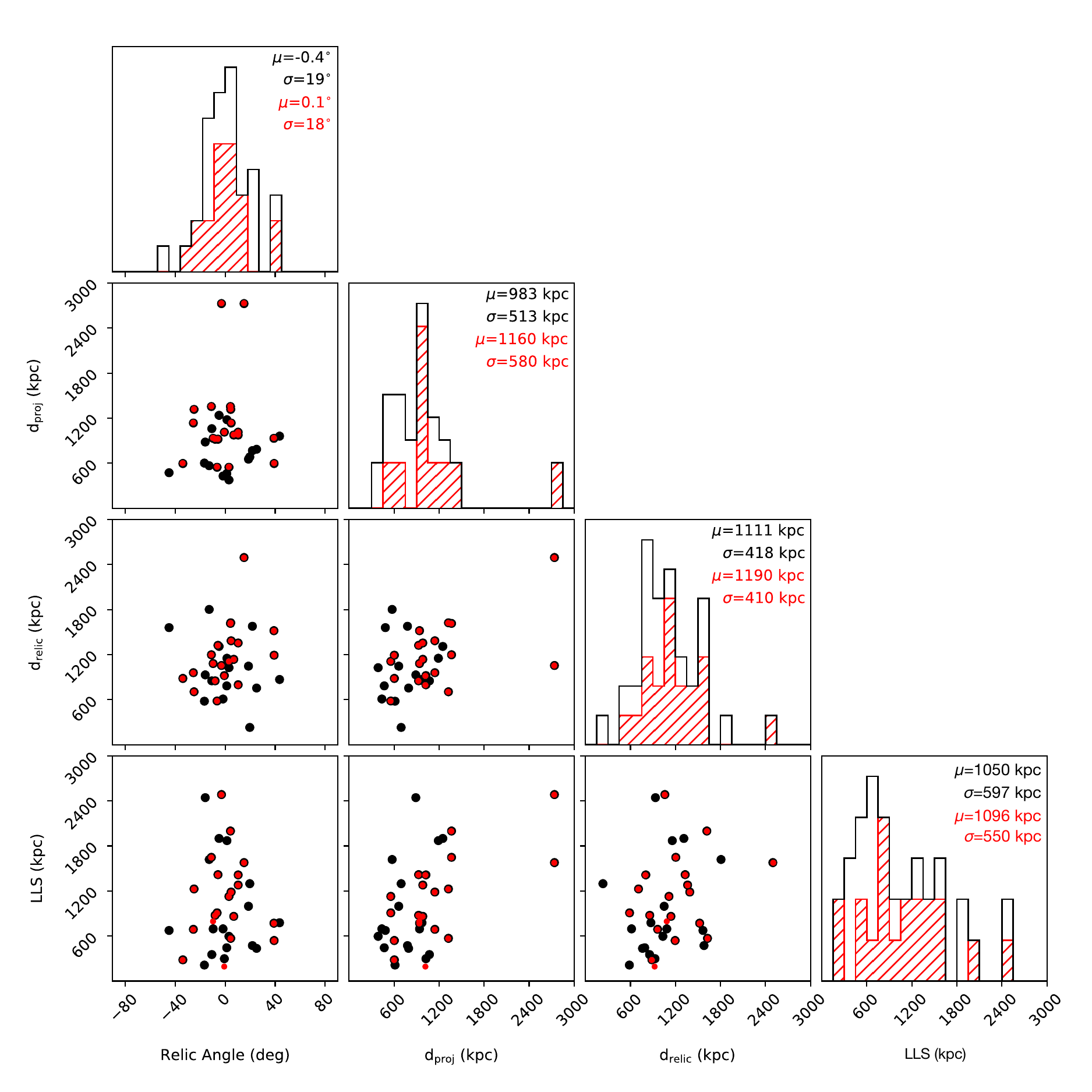}
\caption{Radio relic geometry for 35 radio relics in 24 merging galaxy clusters. Relics tend to be well aligned with the axis connecting BCGs of merging subclusters. By convention here, negative angles imply clockwise angles from the BCG axis toward the relic position vector. The black distribution is for all relics, and the red distribution is for double relic clusters only. See Figure \ref{fig:geometry} for a schematic representation of the defined geometry.}
\label{fig:relics}
\end{figure*}

In Figure \ref{fig:relics} we present the distribution of defined geometrical quantities for the 35 relics across 24 merging clusters. Radio relics are generally very well aligned with the BCG separation axis. Furthermore, with the exception of PLCKG287, radio relic systems have a narrow range of projected BCG separations (d$_{proj}$) and relic distances (from the center of mass). This makes PLCKG287 a particularly interesting system to model as it is clearly an outlier among radio relic systems. Without PLCKG287, the BCG separation and relic distance distributions tighten to 877$\pm$288 kpc and 1070$\pm$349 kpc. 

In the middle panel of Figure \ref{fig:relics}, a weak trend between BCG projected separation and relic distance is apparent. This is expected for the outbound phase of the merger; however, post-apocenter, the BCGs return to smaller projected separations while the radio relics continue to propagate outwards. Detailed analysis is required to determine if a given system is in the outbound or returning phase \citep[see e.g.][]{Ng:2015}. This analysis is beyond the scope of this paper. However, a simple picture of shock propagation does elucidate the methods of \citet{Ng:2015}. The shock propagation speed is proportional to the mass of the opposite subcluster \citep{springel2007}. The canonical example is that of the Bullet Cluster, where the lower mass ``bullet'' has merged from east to west at a high velocity, and the shock front is just ahead of it. The shock on the reverse side of the cluster is much closer to the center of mass. A good analogy is a piston punching through the ICM of the main halo driving the shock ahead, while the more massive cluster is moving slower with respect to the center of mass and thus doesn't propel its shock as fast. 

\begin{figure}
\centering
\includegraphics[width=\columnwidth]{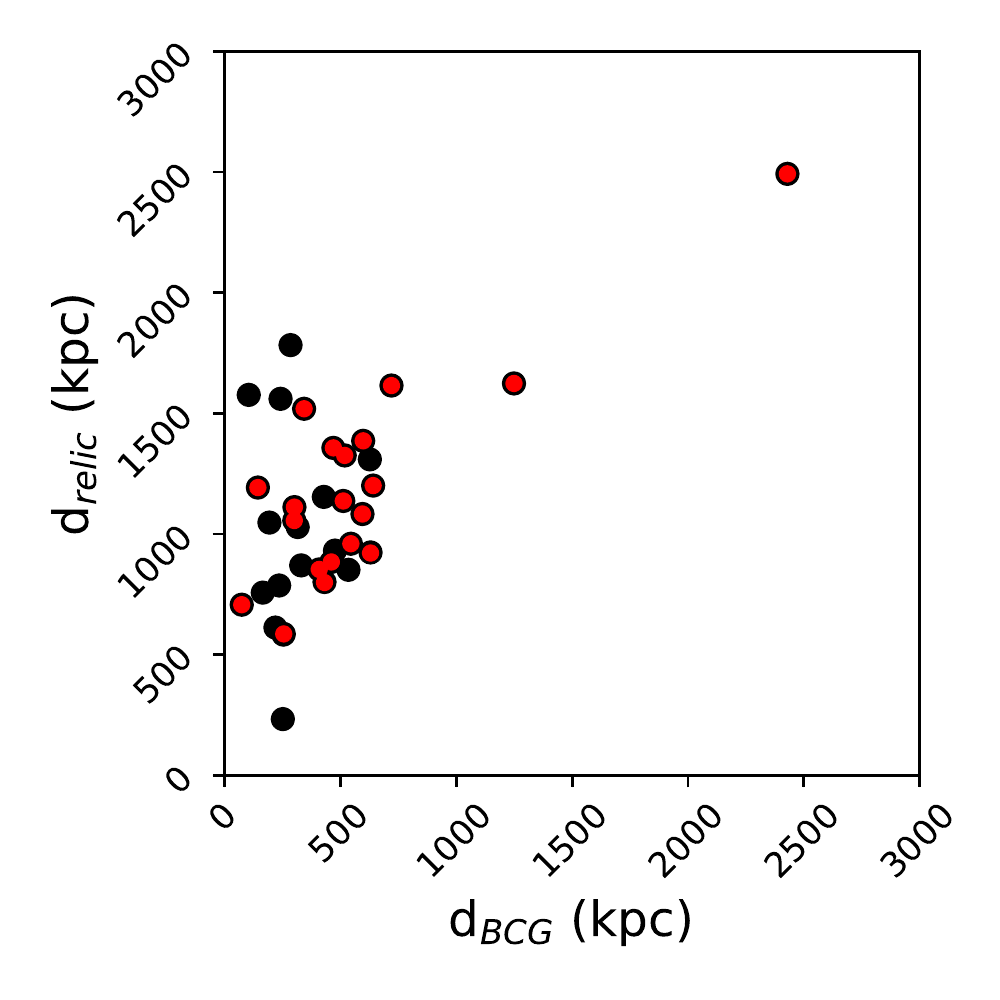}
\caption{Projected distances from the center of mass to radio relics versus projected distance from the center of mass to the opposite subcluster BCG for 35 radio relics from 24 merging clusters (black) and 22 radio relics from 11 double radio relic systems (red). We find a correlation between these two parameters as expected for a model of relic propagation in the center of mass frame.}
\label{fig:COM}

\end{figure}

To check for this, we compare the projected distance to the relic and the BCG in the center of mass frame for all 35 relics in Figure \ref{fig:COM}. If all subclusters are still outbound, we expect a correlation where BCGs further from the center of mass will be on the side of relics that have propagated further. We use the Spearman rank-ordered correlation test to check for correlation between the relic and BCG distances to the center of mass. We find a correlation in both the full relic sample and the double relic sample ($p=0.012$ and $p=0.0046$, respectively). Here, the p-value gives the chance that two uncorrelated datasets would give as strong a correlation. 

\subsection{Global Cluster Relations}

In this subsection, we compile global cluster measures of temperature, X-ray luminosity, and cluster mass (as estimated by the Planck Collaboration) from the literature. We compare each of these measures against each other as well as the global velocity dispersion estimated from our spectroscopic survey. Each of these measures is known to be biased during a merger, but by comparing them, we may get a sense of how they are biased with respect to one another. The X-ray luminosities are quoted in the ROSAT 0.1-2.4 keV band. The Planck mass is taken from the PSZ1 catalog \citep{PlanckMass}. A few clusters are missing from this catalog. For these, the mass is estimated using the X-ray luminosity and the scaling relations of \citet{Pratt:2009} following \citet{deGasperin:2014}.

\begin{figure*}
\centering
\includegraphics[width=\textwidth]{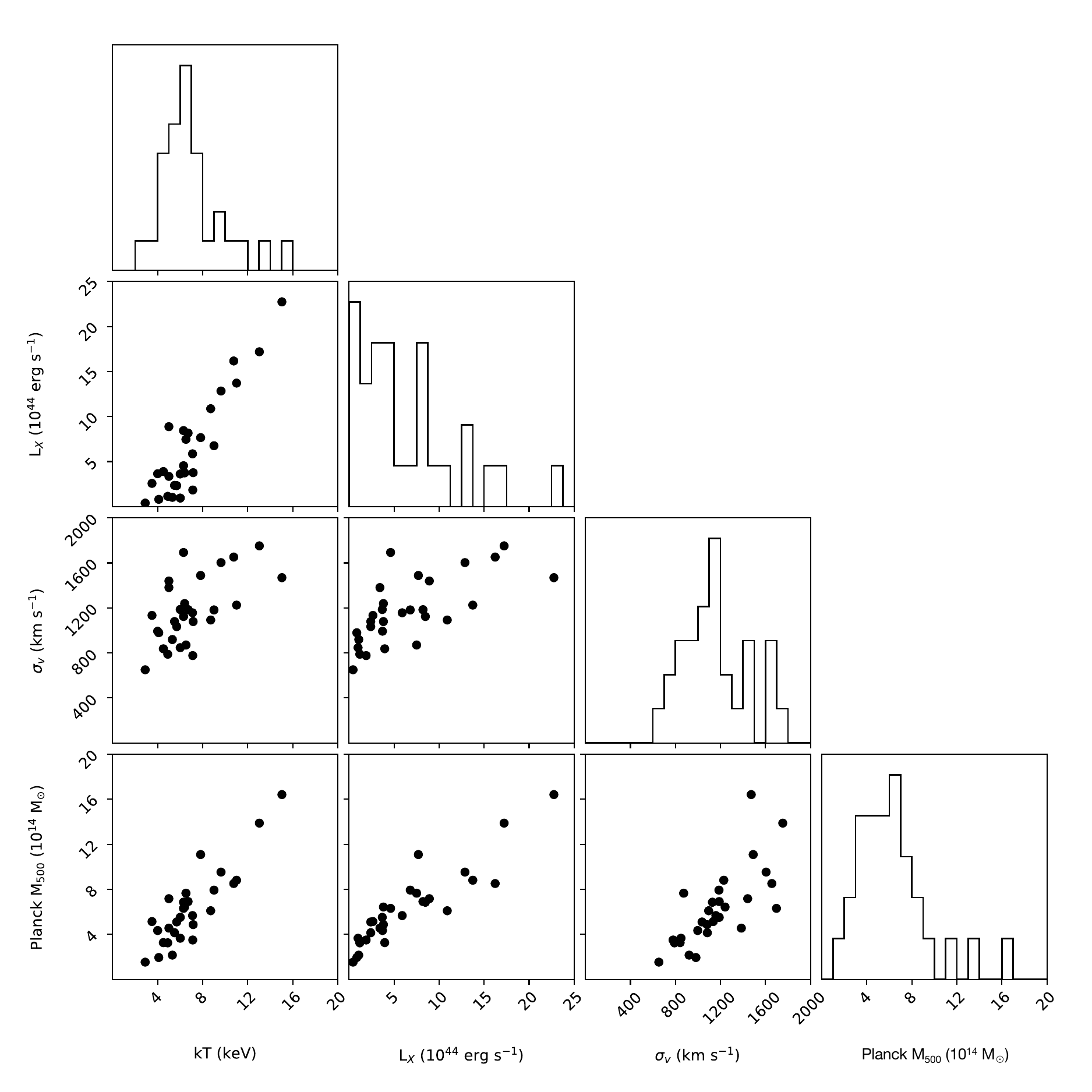}
\caption{Relationships between four global measures for 29 merging clusters. The clear positive correlation between each suggests that, measuring these values gives a first order understanding of the mass of a merging system. Furthermore, with further lensing and dynamical study, we may understand why given systems have higher $\sigma_{v}$ with respect to merger dynamics.}
\label{fig:global}

\end{figure*}

In Figure \ref{fig:global}, we present the relationship between each of these global measures. There is a clear positive relationship between each. Interestingly, the temperature, X-ray luminosity, and Planck mass are related much more tightly with each other than the velocity dispersion. This is explained by the fact that each of these measures is based on the ICM; whereas, the velocity dispersion is traced by the galaxies, which more closely follow the DM distribution. Despite the scatter, there is still clear evidence of a positive correlation between each of these measures, which speaks to the usefulness of quoting these parameters despite the merger induced biases. Furthermore, if the mass from gravitational lensing is known, these ICM indicators potentially give additional information about the merger parameters \citep{Randall:2002}.


\section{Discussion}\label{sec:discussion}

We have presented an optical and spectroscopic survey of 29 merging galaxy clusters featuring $\sim$40 hours of optical imaging and $\sim$5500 spectra of member galaxies. We have coupled these extensive optical observations with $\sim$4000 ks of X-ray integration time and $\sim$250 hr of radio integration time. This comprehensive set of data has allowed us to characterize the redshift, velocity dispersion, minimum number of subclusters, and most likely merger scenario for each system. In 21 of 29 systems, we link the radio relic(s) to galaxy substructure identified by our GMM subclutering analysis on the spectroscopic cluster populations. In six additional systems, the spectroscopic survey is insufficient to detect substructure that is aligned with the radio relic(s), but there is clear substructure in the two dimensional red sequence luminosity distribution and additional evidence of a merger in the X-ray surface brightness map. In the two remaining systems either an extremely bright star (A746) or complex merger scenario (A2255) hinder our ability to decipher the merger that induced the respective radio relics.

On an individual system basis, we have compiled findings from the literature pertaining to X-ray, radio, galaxy, and lensing analyses of each system. We have contextualized the diverse literature with our galaxy analysis, characterized each system based on our substructure analysis, and identified the most likely merger scenario. 

Our chief ensemble finding from the spectroscopic survey and subcluster analysis is the strong evidence for transverseness. The line of sight velocity difference between merging subclusters is generally low with a majority consistent with zero line of sight velocity difference at $3\sigma$. Secondly, radio relics are a robust indicator of the merger axis. The relics are generally positioned within $20\degr$ of the axis connecting the BCGs of merging subclusters. Furthermore, double radio relics are a strong indicator of simple geometries. Clusters with double radio relics have an average of 2.4 identified red sequence luminosity peaks; whereas, single relic clusters have an average of 3.1 galaxy luminosity peaks. 

In the following subsections, we discuss general considerations that highlight specific aspects of our analysis, results, and future directions for merging cluster studies. 

\subsection{Gold sample}

Merging clusters offer a unique setting to study DM, the ICM, and galaxy formation during the most extreme periods of structure formation. These events occur over the course of several billion years, and radio relic selection has allowed us to identify systems in the period between the first and second collision of major mergers. In order to study the physics within the main components of these systems, we must be able to place meaningful constraints on the age and dynamics to control for time varying processes. The only systems we can accurately describe in this context are simple, bimodal systems. Here we reduce our ensemble of 29 systems to a \emph{gold sample} of eight systems that we propose be studied in greater detail individually and together in precise ensemble analyses. Several of these systems have been featured in these types of studies by the Merging Cluster Collaboration to date. The primary feature of these systems is their simple geometry, often highlighted by double relics. Secondary considerations include high mass, redshift, small line of sight velocity difference, and large projected separation.

The Merging Cluster Collaboration \emph{gold sample}:
\begin{itemize}
\item A1240
\item A3411
\item CIZAJ2242
\item MACSJ1149
\item MACSJ1752
\item RXCJ1314
\item ZwCl0008
\item ZwCl1856
\end{itemize}

\subsection{Future Work}

In this subsection, we highlight a few projects that are currently underway to extend the analysis presented here. This is not an exhaustive list, but it offers readers a list of studies to look for in the future. 

\subsubsection{Merger Dynamics}

The techniques developed to understand and quantify the dynamics of merging clusters have progressed rapidly, especially recently, with the rise of fast computational techniques \citep[e.g. MCMAC:][]{MCMAC} as well as powerful, large scale simulations including rich physics. These techniques have greatly enhanced our understanding of these systems. We hope our results here can offer simulators a constrained set of initial conditions with which to simulate these systems. In subsequent dynamical analyses, we will further constrain the dynamics of the eight, gold-sample clusters including timescales since pericenter, the observed phase of the merger, relative velocities, and merger geometry. Furthermore, we will match our observations to analog systems in large scale N-body simulations \citep[e.g., Dark Sky:][]{DarkSky}. Our substructure models coupled with velocity information allow us to identify systems in cosmological N-body simulations that can be easily (re)simulated at higher resolution. This allows simulators to maintain realistic formation histories, accretion, and substructure while also satisfying the dynamical constraints of our results here. This technique will offer a substantial step forward in the simulation of these systems and greatly improve astrophysical inference. 

\subsubsection{BCGs}\label{subsubsec:bcgs}

Recent studies of cosmological simulations \citep{Ng:2017} and simulations of merging clusters with SIDM \citep{Kim:2017, Robertson:2017} have shown that traditional offset measurements between galaxies and DM distributions may be too noisy to constrain DM. On the other hand, the inner most region of the DM profile may offer alternate signals of SIDM. First, if DM halos have large cores, infalling satellite galaxies may survive without falling completely into the BCG. Several clusters in our sample feature strongly lensed source galaxies in the vicinity of the BCG, which may be utilized to measure the surface mass density in the inner region and compare to the number of luminosity peaks in the inner regions. Additionally, BCG alignments have been studied in a range of clusters and have been shown to consistently trace the global cluster profile \citep[e.g.,][]{West:2017}. The BCGs in actively merging clusters may help elucidate during which epoch of cluster formation this alignment is established. We look to answer this question in a forthcoming paper. Visual inspection of the BCG thumbnails reveal several multi-modal BCGs and strong lensing features, which speaks to the utility of these alternate potential SIDM signals.

\subsubsection{Lensing Studies}

In this paper, we presented the red sequence luminosity distribution of each system. In 27 of the 29 clusters, this was based on multi-band Subaru SurpimeCam imaging. This imaging is uniformly deep and wide enough for robust weak lensing studies. We have completed a number of such analyses already \citep{Jee:2015,Jee:2016,Golovich:2017,Benson:2017}, and we will continue to map the total mass profiles of all systems in the gold sample. We have specifically selected eight clusters in the gold sample to carry out a detailed study of offsets between the galaxy and mass distributions. This analysis will result in a constraint on the DM self-interaction cross section.

\section{Acknowledgments}

The authors would like to make special thanks to the generous sharing of a multitude of spectroscopy, radio, and X-ray images. These data have contributed greatly to this project.
We would like to thank the broader membership of the Merging Cluster Collaboration for their continual development of the science motivating this work, for useful conversations, and for diligent proofreading, editing, and feedback. 
This material is based upon work supported by the National Science Foundation under Grant No. (1518246).
This material is based in part upon work supported by STSci grant HST-GO-13343.001-A.
Part of this was work performed under the auspices of the U.S. DOE by LLNL under Contract DE-AC52-07NA27344.
RJW was supported by NASA through the Einstein Postdoctoral grant number PF2-130104 awarded by the Chandra X-ray Center, which is operated by the Smithsonian Astrophysical Observatory for NASA under contract NAS8-03060.
GAO acknowledges support by NASA through a Hubble Fellowship grant HST-HF2-51345.001-A awarded by the Space Telescope Science Institute, which is operated by the Association of Universities for Research in Astronomy, Incorporated, under NASA contract NAS5- 26555.
Some of the data presented herein were obtained at the W.M. Keck Observatory, which is operated as a scientific partnership among the California Institute of Technology, the University of California and the National Aeronautics and Space Administration. The Observatory was made possible by the generous financial support of the W.M. Keck Foundation.
Based in part on data collected at Subaru Telescope, which is operated by the National Astronomical Observatory of Japan.
Funding for the Sloan Digital Sky Survey IV has been provided by the Alfred P. Sloan Foundation, the U.S. Department of Energy Office of Science, and the Participating Institutions. SDSS acknowledges support and resources from the Center for High-Performance Computing at the University of Utah. The SDSS web site is www.sdss.org.
The Digitized Sky Surveys were produced at the Space Telescope Science Institute under U.S. Government grant NAG W-2166. The images of these surveys are based on photographic data obtained using the Oschin Schmidt Telescope on Palomar Mountain and the UK Schmidt Telescope. The plates were processed into the present compressed digital form with the permission of these institutions.
Funding for the DEEP2/DEIMOS pipelines has been provided by NSF grant AST-0071048. 
The DEIMOS spectrograph was funded by grants from CARA (Keck Observatory) and UCO/Lick Observatory, a NSF Facilities and Infrastructure grant (ARI92-14621), the Center for Particle Astrophysics, and by gifts from Sun Microsystems and the Quantum Corporation.
This research has made use of the NASA/IPAC Extragalactic Database (NED) which is operated by the Jet Propulsion Laboratory, California Institute of Technology, under contract with the National Aeronautics and Space Administration.
This research has made use of NASA's Astrophysics Data System.
Based in part on data collected at Subaru Telescope and obtained from the SMOKA, which is operated by the Astronomy Data Center, National Astronomical Observatory of Japan.
This research made use of Montage. It is funded by the National Science Foundation under Grant Number ACI-1440620, and was previously funded by the National Aeronautics and Space Administration's Earth Science Technology Office, Computation Technologies Project, under Cooperative Agreement Number NCC5-626 between NASA and the California Institute of Technology.
This research made use of APLpy, an open-source plotting package for Python hosted at {\tt http://aplpy.github.com}.
\\

{\it Facilities:} Keck (DEIMOS), INT (WFC), Subaru (SuprimeCam), VLT (VIMOS), WSRT, GMRT, VLA, Chandra (ACIS), XMM-Newton (EPIC), ROSAT (PSPC)

\bibliographystyle{aasjournal.bst}
\bibliography{mcc}
\end{document}

%% file: table1.tex
\begin{tabular}{lllllll}
Cluster				&	Short name 	& RA 		&	DEC 		&	Redshift	&	Discovery band		& \S \\
\hline 
1RXS J0603.3+4212		& 	1RXSJ0603	& 06:03:13.4 	&	+42:12:31 	&	0.226	&	Radio			& \ref{subsec:1RXSJ0603}\\
Abell 115				& 	A115			& 00:55:59.5 	&	+26:19:14 	&	0.193	&	Optical			&\ref{subsec:A115}\\
Abell 521 				&	A521			& 04:54:08.6 	&	-10:14:39 		&	0.247	&	Optical			&\ref{subsec:A521}\\
Abell 523 				&	A523			& 04:59:01.0	&	+08:46:30 	&	0.104	&	Optical			&\ref{subsec:A523}\\
Abell 746 				&	A746			& 09:09:37.0	&	+51:32:48 	&	0.214	&	Optical			&\ref{subsec:A746}\\
Abell 781 				&	A781			& 09:20:23.2	&	+30:26:15 	&	0.297	&	Optical			&\ref{subsec:A781}\\
Abell 1240		 	&	A1240		& 11:23:31.9	&	+43:06:29 	&	0.195	&	Optical			&\ref{subsec:A1240}\\
Abell 1300 			&	A1300		& 11:32:00.7	&	-19:53:34 		&	0.306	&	Optical			&\ref{subsec:A1300}\\
Abell 1612 			&	A1612		& 12:47:43.2	&	-02:47:32 		&	0.182	&	Optical			&\ref{subsec:A1612}\\
Abell 2034 			&	A2034		& 15:10:10.8	&	+33:30:22		&	0.114	&	Optical			&\ref{subsec:A2034}\\
Abell 2061 			&	A2061		& 15:21:20.6	&	+30:40:15 	&	0.078	&	Optical			&\ref{subsec:A2061}\\
Abell 2163 			&	A2163		& 16:15:34.1	&	-06:07:26 		&	0.201	&	Optical			&\ref{subsec:A2163}\\
Abell 2255 			&	A2255		& 17:12:50.0	&	+64:03:11 	&	0.080	&	Optical			&\ref{subsec:A2255}\\
Abell 2345 			&	A2345		& 21:27:09.8	&	-12:09:59 		&	0.179	&	Optical			&\ref{subsec:A2345}\\
Abell 2443 			&	A2443		& 22:26:02.6	&	+17:22:41 	&	0.110	&	Optical			&\ref{subsec:A2443}\\
Abell 2744 			&	A2744		& 00:14:18.9	& 	-30:23:22 		&	0.306	&	Optical			&\ref{subsec:A2744}\\
Abell 3365 			&	A3365		& 05:48:12.0	& 	-21:56:06 		&	0.093	&	Optical			&\ref{subsec:A3365}\\
Abell 3411 			&	A3411		& 08:41:54.7	& 	-17:29:05 		&	0.163	&	Optical			&\ref{subsec:A3411}\\
CIZA J2242.8+5301		& 	CIZAJ2242	& 22:42:51.0	& 	+53:01:24 	&	0.189	&	X-ray			&\ref{subsec:CIZAJ2242}\\
MACS J1149.5+2223	& 	MACSJ1149	& 11:49:35.8	& 	+22:23:55 	&	0.544	&	X-ray			&\ref{subsec:MACSJ1149}\\
MACS J1752.0+4440	&	MACSJ1752	& 17:52:01.6	& 	+44:40:46 	&	0.365	&	X-ray			&\ref{subsec:MACSJ1752}\\
PLCKESZ G287.0+32.9 	&	PLCKG287	& 11:50:49.2	& 	-28:04:37 		&	0.383	&	SZ				&\ref{subsec:PLCKG287}\\
PSZ1 G108.18-11.53 	&	PSZ1G108	& 23:22:29.7	&	 +48:46:30	&	0.335	&	SZ				&\ref{subsec:PSZ1G108}\\
RXC J1053.7+5452 		&	RXCJ1053	& 10:53:44.4	& 	+54:52:21 	&	0.072	&	X-ray			&\ref{subsec:RXCJ1053}\\
RXC J1314.4-2515 		&	RXCJ1314	& 13:14:23.7	& 	-25:15:21 		&	0.247	&	X-ray			&\ref{subsec:RXCJ1314}\\
ZwCl 0008.8+5215 		&	ZwCl0008		& 00:08:25.6	&	+52:31:41 	&	0.104	&	Optical			&\ref{subsec:ZwCl0008}\\
ZwCl 1447+2619 		&	ZwCl1447		& 14:49:28.2	&	+26:07:57 	&	0.376	&	Optical			&\ref{subsec:ZwCl1447}\\
ZwCl 1856.8+6616		& 	ZwCl1856		& 18:56:41.3	&	+66:21:56.0	&	0.304	&	Optical			&\ref{subsec:ZwCl1856}\\
ZwCl 2341+0000 		&	ZwCl2341		& 23:43:39.7	&	+00:16:39 	&	0.270	&	Optical			&\ref{subsec:ZwCl2341}\\
\end{tabular}

%% file: table5.tex
\begin{tabular}{llllc}
Cluster 		&	Radio Image		&	Radio Reference			&	X-ray Telescope	&	Exposure (ks) \\
\hline
1RXSJ0603	&	GMRT 610 MHz	&	\citet{vanWeeren:2012}		&	Chandra			&	250\\
A115			&	VLA 1.4 GHz		&	\citet{Govoni:2001b}			&	Chandra			&	360\\
A521			&	GMRT 240 MHz	&	\citet{Venturi:2013}			&	Chandra			&	170\\
A523			&	VLA 1.4 GHz		&	\citet{vanWeeren:2011}		&	XMM				&	30\\
A746			&	WSRT 1382 MHz	&	\citet{vanWeeren:2011}		&	Chandra			&	30\\
A781			&	GMRT 325 MHz	&	\citet{Venturi:2013}			&	Chandra			&	48\\
A1240		&	VLA 1.4 GHz		&	\citet{Bonafede:2009}		&	Chandra			&	52\\
A1300		&	GMRT 325 MHz	&	\citet{Venturi:2013}			&	Chandra			&	100\\
A1612		&	GMRT 325 MHz	&	\citet{vanWeeren:2011}		&	Chandra			&	31\\
A2034		&	WSRT 1382 MHz	&	\citet{vanWeeren:2011}		&	Chandra			&	261\\
A2061		&	WSRT 1382 MHz	&	\citet{vanWeeren:2011}		&	Chandra			&	32\\
A2163		&	VLA 1.4 GHz		&	\citet{Feretti:2001}			&	Chandra			&	90\\
A2255		&	WSRT 350 MHz	&	\citet{Pizzo:2009}			&	XMM				&	42\\
A2345		&	VLA 1.4 GHz		&	\citet{Bonafede:2009}		&	XMM				&	93\\
A2443		&	VLA 325 MHz		&	\citet{Cohen:2011}			&	Chandra			&	116\\
A2744		&	GMRT 325 MHz	&	\citet{Venturi:2013}			&	Chandra			&	132\\
A3365		&	VLA 1.4 GHz		&	\citet{vanWeeren:2011}		&	XMM				&	161\\
A3411		&	GMRT 325 MHz	&	\citet{vanWeeren:2017}		&	Chandra			&	215\\
CIZAJ2242	&	WSRT 1382 MHz	&	\citet{vanWeeren:2010}		&	Chandra			&	206\\
MACSJ1149	&	GMRT 323 MHz	&	\citet{Bonafede:2012}		&	Chandra			&	372\\
MACSJ1752	&	WSRT 1.4 GHz		&	\citet{vanWeeren:2012b}		&	XMM				&	13\\
PLCKG287	&	GMRT 325 MHz	&	\citet{Bonafede:2014}		&	Chandra			&	200\\
PSZ1G108	&	GMRT 323 MHz	&	\citet{deGasperin:2015}		&	Chandra			&	27\\
RXCJ1053	&	WSRT 1382 MHz	&	\citet{vanWeeren:2011}		&	Chandra			&	31\\
RXCJ1314	&	VLA 2-4 GHz		&	N/A						&	XMM				&	110\\
ZwCl0008	&	WSRT 1382 MHz	&	\citet{vanWeeren:2011c}		&	Chandra			&	411\\
ZwCl1447	&	VLA 1.4 GHz		&	\citet{Govoni:2012}			&	RASS			&	21\\
ZwCl1856	&	WSRT 1382 MHz	&	\citet{deGasperin:2014}		&	XMM				&	12\\
ZwCl2341	&	GMRT 610 MHz	&	\citet{vanWeeren:2009}		&	Chandra			&	227\\
\end{tabular}

%% file: 1RXSJ0603.tex
\subsubsection{Literature Review}\label{sssec:sample_toothbrush_review}

1RXSJ0603 (a.k.a. the Toothbrush---named after the radio relic morphology) was first cataloged by RASS as an extended and bright X-ray source ($L_{X,\,0.2-2.4\,\mathrm{keV}}\sim10^{45}$ erg s$^{-1}$). It was confirmed as a galaxy cluster through cross-correlating RASS catalogs with the Westerbrook Northern Sky Survey \citep[WENSS:][]{Rengelink:1997}, the NRAO VLA Sky Survey \citep[NVSS:][]{Condon:1998}, and the VLA Low-Frequency Sky Survey \citep[VLSS:][]{Cohen:2007} in conjunction with inspection of the Two-Micron  All-Sky Survey \citep[2MASS:][]{2mass} images \citep{vanWeeren:2011,vanWeeren:2011c}. \citet{vanWeeren:2012} followed-up with WSRT and GMRT radio observations between 147 MHz and 4.9 GHz and enabled the identification of a large (1.9 Mpc) and bright linear radio relic in the north, a connected and elongated $\sim$2 Mpc radio halo, and two smaller and fainter radio relics to the east and southeast. The number of diffuse radio features suggests a complex merger. A variable polarization fraction along the northern relic between 10 and 60\% was measured, and the radio spectral index was used to estimate a Mach number of 3.3--4.6.  The northern ``toothbrush'' relic and halo are connected by a region with a spectral index slope of $\alpha$ $<$ -2, but then flattens to $\alpha \sim\,-1.2$, and increases slightly towards the center of the halo. \citet{vanWeeren:2012} indicate the possibility that previously accelerated electrons by the shock front are reaccelerated by merger induced turbulence, or even perhaps the halo region is in fact a relic viewed face on. With five spectroscopic redshifts of cluster members they estimated a cluster redshift of 0.225.

\citet{Bruggen:2012} performed hydrodynamical N-body simulations to replicate the elongated nature of the toothbrush relic. They found that they reproduce the relic morphology with a triple merger between two equal mass subclusters merging along the north--south axis with a third smaller subcluster in-falling from the southwest toward the northeast.

\citet{Ogrean:2013c} observed 1RXSJ0603 with XMM-Newton and found X-ray gas density and temperature discontinuities at the locations of the three relics identified by \citet{vanWeeren:2012} indicating that the relics likely trace underlying shocks in the ICM. However they only estimate a Mach number $<$ 2 for the northern shock (inconsistent with the radio Mach number of 3.3--4.6). They find that the X-ray shock is in part spatially offset from the radio emission in both distance and position angle. This runs counter to the DSA assumption that particle acceleration occurs at the shock.

\citet{Stroe:2013,Stroe:2015} present a narrow band H$\alpha$ survey or the 1RXSJ0603 field. They found no enhanced H$\alpha$ emitters are distributed near the northern shock, although there are at least six near the eastern relic candidate. The H$\alpha$ luminosity of 1RXSJ0603 is consistent with the low star formation rates of blank field galaxies at $\text{z}=0.2$. They propose the system to be well evolved given the findings. 

\citet{vanWeeren:2016} present new deep LOFAR low frequency radio observations and combine with VLA and GMRT data to provide a wide view of the radio spectrum. Spectral steepening from $\alpha=-0.8\pm0.1$ at the northern edge to $\alpha\sim-2$ towards the south. The spectral index suggests a Mach number of $\mathcal{M}=2.8^{+0.5}_{-0.3}$ assuming DSA. This confirms the discrepancy with the X-ray data from Chandra ($\mathcal{M}\lesssim1.5$) which suggests that the relic emission is supported by pre-accelerated particles from a source such as radio galaxies. 

\citet{Kierdorf:2016} observed the radio relic at high frequency (4.85 and 8.35 GHz with 0.5 and 1.1 GHz bandwidths, respectively) with the Effelsberg 100-m telescope. They find a variable polarization along the radio relic with values up to 45$\pm$7\%. Interestingly, the angle of the polarization vectors is variable along the relic, which is unexpected for a single shock front. Internal shock structure or foreground Faraday rotations are possible explanations.

\citet{Jee:2015} present the only lensing analysis with a joint WL analysis on HST ACS/WFC3 and the Subaru SuprimeCam data presented in Paper 1 and analyzed in this paper. Four significant lensing peaks are discovered with good agreement between the lensing and cluster luminosity. Two of the four lensing peaks are situated in the north of the cluster toward the ``toothbrush'' relic, one is far to the southeast, and the fourth is to the west. The two most significant peaks are oriented north--south and a merger between the two appears to have caused the elongation of the X-ray surface brightness profile as well as the ``bristle'' portion of the relic. \citet{Jee:2015} fit a bimodal NFW profile \citep{NFW} and estimate $M_{200} = 6.29^{+2.24}_{-1.62} \times 10^{14}\,M_{\odot}$ and $1.98^{+1.24}_{-0.74}\times10^{14}\,M_{\odot}$ for the main north and south subclusters, respectively. The total mass is estimated to be $\sim10^{15}\,M_{\odot}$.

\subsubsection{Results}

In Paper 1, we described our observations and found 242 cluster members. The one dimensional redshift analysis of spectroscopic cluster members indicates that 1RXSJ0603 is a massive and rich cluster at $z=0.22631\pm0.00038$. The global redshift histogram is well fit with a single Gaussian despite the multimodality identified by \citet{Jee:2015}. This could indicate that the merger is mostly in the plane of the sky (i.e. there is little peculiar velocity difference along the line of sight) and/or the subclusters are near apocenter (i.e. the relative \emph{3D} velocity is small). 

In Figure \ref{fig:dstest} we present the red sequence luminosity map (r band), which is smoothed with a 41\arcsec\,two dimensional Gaussian kernel. We confirm the results of \citet{Jee:2015} finding up to four galaxy subclusters that are consistent with the lensing peaks. The DS-test circles in Figure \ref{fig:dstest} indicate a complex merger is underway between the various components. 

In the northern portion of the cluster, there is an east--west elongation. On the east side, the galaxies are at a redshift of $z\sim0.23$ while on the west side, the galaxies are at a redshift of $z\sim0.22$. Further south there is a bright star that affected our ability to target galaxies for spectroscopy. However, \citet{Jee:2015} were able to successfully model and subtract the star from the Subaru/SuprimeCam images. Note that we made use of the \citet{Jee:2015} photometry in our analysis here. For details on the stellar profile subtraction see \S2.1 of \citet{Jee:2015}. We detect a subcluster behind the star populated by a handful of cluster members at $z\sim0.235$ in this region. Finally, in the south of the cluster, a large and well separated subcluster with $z\sim0.22$ is aligned along the axis connecting the ``head'' of the Toothbrush relic with the Eastern half of the northern group of galaxies. 

These four prospective subclusters are input into our MCMC-GMM analysis with input priors outlined in Table \ref{tab:subclusters}. In total, twelve GMM models are run (see Eq. \ref{eq:models}). Details of these twelve models are listed in Table \ref{tab:mcmcgmm}. We select the model with the lowest BIC score, which for 1RXSJ0603 is a four halo model including all four potential halos described above. The four subclusters are presented in projected RA/DEC space in Figure \ref{fig:scatter}. The subclusters are in agreement with the red sequence luminosity density map, and the substructure is roughly aligned with the radio relic \citep[the contours in Figure \ref{fig:scatter} depict 610 MHz GMRT data from][]{vanWeeren:2012}. 

The subcluster redshift histogram is presented in Figure \ref{fig:subclusthist}. The head portion of the toothbrush radio relic is connected along an axis with the two most massive subclusters (2 and 4 in orange and pink, respectively); however, we cannot rule out more complicated merger scenarios. The brightness of the relic suggests that an extremely energetic merger occurred, which indicates that the primary collision occurred between subclusters 2 and 4. Furthermore, the X-ray surface brightness map shows a bright ridge of gas connecting these two regions of the cluster lending further support for this scenario (see Figure \ref{fig:xray}). To fully explore the range of merger scenarios, detailed simulations are necessary covering a range of initial conditions. 

In Figure \ref{fig:bcg}, we present a 150 kpc $\times$ 150 kpc cutout centered on the cluster BCG, which is associated with subcluster 2. Analogous figures for each cluster are presented in the respective subsections. Followup studies related to BCG images are discussed in \S\ref{sec:discussion}. 

Finally, in Figure \ref{fig:cartoon}, we present our preferred merger scenario for 1RXSJ0603. This represents one of the most complicated merger scenarios of our ensemble.  Analogous figures for each cluster are presented in the respective subsections. These cartoons should be viewed as possible explanations of the data instead, and especially in the more complex systems, a range of cartoons could be drawn. For example, our representation of 1RXSJ0603 is contradicted by the scenario proposed by \citet{Bruggen:2012}.

%% file: A115.tex
\subsubsection{Literature Review}

A115 has been extensively studied since its discovery as a double X-ray peak by \citet{Forman:1981}. \citet{White:1997, Shibata:1999, Gutierrez:2005} have also studied its X-ray properties. \citet{Giovannini:1987} conducted a 4.9 GHz VLA observation of the cluster primarily studying the bright radio galaxy source near the northern radio relic. \citet{Govoni:2001b} followed this with a 1.4 GHz VLA observation and discovered the diffuse radio emission north of the cluster believed to be a radio relic. They note that there are two radio bright cluster members embedded in the diffuse relic emission. The relic is highly extended ($\sim$2.5 Mpc $\times$ 0.5 Mpc) perpendicular to the supposed merger axis.

More recent X-ray analysis based on Chandra X-ray observations such as \citet{Gutierrez:2005} (50 ks) find that the northern and southern subclusters have cool cores ($\sim$3 keV) that are significantly cooler than the surrounding gas ($\sim$5 keV) and the hottest gas ($\sim$8 keV), which is located between the two subclusters. It is atypical for there to be two cool cores in a post merger system. \citet{Rossetti:2017} found $\sim$30$\%$ of Planck SZ clusters to have a cool-core, thus only $\sim$10$\%$ of bimodal major mergers host two cool-cores pre-pericenter. It is unlikely that both would survive a head on collision, which perhaps suggests the subclusters did not pass through one another and instead merged with a large impact parameter.

\citet{Botteon:2016} analyzed deep Chandra observations (334 ks) finding evidence for a shock in both the density and temperature profiles of the ICM. Both methods suggest a low mach number of $\mathcal{M}\sim1.4-2.0$. The shock is spatially coincident with the western end of the radio relic. They argue that the location of the shock is explained by an off axis merger between unequal mass subclusters based on simulation results \citep{Ricker:2001}.

The picture of an unequal mass merger is substantiated by \citet{Barrena:2007} with a spectroscopic survey obtaining 88 cluster member spectra, following up on an earlier survey by \citet{Beers:1983}, which resulted in 29 redshifts. In addition to clearly detecting the northern and southern subclusters, \citet{Barrena:2007} also find evidence for two smaller subclusters to the east ($\sim$0.5 Mpc). \citet{Barrena:2007} interpret the system as being a pre-merger based on the agreement in location of the BCG's and two X-ray peaks.  However, \citet{Gutierrez:2005,Botteon:2016} suggest that there has been a glancing merger. This is supported by the presence of the radio relic in the north. \citet{Barrena:2007} find a large line-of-sight velocity difference between the north and south subclusters of 1646 km s$^{-1}$ and a projected separation of 0.89 Mpc. They estimate that the merger is 100 Myr from core passage and is occurring within 20$\degr$ of the plane of the sky. This would require a substantial 3D merger speed of $\sim$4500 km s$^{-1}$, which is likely too large even for free-falling 10$^{15}$ M$_{\odot}$ subclusters from infinity.

\citet{Okabe:2010} performed a weak lensing analysis of A115 based on Subaru/SuprimeCam imaging and detected the north and south subclusters at a significance of $\sim$4 and 5 sigma respectively.  Interestingly both the north and south lensing peaks are offset from the corresponding galaxy density peaks (both luminosity and number density peaks) substantially. \citet{Okabe:2010} only noted the complexity of this merger in passing stating that it is the reason for it being an outlier in their one-dimensional lensing analysis. \citet{Oguri:2010} conducted a second lensing analysis of this sample but made no significant note of the system.

\subsubsection{Results}

In Paper 1 we describe our observations and archival imaging used. The spectroscopic catalog consists of 198 galaxies confirmed at the cluster redshift. The one dimensional redshift analysis indicates that A115 is a massive system at $z=0.19285\pm0.00040$. The global redshift distribution is very well fit by a single Gaussian with a rest frame velocity dispersion of 1439$\pm$79 km s$^{-1}$. 

The red sequence luminosity (i band) is smoothed with a 52\arcsec\, two dimensional Gaussian kernel. We confirm a bimodal distribution with a north to south orientation. The DS-test circles indicate that the south subcluster is at a redshift of $\sim$0.1925 and the northern subcluster is at a slightly higher redshift of $\sim$0.195, which is also true for the two BCGs.

These two potential subclusters are input into our MCMC-GMM analysis with input priors outlined in Table \ref{tab:subclusters}. Details of the two models run are presented in Table \ref{tab:mcmcgmm}. We select the second model with two halos based on the substantially lower BIC score. Resultant subclusters 1 and 2 (north to south) are presented in projected space in Figure \ref{fig:A1}. There is good agreement between the red sequence luminosity and spectroscopy with a slight offset in subcluster 1 possibly stemming from uneven spectroscopic sampling. There are far more galaxies detected spectroscopically to the north of the BCG, which biased the GMM results. The two subclusters are roughly aligned with the western end of the radio relic \citep[1.4 GHz VLA contours in Figure \ref{fig:A1} from][]{Govoni:2001b} and the associated shock \citep{Botteon:2016}. The two subclusters have a line of sight velocity difference of 256$\pm$162 km s$^{-1}$. The more complete data rules out the velocity difference of $\sim$1600 km s$^{-1}$ previously reported.

The X-ray surface brightness map shows clear bimodality and good agreement between the galaxies and X-ray peaks. There is clear evidence of a glancing merger in the swirl of gas to the east of the line connecting the two subclusters. Inspection of the wake features trailing the two gas cores suggest that the northern (southern) core is traveling southwest (northeast). This suggests that the subclusters have reached apocenter and the gas cores are spiraling around as gravity pulls them back.  Simulations exploring a range of impact parameters will be necessary to fully understand the generation of a radio relic in glancing merger, but it seems that the merger phase can be well constrained given the present X-ray morphology.

%% file: A521.tex
\subsubsection{Literature Review}

Abell 521 has been extensively studied since its first detailed study by \citet{Arnaud:2000}, who noted a northwest to southeast bimodal X-ray distribution in the ROSAT/HRI imaging and an `X' shaped galaxy distribution with one of the axes corresponding closely to the X-ray emission. They suggested that A521 might be a young cluster forming at the crossing of two filaments with one pointing towards A517 (z=0.2244, projected 82$\arcmin$ away) and the other in the direction of A528 (z=0.2896, projected 106$\arcmin$ away) and A518 (z=0.1804, projected 50$\arcmin$ away). This study was also the first to note the existence of the large diffuse radio relic in the southeast from NVSS imaging.

\citet{Maurogordato:2000} used 41 spectroscopic redshifts and found evidence of three subclusters and a large velocity dispersion ($\sim$1400 km s$^{-1}$) for the system. The core of the brightest cluster galaxy showed clear signs of interaction effects, including the presence of multiple nuclei. They noted an arc-like feature to the north of the BCG, but the spectra placed on the arc did not detect the redshift. \citet{Ferrari:2003} obtained 125 spectroscopic redshifts of cluster galaxies and found a similarly high velocity dispersion for the system. They claim that the system is composed of 7 groups and a ridge.

\citet{Ferrari:2006} observed A521 for 38 ks with Chandra and confirmed the general finding of \citet{Arnaud:2000} but with the higher resolution were able to identify two dominant clumps, a diffuse southern clump and a denser northern clump, which appear to have two sub-peaks. Additionally they found evidence for smaller clumps in the northeast and west.  The two dominant X-ray peaks have a northwest to southeast axis consistent with the elongated galaxy axis. The radio relic is found to lie at the periphery of the X-ray emission which has been confirmed by \citet{Giacintucci:2008} and \citet{Bourdin:2013}.

A521 has been observed in the radio from 153 MHz - 5 GHz; over this range the radio relic has a spectral index of $\alpha = -1.45$ \citep{Giacintucci:2008,Brunetti:2008,Dallacasa:2009, Macario:2013,Venturi:2013}.  \citet{Giacintucci:2008} and \citet{Macario:2013} noted that these results support the scenario where the radio relic is the result of DSA. \citet{Macario:2013} also note that the radio halo of A521 has a very steep radio spectrum with a slope $\alpha\sim-1.8$, which suggests a large energy budget in the form of relativistic electrons. \citet{Giacintucci:2008} estimates $\mathcal{M}\sim2.3$ based on the radio observations, which is consistent with the X-ray estimate of $\mathcal{M}=2.4\pm0.2$ \citep{Bourdin:2013}.

\citet{Okabe:2010} performed a weak lensing analysis of A521 noting that the radial shear profile of the cluster is poorly fit by an singular-isothermal-sphere profile and acceptably fit by an NFW profile, for which the estimate a total $M_{200} = \left(4.58 \pm 1.0\right) \times 10^{14}$ M$_\odot$. \citet{Richard:2010} observed the central region with HST WFPC2 and Keck LRIS and confirmed the arc identified by \citet{Maurogordato:2000} to be three images of a face on spiral galaxy at $z=1.043$. They complete a strong lensing analysis and estimate the mass within 250 kpc of $6\times10^{13}$ M$_{\odot}$. They noted that since the strong lensing information is only on one side of the BCG the location of the peak of the mass distribution is not well constrained.

\subsubsection{Results}

We made use of archival Subaru/SuprimeCam imaging (V and i bands) and 120 cluster member redshifts from \citet{Maurogordato:2000,Ferrari:2003}. The global cluster redshift is $0.24676\pm0.00041$ and a velocity dispersion is $1100\pm76$ km s$^{-1}$. This redshift distribution is well fit by a single Gaussian (see Paper 1).

The red sequence luminosity distribution has at least three distinct peaks (Figure \ref{fig:A2}). The central peak is also elongated along the same northwest to southeast axis. Each of these regions is dominated by a distinct bright galaxy; however, the cluster-wide BCG is in the northern half of the middle luminosity peak. We placed subcluster priors centered on each galaxy overdensity and used conservative redshift windows (see Table \ref{tab:subclusters}) centered on the redshift of the brightest galaxy of each halo. The preferred model from the BIC analysis has two subclusters stemming from priors b and d in Table \ref{tab:subclusters}. Resultant subclusters 1 and 2 are presented in projected space in Figure \ref{fig:A2}. There is good agreement with two of the three red sequence luminosity peaks. The NW red sequence peak was under-sampled spectroscopically, which likely contributed to our GMM's preference to discard it. More confirmed cluster members in this region are necessary to resolve this structure. The same is true of the southern extension of subcluster 1 (the dashed ellipse drawn inside subcluster 1 in Figure \ref{fig:A521}. More redshifts would be required to explore this further. Subclusters 1 and 2 are well aligned with the radio relic \citep[see Figure \ref{fig:A2}, contours based on GMRT 240 MHz from][]{Venturi:2013}. The subcluster redshift histogram shows that subcluster 1 is more massive than subcluster 2; and their line of sight velocity difference is very small. Subcluster 2 is not well fit by a single Gaussian. Inspecting the DS-circles in Figure \ref{fig:A2} shows both blue and green circles in the central region of this subcluster, which could indicate composite line of sight structure or a lack of spectroscopy sampled from the underlying distribution. 

The X-ray surface brightness map shows a bright peak of X-rays associated with subcluster 1, and a diffuse region near subcluster 2. This is indicative of a major merger between the two subclusters, along with the radio relic and galaxy distributions, which all lie colinearly.  The third red sequence luminosity peak, in the northwest of the cluster has no associated X-ray surface brightness peak, which suggests it has already undergone core passage. There is an extended smudge of gas to the south of this region suggesting it may have passed through this region. 

A521 may be as complex as a four way merger, but the available spectroscopy is unable to overcome the added parameters for more complex models than a simple bimodal model. That being said, the two preferred halos are likely the two that merged and launched the radio relic given the coliniarity between the two subclusters, the X-ray extension, and the radio relic. A521 is a prime candidate for deeper optical imaging and refined strong and weak lensing analyses to help map the dark matter distribution and better understand this system.

%% file: A523.tex
\subsubsection{Literature Review}
A523 has been the subject of three detailed studies. \citet{Giovannini:2011} were the first to note the diffuse radio emission associated with the cluster. They found a north--south bimodal distribution of galaxies with RASS X-ray emission between the two galaxy over-densities and also elongated in a north to south direction. They classify the diffuse radio emission as a radio halo but note that the radio luminosity is abnormally high given the X-ray luminosity. They also noted that the elongation of the radio halo perpendicular to the elongated distribution of gas and galaxies is atypical of radio halos. 

\citet{vanWeeren:2011} confirm the diffuse radio emission and bimodal galaxy distribution with deeper INT/WFC imaging. However, they suggest that the diffuse radio emission should be considered as a possible radio relic due to its patchy morphology and perpendicular orientation with respect to the cluster gas.

\citet{Girardi:2016} recently presented a detailed multi-wavelength study of the system. With new data including 132 spectroscopically confirmed galaxies (80 cluster members), INT imaging, and a 27 ks Chandra exposure, it is found that A523 is a massive system ($\sim$7-9$\times$10$^{14}$ M$_\odot$) based on $\sigma_{v}$-mass scaling relations. Two subclusters are identified along the same axis evident from previous studies. They find disturbance in in the ICM with elongation toward the north/northeast direction in agreement with the RASS X-ray image. They confirm the presence of the radio feature and classify it as an extended radio halo. A merger of a bimodal head-on collision is discussed; however, more complicated scenarios are not ruled out. In addition, directly in the background a cluster present at $z\sim0.14$ with an east--west bimodal distribution.

\subsubsection{Results}

We observed A523 with Subaru/SuprimeCam in g and r bands, and with three slitmasks with Keck DEIMOS. Our spectroscopic survey overlapped that of \citet{Girardi:2016} with an RMS velocity difference of $90 (1+z) $ km s$^{-1}$ based on 59 duplicate redshifts. Checking the consistency of the two surveys, we find 59 duplicate redshifts with a mean redshift difference of 0.00035, or 96 km/s rest-frame. This suggests a difference in wavelength calibration between the two surveys, which is significant in a statistical sense (the standard error in the mean is 15 km/s rest-frame) but a factor of 50 smaller than the velocity dispersions of the relevant subclusters discussed below. Furthermore, this difference does not affect the subcluster relative velocity because each subcluster is well sampled in both surveys. Therefore, we adopt the \citet{Girardi:2016} redshifts for galaxies not in our survey. The standard deviation of the repeat measurements is 116 km/s in the rest frame, again much smaller than the velocity dispersions of the relevant subsclusters. Our spectroscopic survey adds 175 unique galaxy redshifts and brings the total cluster member population to 149 galaxies.  From our one dimensional redshift analysis in Paper 1, the global redshift is 0.10389$\pm$0.00027 and the velocity dispersion is 921$\pm$53 km s$^{-1}$ in agreement with \citet{Girardi:2016}.  

Additionally, we bring the number of background cluster members to 53 galaxies. Upon closer inspection, these 53 galaxies are split in velocity space into two groups; 37 galaxies with z$\sim$0.14 are spatially dispersed behind Abell 523 with two over densities noted by \citet{Girardi:2016}. These galaxies have an average redshift and velocity dispersion of 0.13984$\pm$0.00032 and 537$\pm$66 km s$^{-1}$, respectively. Additionally, 16 galaxies are tightly bunched in velocity space with a redshift of $\sim$0.15, but this over density of galaxies in velocity space is highly diffuse in projected space. Together, the multiple substructures along the line of sight suggest A523 is part of large scale structure, perhaps a filament oriented along the line of sight with several clusters embedded. These systems are not dynamically interacting; the physical distance between redshift 0.104 and 0.14 is $\sim$100 Mpc. 

We confirmed a bimodal distribution of cluster galaxies with a north-south orientation. The DS-test circles indicate the northern subcluster has a redshift around 0.103, and the southern subculture has a redshift around 0.105. These two potential subclusters are input into our GMM analysis with prior distributions listed in Table \ref{tab:subclusters}. The two component model is strongly preferred over the null model with a single halo. Details of the two models are presented in Table \ref{tab:mcmcgmm}. Resultant subclusters 1 and 2 are presented in projected space in Figure \ref{fig:A3}. The two subclusters are in good agreement with the red sequence luminosity distribution. The subcluster redshift distributions are presented in Figure \ref{fig:A3}. Both subclusters are well fit by single Gaussians and suggest a merger in the plane of the sky with a line of sight velocity difference of 420$\pm$125 km s$^{-1}$. 

Interestingly the diffuse radio feature is positioned nearly coincident with the northern subcluster, which poses a challenge for a bimodal merger scenario between subclusters 1 and 2 that launched a shock traveling north generating the radio relic. There have been conflicting classifications for this feature, despite its elongated morphology situated perpendicular to the merger axis. The X-ray surface brightness map (see Figure \ref{fig:A3}) shows an elongated and disturbed ICM situated between the two subclusters, which is a clear indication of a major merger between subcluster 1 and 2; however, as noted above, the radio relic's position is a challenge to explain as the result of a shock launched by this merger. Deeper X-ray and radio data may be required to resolve this issue. Furthermore, a weak lensing analysis may help by indicating the location of dark matter within this cluster. 

%% file: A746.tex
\subsubsection{Literature Review}

A746 is a very little studied cluster. \citet{vanWeeren:2011} identified a radio relic candidate in NVSS and confirmed it with deeper WSRT 1.4 GHz imaging. They found that the relic is located 1.7 Mpc to the northwest of the cluster center and has a physical extent of 1.1 Mpc$\times$0.3 Mpc. They also measured a large polarization fraction of $\sim$50\% (suggesting the merger axis is close to the plane of the sky). Their galaxy number density contours show a slight elongation along the north-south direction, which is roughly offset 45$\degr$ relative to the direction of the relic. However, optical information is highly contaminated by the bright star f UMa ($V_{mag}=4.5)$ located near the radio relic. 

\subsubsection{Results}

We observed A746 with one Keck DEIMOS slitmask and two bands (g and r) of Subaru SuprimeCam. \citet{Jee:2015} demonstrated the success of subtracting the stellar profile from optical imaging for 1RXSJ0603, which will be required before detailed analysis may be completed, but this is beyond the scope of this paper.  

The DEIMOS observation resulted in 66 cluster members and a cluster redshift of 0.21434$\pm$0.00059 and a cluster velocity dispersion of 1187$\pm$111 km s$^{-1}$. The red sequence luminosity density is complex with substructure aligning north to south and east to west. The bright star may be limiting the ability to detect objects within the stellar halo where multiple galaxies are visible, but they are very highly contaminated. There is an additional group of red sequence galaxies $\sim$7\arcmin\, to the southeast of the cluster center nearly colinear with the radio relic (see Figure \ref{fig:A4}). However, our spectroscopic survey did not cover this region. Furthermore, the X-ray surface brightness does not appear to be disturbed in this direction (see Figure \ref{fig:A4}). 

We conclude that the large amount of substructure in the red sequence luminosity distribution along with the clearly disturbed X-ray morphology indicate merger activity that generated the radio relic. However, given the extreme stellar contamination, we can not draw conclusions to as where merging substructure is located. We will not use A746 for any further ensemble analysis. 

%% file: A781.tex
\subsubsection{Literature Review}
A781 is a well studied system, that is actually composed of multiple clusters at two different redshifts ($z=0.3\,\text{and}\,0.43$ separated by $\sim$240 Mpc along the line of sight). \citet{Jeltema:2005} studied it as part of a Chandra archival cluster survey. They briefly noted that it is a complex cluster with multiple peaks. \citet{Wittman06} obtained deep (R$\sim$27) multi-band optical imaging of the A781, since it is in Field 2 of the Deep Lens Survey \citep[DLS:][]{DLS}. They used this imaging to map the mass of the system via a weak lensing analysis.  The mass map showed good agreement with the Chandra X-ray map. However, the western most subcluster was not discussed (since it fell outside the Chandra footprint).There have been multiple mass estimate analyses performed on A781 including weak lensing, X-ray, and galaxy velocity dispersion \citep{Abate:2009,Kubo:2009,Cook:2012,Wittman:2014, Miyazaki:2015}. Mostly recently, \citet{Schneider:2017} implements a Gaussian Process based lensing technique and show good agreement with an analogous method to \citet{Wittman:2014}. 

In an overlapping spectroscopic survey of DLS Field 2, \citet{Geller:2005} found that the two subclusters at $z\sim0.3$ had redshifts of 0.302 and 0.291, respectively, and they found that the easternmost subcluster is at a redshift of 0.4265. \citet{Sehgal:2008} conducted two detailed X-ray analyses of the system using Chandra and XMM data, with their focus being on comparing X-ray and weak lensing mass estimates. They confirmed the previously reported subclusters \citep{Geller:2005, Wittman06}, as well as identified the West cluster in the larger field-of-view XMM image. In a later study \citet{Geller:2010} also measured the redshift of this subcluster to be 0.4273 thus confirming that the A781 field consists of at least two subclusters each at two redshifts. In total they have obtained $\sim$400 spectroscopic redshifts of the A781 system.  While not specifically isolated in their analysis the substructure within the main $z\sim0.3$ subcluster appears to be at the same redshift, \citep[see Figure 13 of][]{Geller:2010}.

\citet{Venturi:2008} were the first to carry out a detailed radio study of A781 as part of their GMRT radio survey. In addition to a number of extended radio galaxies, they noted a low surface brightness radio feature southeast of the Main cluster, without a matching optical counterpart, which they referred to as ``peripheral radio emission''. \citet{Cassano:2010} find A781 to be peculiar in that it is disturbed, yet they find no radio halo, which is in contrast with most of their sample. This is contradicted by \citet{Govoni:2011}, who claim to find evidence for a radio halo, as well as suggest that the radio emission identified by \citet{Venturi:2008} is a radio relic on the grounds of its radial steepening. \citet{Venturi:2011} question the claim of a radio halo, arguing that it would require an unusually flat radio spectrum since they do not detect it in their 325 and 610 MHz observations. 

Most recently, A781 was observed by LOFAR. Botteon et al. (in prep) present 140 MHz images and do not find evidence for the radio halo in the main cluster center. They constrain the 143 MHz flux to below 50 mJy.

\subsubsection{Results}

We utilize the redshift catalog of the SHELS redshift survey, which covers the DLS Field 2 with complete spectroscopic coverage to a limiting magnitude of $R=20.6$ using Hectospec \citep{Geller:2014}.  We combine known redshifts from the literature with the SHELS survey data to study A781. We identified 79 cluster members for the East group at with a redshift of 0.42645$\pm$ 0.00040 and a velocity dispersion of 766$\pm$64 km s$^{-1}$. We also identified 54 cluster members for the West group at a redshift of 0.42769$\pm$0.00048 and a velocity dispersion of 764$\pm$79 km s$^{-1}$. The east and west groups are separated by 21\arcmin\, in projection ($\sim$7 Mpc at $z=0.43$). We will not discuss these groups further since their X-ray surface brightness distributions indicate that they are isolated halos and are uninvolved in a merger. 

Meanwhile, there are 430 cluster members at the $z\sim0.3$ cluster redshift. The global velocity distribution is very poorly fit by a single Gaussian indicating large velocity differences between the substructure along the line of sight (see Paper 1). The global redshift and velocity dispersion are 0.29748$\pm$0.00034 and 1692$\pm$54 km s$^{-1}$, respectively. We note that this extreme velocity dispersion not indicative of an extremely massive system, but rather of a system with large line of sight velocity differences. 

We made use of archival Subaru/SuprimeCam imaging (V and i bands) and present the red sequence luminosity distribution (smoothed with a 40\arcsec\, Gaussian kernel) in Figure \ref{fig:A5} along with the DS-test for the 430 cluster members. Focusing in on the z$sim$0.3 galaxies, the source of the bimodal global redshift distribution is immediately evident with the a significant line of sight velocity difference between the east ($z\sim0.29$) and west subclusters ($z\sim0.3$). Additionally, there is composite structure within the west subcluster with at least three groups. The luminosity peaks in the middle with extended light toward the north and west. Each of these extensions in the red sequence luminosity distribution hosts a bright BCG. We center the priors for our GMM on these galaxy locations and select a conservative redshift prior straddling the individual BCG redshifts for each group and a conservative prior range for the Middle cluster as well (see Table \ref{tab:subclusters}). Given the large line of sight velocity difference and large physical separation, our code should have no problem separating these structures. We note that the radio relic is situated to the southeast of the Main cluster. From the DS-test and redshift sequence luminosity distribution, it is ambiguous which subclusters provide the merger to create the radio relic. 

Our GMM analysis of twelve models varying from one to four subclusters results in a four component model with the lowest BIC score. In addition, a three halo model without the halo b (see Table \ref{tab:subclusters}) was nearly as good a fit. We focus on the four halo model because we can identify four distinct BCGs in the optical image as well as evidence for four distinct subclusters in the X-ray surface brightness profile (see Figure \ref{fig:A5}). Subclusters 1, 2, and 3 each are at $z\sim0.3$ and subcluster four is at $z\sim0.29$. The projected location of the three subclusters at $z\sim0.3$ suggest a three way merger. The radio relic is situated in the southeast, which could indicate a merger between subclusters 1 and 3 or subclusters 2 and 3 that caused the radio relic. Inspection of the X-ray surface brightness shows that the gas is clearly more disturbed between subclusters 1 and 3 (green and purple in Figure \ref{fig:A5}).  In fact, the ICM of subcluster 2 is situated coincident with the red sequence luminosity peak and appears to have a generally undisturbed profile; although, more X-ray integration will be necessary to confirm. There are relatively small line of sight velocity differences between these two subclusters, which are fit with individual Gaussians and presented in Figure \ref{fig:A5}. 

In summary, these data provide a good example of the tremendous value of magnitude limited spectroscopy and combining it with multi-wavelength observations. Taking all evidence into account, the merger likely occurred between subclusters 1 and 3 launching the shock that created the radio relic to the southeast. Subcluster 2 is yet to reach core passage (although very near), and subcluster 4 is positioned along the line of sight with a large peculiar velocity difference. These four subclusters will likely eventually coalesce and form a single very massive cluster in the future.

%% file: A1240.tex
\subsubsection{Literature Review} 

\citet{Kempner:2001} first suggested that A1240 hosts a radio relic based on diffuse emission detected in WENESS and NVSS images and an elongated X-ray emission from a pointed ROSAT observation. They suggested a non-zero impact parameter due to the slightly off-axis location of the relics. \citet{Bonafede:2009} confirmed two relics with VLA observations at 325 MHz and 1.4 GHz, and found average spectral index values of $-1.2\pm 0.1$ and $-1.3\pm 0.2$ for the northern and southern relics, respectively. They also measured integrated fractional polarizations of 26\% and 29\% for the northern and southern relics, respectively, with both having values ranging up to 70\%. The most recent radio study was presented by \citet{Hoang:2018}. They present radio images across a large range of frequencies and radio telescopes including LOFAR 120-168 MHz, GMRT 595-629 MHz, and VLA 2-4 GHz. They show that the radio relic spectra steepen from their outer edges toward the cluster center and the electric field vectors are well-aligned with the radio relic major axes suggesting a high polarization, which they use to constrain the viewing angle of the merger to within 40$\degr$ of the plane of the sky.

\citet{Barrena:2009} analyzed SDSS and INT WFC imaging and found a bimodal galaxy distribution---like the relics, separated along a north to south axis.  They used 95 member redshifts to find a line of sight velocity difference between the two subclusters of $\sim 400$ km s$^{-1}$, and using a two-body model, they estimated a rest-frame 3D velocity difference of $\sim$2000 km s$^{-1}$.  They measured a LOS velocity dispersion for the system of $\sim 870$ km s$^{-1}$ and a corresponding mass of $\sim 0.9-1.9\times10^{15}$ M$_\odot$. \citet{Barrena:2009} also analyzed archival Chandra data (51 ks) and derived a temperature of $\sim$ 6 keV confirming a massive system. They also noted A1237 $\sim 10 \arcmin\,$ to the south at a similar redshift; however, it is likely uninvolved in the merger in the observed state based on the position of the radio relics. 
 
\subsubsection{Results} 

We observed two additional slitmasks for  A1240 with Keck DEIMOS bringing the total number of confirmed cluster members to 146. The cluster redshift is 0.19458$\pm$0.00029, the velocity dispersion is 853$\pm$62 km s$^{-1}$, and the redshift distribution is well fit by a single Gaussian. We also identify 24 cluster members for A1237 located $\sim$2 Mpc to the southwest of A1240.  A1237 has a redshift of of 0.19374$\pm$0.00053 and a velocity dispersion of 721$\pm$83 km s$^{-1}$. 

We observed A1240 in g and r bands with Subaru/SuprimeCam. The red sequence luminosity distribution is smoothed with a 52\arcsec\, Gaussian kernel and presented in Figure \ref{fig:A6}. A bimodal distribution aligned north-south between the radio relics is revealed suggesting a simple geometry. We select priors for the two subclusters for our GMM analysis based on conservative windows around the BCG redshifts and locations for these two subclusters (see Table \ref{tab:subclusters}). The two halo model is strongly favored by the BIC score (see Table \ref{tab:mcmcgmm}). The two inferred subclusters are well aligned with the red sequence luminosity distribution and between the two radio relics (see Figure \ref{fig:A6}). The redshift distributions for the two subclusters are presented in Figure \ref{fig:A6}. The two are very similar suggesting a 1:1 mass ratio in the plane of the sky. The line of sight velocity between the two subclusters is 394$\pm$117 km s$^{-1}$, in good agreement with the number reported by \citet{Barrena:2009} with fewer redshifts.

Inspection of the X-ray surface brightness distribution in Figure \ref{fig:A6} further suggests a major merger aligned north to south making A1240 among the cleanest bimodal mergers known and is a good candidate for further study. The fact that the radio relics straddle subclusters 1 and 2, and A1237 is located 2 Mpc south east strongly suggest it has not interacted in this merger; although, it likely will merger with A1240 in the future.  

%% file: A1300.tex
\subsubsection{Literature Review}

\citet{Pierre:1994} studied A1300 with X-rays, as part of RASS, and in the optical with the ESO 3.6 m Telescope. Of their 42 cluster RASS sample they found that it was the most luminous at $z>0.3$. The cluster was found to be optically rich and extended suggesting merger activity. \citet{Pierre:1997} conducted a spectroscopic and photometric (B and R) follow-up of the cluster. With 52 spectroscopic redshifts of cluster members they measured a velocity dispersion of 1210 km s$^{-1}$. While they noted apparent substructure in 2D projected space, the velocity histogram was well modeled by a single Gaussian. \citet{Lemonon:1997} presented deeper ROSAT observations of the cluster as well as added 10 more spectroscopic cluster member redshifts from the previously non-surveyed northern region. They argue that the multiple wavelength analysis suggests that A1300 is a major cluster merger that occurred at an early epoch. They argue that it is a merger where the clusters have already returned after their first pass through, since they do not detect an offset between the X-ray gas and the galaxies, nor do they detect evidence for substructure in the velocity histogram; although, this latter observation would be satisfied in a plane of sky merger at any epoch.

\citet{Giovannini:1999} included A1300 in a list of radio halo and relic candidates in NVSS. \citet{Reid:1999} discovered a number of diffuse radio sources in the vicinity of the cluster. They detect a number of tailed radio galaxies along the north-south axis of the cluster and one to the northwest. Additionally, they argue that the diffuse emission to the southwest is a radio relic and that the central diffuse radio source coincident with ROSAT detected diffuse X-ray emission is a radio halo. \citet{Venturi:2013} conducted GMRT follow-up observations of A1300 and largely confirm the \citet{Reid:1999} interpretation. However, they discovered another possible radio relic candidate to the northwest. Unlike most mergers, the two relics are discovered at 90\degr\, about the cluster center, and the radio halo is offset considerably from the bulk of the X-ray gas emission.

\citet{Ziparo:2012} conducted a detailed optical and X-ray analysis of the system utilizing WFI imaging, VIMOS spectra, and XMM X-ray spectral imaging. They note that the X-ray surface brightness map of A1300 appears disturbed and exhibits the signature of a shock consistent with the radio observations. They also claim the detection of a large-scale filament in which the main cluster is embedded. \citet{Ziparo:2012} compare their observations with existing simulations and argue that A1300 is a complex system where a major merger occurred $\sim$3 Gyr ago with continual minor mergers occurring as filamentary groups are feed into the cluster. Note that the galaxy density extensions to the northeast and southeast of the cluster center are inline with the two radio relics discovered by \citet{Reid:1999} and \citet{Venturi:2013}. However, it is unclear that minor mergers could produce bright radio relics.

\subsubsection{Results}

We observed A1300 with Subaru/SuprimeCam in g and r bands. We utilize 227 cluster member galaxies from \citet{Ziparo:2012}. We find a cluster redshift of 0.30550$\pm$0.00034 and a velocity dispersion of 1227$\pm$55 km s$^{-1}$ in agreement with previous studies. The red sequence luminosity distribution has a primary elongation along a northeast to southwest axis aligned with the radio relic. Additionally, there is an extension to the north that could explain the radio relic candidate (see Figure \ref{fig:A7}). The DS-test shows a group of large orange circles in the northeast; however, these galaxies have too low luminosity to stand out in the red sequence luminosity map. They are likely a well sampled low mass group that is too far from the cluster center to be dynamically important. Visual inspection of the SuprimeCam images do not suggest a rich subcluster. 

We place potential subcluster priors centered on the BCG of the three extended regions of the luminosity distribution, and center the redshift prior on the respective BCG redshifts (see Table \ref{tab:subclusters}). In total, five GMM models are run and compared with the BIC score. The single halo model is preferred. Inspecting the projected extent and density of the spectroscopic survey (see Figure \ref{fig:A7}) suggests that the three subclusters are not well separated in projected space. Furthermore, the spectroscopic survey has not densely sampled the inner regions of the subclusters instead sampling the global cluster fairly uniformly over an area $\sim3\times3$ Mpc. Thus, the GMM does not overcome BIC penalties of more complex models. 

The X-ray surface brightness profile (see Figure \ref{fig:A7}) is suggestive of a complex merger scenario. The main red sequence luminosity peak is associated with the core of the ICM; however, the ICM extends toward the north, east, and southeast. The core of the cluster (the highest luminosity peak in the south) has the brightest X-ray emission. While this portion of the cluster is aligned with the overall red sequence luminosity major axis and radio relic in the southwest, the ICM and BCG (see lower left panel of Figure \ref{fig:A7}) are aligned southeast to northwest, which further complicates the merger scenario.

Since the galaxy data alone do not support more complex subcluster models, we conclude with a mention of the BCGs for each potential subcluster. We listed our three proposed priors for the potential subclusters in Table \ref{tab:subclusters}. The redshifts of the three BCGs are 0.3082, 0.3016, and 0.3077, respectively. The confirmed radio relic to the southwest was likely launched by a merger between the second and third listed, which have a large line of sight velocity difference. This could explain the small angular projection between subclusters; i.e., the merger is occurring substantially along the line of sight. In order to explore this further, the cluster cores should be more densely sampled spectroscopically. 

%% file: A1612.tex
\subsubsection{Literature Review} 
A1612 is a little studied cluster.  \citet{vanWeeren:2011} provided the first targeted analysis of the system finding evidence for it being a post merger radio relic cluster. From ROSAT X-ray imaging they find an elongated gas distribution that is in agreement with the elongated (slightly bimodal) galaxy distribution based on SDSS imaging. With GMRT observations they find diffuse radio emission for which they cannot find an optical counterpart and classify it as a likely relic. The relic sits slightly off axis to the south of the cluster X-ray emission.  

More recently, \citet{Kierdorf:2016} observed the relic at 4.85 and 8.35 GHz with the Efelsberg 100-m telescope. Unfortunately, the large beam and nearby bright radio galaxy washed out the relic. A polarization of 20$\pm$7\% is estimated at 8.35 GHz near the relic location. Furthermore, Chandra X-ray data (29 ks) are presented, and the smoothed X-ray surface brightness map suggests a merger along a northwest to southeast axis, and X-ray point sources are detected in association with the  radio point sources. Each point source has an optical counterpart in SDSS and appears to be the BCG for each merging subcluster. A global X-ray temperature of $T_{X}=5.5^{+0.4}_{-0.3}$ keV and 0.1-2.4 keV luminosity of $L_{X}=1.8\times10^{44}$ erg s$^{-1}$ are estimated. 

\subsubsection{Results} 

We obtained one slitmask of Keck/DEIMOS spectroscopy, which when combined with spectra from SDSS \citep{sdss14} results in 73 cluster member galaxies. We made use of archival Subaru/SuprimeCam imaging in V and i bands. The global redshift distribution is well fit by a single Gaussian of redshift 0.18229$\pm$0.00049 and velocity dispersion 1081$\pm$96 km s$^{-1}$. Meanwhile, the red sequence luminosity distribution is bimodal along an axis from southeast to northwest. The radio relic is slightly south of this axis (see Figure \ref{fig:A8}). We place conservative priors on the peaks of the red sequence luminosity distribution (see Table \ref{tab:subclusters}) and the two halo model is preferred by the BIC score.  

The two subclusters are displayed in projected space in Figure \ref{fig:A8} showing good agreement with the red sequence luminosity distribution. In velocity space, the two subclusters appear to be similarly massive and have a small line of sight velocity difference of 167$\pm$198 km s$^{-1}$. Inspection of the X-ray surface brightness distribution shows elongation between the two subclusters. In summary, A1612 is undergoing a bimodal merger between similarly massive subclusters in the plane of the sky. The offset radio relic suggests a possible non-zero impact parameter; however, additional X-ray data, spectroscopy, and radio imaging will be necessary to understand this merger. 

%% file: A2034.tex
\subsubsection{Literature Review}

A2034 is a well studied system at all wavelengths. \citet{Kempner:2001} noted faint extended WENSS radio emission north of the cluster center coincident with a discontinuity in the RASS X-ray image. \citet{Kempner:2003} used Chandra to identify a cold front near this location and confirm that A2034 is a merging cluster. \citet{Giovannini:2009} identified diffuse radio emission coinciding with the center of the cluster but noted that it appears elongated and irregular with respect to other radio halos. They also confirmed the diffuse radio emission north of the cluster (but connected to the other diffuse emission) and its coincidence with the X-ray cold front. With deeper Chandra data \citet{Owers:2014} show that this cold front is actually a shock with a Mach number of 1.59$\pm$0.06 corresponding to a shock velocity of $\sim$2000 km s$^{-1}$. \citet{Kempner:2003} also argued that the observed southern X-ray excess is not associated with the cluster and is likely a background cluster between $0.3 < z < 1$, however \citet{Owers:2014} infer that the excess is consistent with it being gas stripped during the merger.

With WSRT \citet{vanWeeren:2011} observed the diffuse emission to brighten near the position of the shock. They note that it is debatable whether this really is a radio relic. They also identify a small candidate radio relic ($\sim$200 kpc$\times$75 kpc) to the west of the cluster center. \citet{vanWeeren:2011} also note that the galaxy distribution appears bimodal, with two relatively equal distributions. However, \citet{Okabe:2008} find a more complex galaxy distribution based on luminosity density maps using Subaru imaging. \citet{Owers:2014} use 328 spectroscopically confirmed cluster members to show that there is a substructure located at the front edge of the shock and that the galaxy distribution is more complex. Given the redshifts of this northern and main structures as well as the X-ray shock velocity they estimate the merger is occurring along an axis within $\sim$23 degrees of the plane of the sky and observed $\sim$0.3 Gyr after core-passage. \citet{Okabe:2008} also present the only weak lensing analysis and estimate a $M_{200}=10.24\pm6.14\times10^{14}$ M$_\odot$, which is consistent with the \citet{Geller:2013} spectroscopic velocity caustic estimate of 8.0$\pm$0.1$\times$10$^{14}$ M$_\odot$. Both of these mass estimates treat A2034 as a single system when estimating the mass.

Most recently, \citet{Shimwell:2016} present low frequency LOFAR radio observations and detect a complex network of steep spectrum radio features including a radio halo as well as up to three candidates for radio relics. One of these relics is coincident with the shock detected by \citet{Owers:2014} and another is the same relic candidate detected by \citet{vanWeeren:2011}.

\subsubsection{Results}

We make use of 140 archival spectroscopic cluster members from NED. We also utilized archival g and R band images from Subaru/SuprimeCam. The R band image is in fact the deepest image in our sample. The spectroscopy indicates a global redshift of 0.11381$\pm$0.00033 and a global velocity dispersion of 1080$\pm$55 km s$^{-1}$. The velocity histogram is well fit by a single Gaussian. Meanwhile, the red sequence luminosity distribution indicates at least three subclusters. The two brightest peaks are oriented north to south with the northern group actually slightly north of the diffuse radio emission identified as a radio relic candidate, which is unusual and challenges the picture of the relic tracing the shock generated by this possible north to south merger. The third peak in the red sequence luminosity distribution is situated to the southwest and has lower surface brightness. 

We generated three subcluster priors (Table \ref{tab:subclusters}) and the GMM results in the three subcluster model with the lowest BIC. The three subclusters have relatively large line of sight velocity differences compared to other systems in our sample (1240$\pm$147 km s$^{-1}$ between subclusters 1 and 2 and 778$\pm$149 km s$^{-1}$ between subclusters 1 and 3). Subclusters 1 and 2 are similar in velocity dispersion, and subcluster 3 is a low mass subcluster. Inspection of the X-ray surface brightness distribution (see Figure \ref{fig:A9}) further indicates that the primary merger is between subclusters 1 and 2. Interestingly, the northern subcluster 1 is ahead of the shock, which is hard to explain if the merger is bimodal, but could be explained if the relic is patchy and viewed in projection as the line of sight velocity difference suggests. We are unable to rule out more complex merger scenarios.

%% file: A2061.tex
\subsubsection{Literature Review}

A2061 is a member of the Corona Borealis supercluster, which is collectively at $z\sim0.07$. This supercluster is extensively studied by SDSS thanks to its low redshift and full coverage by the SDSS footprint. A2061 is located $\sim$30$\arcmin$ (2.7 Mpc) southwest of Abell 2067 (hereafter A2067), with a relative line of sight velocity difference of $\sim$600 km s$^{-1}$, and to which it is likely gravitationally bound \citep{Rines:2006}. This velocity difference is significantly lower than the $\sim$1600 km s$^{-1}$ estimate of \citet{Oegerle:2001}. \citet{Rines:2006} note that A2061 is about 4 times more luminous in X-rays than A2067. \citet{Marini:2004} note that the X-ray profile of A2061 is elongated in the northeast to southwest along the same axis as A2067, however \citet{vanWeeren:2011} note that this elongation may be due to observed optical substructures within A2061 rather than an interaction with A2067. They also note that these substructure interactions within A2061 are likely the cause of the observed radio relic to the southwest of A2061 \citep{Kempner:2001, Rudnick:2009,vanWeeren:2011}. 

\citet{Abdullah:2011} carried out an updated velocity dispersion analysis of A2061 finding $\sigma_v = 725 \pm 67$ km s$^{-1}$ corresponding to a viral mass of 5.45$\pm$ 0.19$\times$10$^{14}$ M$_\odot$. \citet{Einasto:2012} carried out a GMM substructure analysis of the A2061/A2067 system and found that it was best fit by a three component mixture, but they do not present the spatial or redshift distributions of these substructures. The Planck Collaboration \citep{Planck_filaments:2013} included the A2061/A2067 system in their study of filaments between interacting clusters but do not note anything specific about this system. \citet{Farnsworth:2013} used Green Bank Telescope observations and found evidence for a possible inter-cluster filament between A2061 and A2067. They also find the first evidence of a radio halo in A2061 at 1.4 GHz.

\subsubsection{Results}

We obtained 157 cluster member redshifts for A2061 from SDSS \citep{sdss14} and NED. We observed the system with Subaru/SuprimeCam in g, r, and i bands. The velocity distribution is well fit by a single Gaussian of redshift 0.07805$\pm$0.00025 and velocity dispersion 841$\pm$55 km s$^{-1}$. The red sequence luminosity is bimodal in a northeast to southwest axis. The northeast bright peak is also elongated along this axis (see Figure \ref{fig:A10}).

We run one and two subcluster models with our GMM; the one halo model is preferred. Similar to A1300, we believe this is due to small projected separation between the two subclusters and also the spectroscopic sampling, which does not densely probing the cores of the two subclusters. Evidence from the bimodal red sequence luminosity distribution, extended X-ray emission, and radio relic in the southwest suggest a merger along this axis; however, the spectroscopic data are unable to further confirm this picture given the additional model parameters necessary. 

%% file: A2163.tex
\subsubsection{Literature Review}

A2163 is one of the richest Abell clusters and has been extensively studied since it was identified as an exceptionally hot system \citep[15 keV;][]{Arnaud:1992}. This cluster has been identified as the most massive in Planck SZ catalog \citep{PlanckMass}. Many X-ray studies have found evidence that the cluster has a non-isothermal gas distribution with strong temperature variations in the center of the cluster \citep{Markevitch:1994, Elbaz:1995,Markevitch:1996, Markevitch:2001, Govoni:2004, Ota:2014}. \citet{Feretti:2001} were the first to identify a radio halo centered on the cluster and a potential radio relic $\sim$2.2 Mpc to the northeast of the cluster center. \citet{Feretti:2004} followed up the initial radio observations and measured the spectral index associated with the radio emission. Note that the hottest part of the X-ray emission \citep[18 keV;][]{Bourdin:2011, Ota:2014} is correlated with the location of the radio relic in the northeast of the main cluster. Many of the aforementioned X-ray studies along with \citet{Rephaeli:2006} have attempted to measure a non-thermal component due to inverse Compton scattering of CMB photons by the relativistic population of particles suspected to be associated with the diffuse radio emission, but no conclusive evidence of non-thermal emission has been observed yet. SZ analyses have been completed as well \citep{Wilbanks:1994,Nord:2009}. The latter study found a low line of sight velocity difference and results consistent with an extremely massive cluster. The most recent radio analysis covered low frequencies between 88 MHz and 200 MHz using the Murchison Widefield Array. The radio relic and halo were detected in all radio images, and the spectral index for the halo and relic between 88 MHz and 1.4 GHz were estimated to be -0.90$\pm$0.19 and -1.05$\pm$0.19, respectively \citep{George:2017}. 

A number of weak lensing analyses of the cluster have been carried out \citep{Squires:1997, Radovich:2008, Okabe:2011, Soucail:2012}. \citet{Okabe:2011} and \citet{Soucail:2012} both find evidence for a bimodal mass distribution with the peak of the X-ray gas located between the two mass peaks. Additionally, \citet{Soucail:2012} identifies a mass peak $\sim$6$\arcmin$ ($\sim$1.2 Mpc) to the north of the main cluster, which corresponds to substructure A2163-B identified in a spectroscopic survey of the area by \citet{Maurogordato:2008}. They argue that this structure will eventually merge with the main subcluster. \citet{Soucail:2012} estimate that the main cluster is composed of two subclusters of masses 7.1 and 2.5$\times$10$^{14}$ M$_\odot$ that have undergone their first pass through and that the infalling A2163-B has a mass of $\sim$2.7$\times$10$^{14}$ M$_\odot$. There is some discrepancy between these and the \citet{Okabe:2011} mass estimates, which suggest a 8:1 to 10:1 mass ratio of the merging subclusters in the main cluster and a total mass for the system that is about 50\% larger than the \citet{Soucail:2012} estimate. \citet{Bourdin:2011} estimate a 4:1 mass ratio between the A2163-A subcluster components and conduct a timing-argument dynamics analysis which they suggest implies that the A2163-A subclusters collided $\sim$0.1-1.0 Gyr ago. All of these weak lensing analyses are based upon ground based weak lensing, and HST data reveals a more unsettled core than a simple bimodal structure (private communication with Doug Clowe). Most recently, \citet{Cerny:2017} produced a strong lensing analysis of the core of the northeast (roughly the same region as the bottom left panel of Figure \ref{fig:A11}). This analysis is part of the HST RELICS campaign. Of five systems in \citet{Cerny:2017}, A2163 has the lowest lensing strength despite the fact that it is the highest mass cluster in the Planck catalog.

\subsubsection{Results}

We base our analysis upon 382 archival spectroscopic galaxies from NED and \citet{Maurogordato:2008} as well as archival Subaru/SuprimeCam imaging in V and R bands. The spectroscopy indicates an extremely massive system with a redshift of 0.20115$\pm$0.00035 and a velocity dispersion of 1469$\pm$57 km s$^{-1}$. The redshift distribution is well fit by a single Gaussian. 

The red sequence luminosity distribution is clearly bimodal between the northern and main components identified by \citet{Soucail:2012}. The main cluster has three brightness peaks aligned east to west. The cluster BCG is in the easternmost subcluster. The redshifts of each subcluster are similar (see Figure \ref{fig:A11}), which poses a challenge for our GMM with small offsets in the projected spatial dimensions as well. We center uniform priors on the RA, DEC, and redshift coordinates of the respective BCGs for the four peaks in the red sequence luminosity distribution and implement our GMM on possible models ranging from a single halo to four (see Tables \ref{tab:subclusters} and \ref{tab:mcmcgmm}). The lowest BIC model is the single halo model.This is not supported by the clear multi-modality in the projected distribution of red sequence light and the clear merger scenario presented in the literature. However, this discrepancy between the literature and our GMM results in to be expected because the cluster is extremely massive, so small velocity differences are hard to detect, which effectively reduces the GMM to a two dimensional test. Furthermore, the projected offsets are very small, so it is more economical to place a single halo covering all of the data than to add model parameters in order to split the data. Finally, as we mentioned in \S\ref{subsec:A1300} and \S\ref{subsec:A2061}, when the spectroscopic survey does not densely sample the peaks of the galaxy populations, the code is more likely to place single halos on the diffuse sampling of galaxies. This is partially by design with the BIC analysis, which identifies the minimum number of subclusters necessary to describe the spectroscopic data. 

Given the completeness of the spectroscopy in A2163 and the clear evidence for a major merger, we altered some of the priors for this cluster. We placed a variable velocity dispersion prior on each subcluster. Table \ref{tab:subclusters} lists the priors for the RA, DEC, and redshift. We allowed halos a, c, and d to have velocity dispersions between 500 and 1000 km s$^{-1}$ while subcluster b was allowed to vary between 800 and 1400 km s$^{-1}$ since it is clearly the core of the main cluster.  We also lowered the minimum subcluster size, defined as the projected radius within which 68$\%$ of the assigned galaxies must reside. For all clusters, this prior has been allowed to vary between 0.25 and 1 Mpc. We lowered the minimum value to 150 kpc for A2163. The single halo model was still preferred, but a three halo model is only slightly disfavored by the BIC. Interestingly, the three-halo model matches well with results in the literature. We present this model hereafter due to the abundant information indicating the cluster is in a major merger, effectively using these findings as a strong prior for this scenario.

The projected locations and velocity distributions of the three subclusters are presented in Figure \ref{fig:A11}. Subclusters 2 and 3 appear to compose the major merger. The radio relic is far to the northeast suggesting a well evolved merger. A possible scenario is that these two subclusters passed through each other long ago allowing the radio relic to reach such a distant position. More recently, the fourth subcluster that has been identified in the X-ray \citep{Bourdin:2011} as a cool-core remnant seems to have plunged through the cluster center (see the merger scenario in the bottom right of Figure \ref{fig:A11}). \citet{Randall:2002} showed that mergers enhance ICM observables and bias high related mass estimates for a short time after core passage. It appears this ongoing merger may be biasing the mass determination associated with SZ observations. The galaxies associated with this subcluster are too near subcluster 2, to be disentangled using the currently available redshift data. In order to offer further interpretation of the cluster merger scenario, much more detailed simulations will be necessary. A full HST mosaic of the cluster may be necessary to deblend the various mass components in the cluster center given the conflicting results from the various ground-based lensing studies.

%% file: A2255.tex
\subsubsection{Literature Review}

\citet{Tarenghi:1976} were the first to discuss A2255 as a potential merging system with fifteen spectroscopic cluster members. They used the velocity dispersion and X-ray emission to show that the system did not follow a $\sigma_{v}$--$L_{X}$ relation. \citet{Hintzen:1980} followed with a spectroscopic analysis of radio galaxies. They showed that the cluster must be $\sim\text{4}\times\text{10}^{14}\,M_\odot$ in order to provide a large enough ICM halo to support the observed radio emission. \citet{Burns:1995} presented the first multi-wavelength analysis with X-ray, radio, and optical observations. The X-ray analysis identified a potential cool core (2 keV) as opposed to the cluster global temperature of $\sim$ 7 kev. Furthermore, the cluster was reported to have a velocity dispersion of 1240 km s$^{-1}$. The two BCGs have a line of sight velocity difference in excess of 2000 km s$^{-1}$. Finally, \citet{Burns:1995} discovered a steep spectrum radio source that had been speculated to be a radio halo; however, this emission sits 0.6 Mpc to the north of the X-ray center. \citet{Burns:1995} preferred merger scenario is a merger between the bulk of the cluster with a galaxy group; this scenario is substantiated by numerical hydrodynamical/N-body simulations, which reproduced X-ray and radio features of the cluster. 

A series of X-ray and radio analyses followed and generally agreed with the previous findings. \citet{Feretti:1997} classified the radio source as a radio relic and found the global X-ray temperature to be lower than previous findings at 3.5$\pm$1.5 keV. \citet{Davis:1998} agree with \citet{Burns:1995} with the infalling group picture; however, they suggest a much more energetic merger between larger components that could create the observed X-ray temperature variations and offsets between the X-ray surface brightness and BCGs. \citet{Miller:2003} present a deep VLA 1.4 GHz analysis and detect an abnormal abundance of radio cluster members, and it is proposed that the merger phase and geometry explain the surplus. \citet{Govoni:2005} present deeper VLA 1.4 GHz data and report strong evidence for ordered magnetic fields in the form of strongly polarized regions extending $\sim$400 kpc. The radio relic sits to the northeast of the cluster center and also exhibits polarized emission. \citet{Sakelliou:2006} present XMM-Newton observations of the cluster scale ICM and reach similar conclusions to previous X-ray studies. The global temperature is 6.9 keV with the east generally cooler than the west. \citet{Pizzo:2009} detect two additional radio relic candidates with one opposite the cluster center of the previously detected northeast relic. Additionally, they study the spectral index of the radio features and find steepening in the northeast radio relic toward the cluster center. Most recently, \citet{Akamatsu:2017} study the relic region with Suzaku and confirm the presence of a shock associated with the relic via density and temperature jumps in upstream and downstream regions relative to the relic location. The temperature ratio suggests a Mach number of $\sim1.4$, which is lower than expected from the radio data suggesting pre-accelerated elections in the shock region. 

\citet{Yuan:2003} present a very detailed study of the galaxy population with SDSS and 13 additional filters between 3000 and 10000\AA. Photometric and spectroscopic observations resulted in more than 500 cluster members that are distributed into a single over density with east--west elongation. Several groups are identified at the periphery suggesting the cluster is still dynamically active. 

\subsubsection{Results}

We obtained 268 spectroscopic cluster members from NED and SDSS \citep{sdss14} and archival Subaru/SuprimeCam imaging in B and R bands. The spectroscopic cluster members are well fit with a single Gaussian with redshift 0.08012$\pm$0.00024 and velocity dispersion 1137$\pm$50 km s$^{-1}$. As reported earlier, the red sequence luminosity distribution is elongated east to west. There are two BCGs in the west and one in the east of this elongation. Inspection of the two BCGs in the west suggests they are moving along the line of sight with relative velocities of $\sim$2000 km s$^{-1}$. Given the presence of galaxies at each corresponding redshift in the area, we place priors for a subcluster at each redshift in projection, and a third in the east centered on the redshift and projected location of the third BCG. A total of five models ranging from a single cluster to these three subclusters are run with the GMM. The BIC score does not support multiple subclusters. We again attribute this to the spectroscopic sampling, which is too sparse in the dense cluster center to pull apart the various subcomponents. 

The radio relic is situated to the northeast, but does not have a clear pair of subclusters with an orientation that easily explains its presence. Given the potential of a three way merger with large line of sight velocities, complex merger scenarios are feasible, but they will require more spectroscopy in the cluster center along with detailed simulations to explore in full. See Figure \ref{fig:A12} for a proposed merger scenario. In the lower right, the two line of sight BCGs are visible in the same frame. 

%% file: A2345.tex
\subsubsection{Literature Review}

\citet{Dahle:2002} performed an optical and weak lensing analysis of Abell 2345 finding that the luminosity and number density distributions of galaxies have a peak near the ``well-defined core" that is dominated by a cD galaxy (also the BCG). However, they note that archival ROSAT X-ray images show a large amount of substructure and suggest that this is evidence that the cluster may be in a dynamically young state. They do find that the highest peak in their weak lensing map is offset east of the cD galaxy by $\sim$1.5$\arcmin$, but that a secondary peak is much closer to the cD galaxy. Note that the \citet{Dahle:2002} analysis only covers the eastern subcluster. \citet{Cypriano:2004} also estimated the mass of the system via weak lensing but did not discuss A2345 in detail. They fit a single isothermal sphere to the convergence map and estimate $\sigma_{SIS}=909\pm138$ km s$^{-1}$, which is in agreement with \citet{Dahle:2002}.

\citet{Giovannini:1999} were the first to identify A2345 as a radio relic cluster in NVSS, identifying two extended sources and making it the second double relic system discovered \citep[the first being A3667,][]{Rottgering:1997}. \citet{Bonafede:2009} confirmed the presence of the two relics with VLA observations at 325 MHz and 1.4 GHz. They note several discrete radio sources near the western relic, but none are expected to produce diffuse radio emission. They estimated average spectral index values of 1.5$\pm$0.1 and 1.3$\pm$0.1 for the eastern and western relics, respectively. \citet{Bonafede:2009} also measured mean fractional polarizations of 14\% and 22\% (with values up to 55\%) for the eastern and western relics, respectively. The ROSAT X-ray image of the system indicates a northwest to southeast elongation that fits the radio relic picture. They also make a rough calculation based on the approximate Mach number that the collision velocity was $\sim$1200 km/s and the merger occurred $\sim$0.4 Gyr before the observed state.  The most recent radio analysis was completed by \citet{George:2017} using the Murchison Widefield Array at low frequency between 88 and 215 MHz. They confirm the presence of both radio relics as well as a number of unresolved radio point sources. Spectral indices of -1.29$\pm$0.07 and -1.52$\pm$0.08 were measured between 88 MHz and 1.4 GHz for the east and west relics, respectively.

\citet{Boschin:2010} conducted a redshift analysis of the system using 98 cluster member redshifts. They claim that there are three separate subclusters (E, NW, SW), based on their photometric and spectroscopic analysis with the NW and SW subclusters having similar mean velocities but $\sim$800 km s$^{-1}$ velocity difference compared to the E subcluster. They claim that the NW-SW subcluster axis is a natural explanation for the western relic while the E-NW-SW subcluster axis roughly coincides with eastern relic. However, they note that if this is the case it is perhaps surprising that there is not a third relic in the south. \citet{Boschin:2010} estimate the system mass to be $\sim$2$\times$10$^{15}$ M$_\odot$, but they note uncertainty due to the close proximity of the subclusters in projected and redshift space. They also performed a rough timing argument dynamics analysis of the two possible mergers. For the E and NW+SW they argue that merger axis is $\sim$40$\degr\,$ with respect to the plane of the sky, and is $\sim$0.35 Gyr since they collided with a velocity of $\sim2000-2800$ km s$^{-1}$. For the NW-SW scenario they estimate that the merger axis is $\sim2-20$$\degr\,$ and that they collided $\sim$0.2 Gyr ago with a velocity of $\sim300-1400$ km s$^{-1}$. This dynamic analysis should be considered a very rough estimate since the system is not well suited to the assumptions implicit with the timing argument (e.g. it is a multi-modal system with overlapping halos).

\subsubsection{Results}

We obtained the spectroscopic redshifts presented in \citet{Boschin:2010} and combined with the available spectra from NED. We also obtained archival V and R band Subaru/SuprimeCam images for our photometric analysis. The redshift distribution is well fit with a single Gaussian of redshift 0.17881$\pm$0.0045 and a velocity dispersion of 1158$\pm$88 km s$^{-1}$ based upon 101 spectroscopic cluster members. This is consistent with the previous results. The red sequence luminosity map has three peaks roughly in agreement with \citet{Boschin:2010}; however, their analysis of photometric galaxies was not weighted by luminosity, and the brightest galaxies in the NW over density are situated nearly coincident with the radio relic in that direction. This is be difficult to explain if the NW subcluster was involved in the merger that launched the western radio relic. 

We place subcluster priors on the three luminosity peaks and center the redshift window conservatively around the respective BCG redshifts. The GMM is run for five models ranging from one to three subclusters. The BIC score prefers the single halo model, which we again attribute to the lack of spectroscopic coverage in the high density regions; although, in this case the overall number of spectroscopic cluster members is small regardless of their position. We conclude that more spectra are needed to complete a full substructure redshift analysis and report only the respective BCG redshifts for ensemble analysis in \S\ref{sec:results_sample}. Finally, we note that the X-ray surface brightness map from XMM-Newton EPIC suggests a complex merger (see Figure \ref{fig:A13}). This is anomalous among most double relic systems, which tend to be bimodal. 

%% file: A2443.tex
\subsubsection{Literature Review}

A2443 was part of the MX Northern Abell Cluster Survey II \citep{Miller:2002}. They obtained 12 spectroscopic redshifts of cluster members and estimate a velocity dispersion of 975 km s$^{-1}$. \citet{Wen:2007} conducted a 14 medium band optical imaging survey of the A2443 where they reached a limiting magnitude of i$\sim$20 (note that the seeing was 5.2$\arcsec$). From photometric redshifts they identified 301 galaxies at the cluster redshift. In the same field they observed ZwCL 2224.2+1651 ($z\sim0.1$) located 15.7$\arcmin$ ($\sim$2.5 Mpc) to the southeast of A2443, but note that there is insufficient evidence to determine whether the two clusters are interacting. A2443 appears to be elongated in the direction of ZwCL 2224.2+1651. \citet{Wen:2007} observe that the blue fraction of galaxies is much larger in ZwCL 2224.2+1651 compared to A2443.

\citet{Cohen:2011} observed A2443 in the radio from VLA at 1425, 325, and 74MHz and find a number of diffuse radio sources. Two head-tail radio galaxies in the northwest show bent tails suggesting motion towards the northwest (consistent with the elongation of the cluster) and possibly suggestive of outgoing subcluster motion post pericenter. In addition to five head-tail radio galaxies they find evidence for a diffuse radio relic with east-west elongation. While there are a number of potential point source contaminants in this region, they note that the spectral index of the diffuse emission is too steep to easily be explained by radio galaxy emission and argue that it is much more likely to be a radio relic.

\citet{Clarke:2013} observed A2443 with 15 ks of Chandra X-ray observations and found evidence that the ICM is highly disturbed and elongated along a northwest to southeast axis consistent with the elongated galaxy distribution. They also find two X-ray surface brightness edges, one along the northeast face and one along the southeast face. The southeast edge is largely coincident with the radio relic candidate from \citet{Cohen:2011}, suggesting that it may be the result of shocked ICM from a merger. 

\subsubsection{Results}

We observed A2443 with two slitmasks with Keck DEIMOS, combined the spectroscopy with archival data, and obtained 157 cluster members. A2443 is the only cluster in our sample without Subaru SuprimeCam imaging; however, the cluster is covered with SDSS and is of sufficiently low redshift such that the SDSS imaging is sufficient to trace the red sequence. The spectroscopy is well fit by a single Gaussian with redshift 0.10979$\pm$0.00023 and velocity dispersion 780$\pm$45 km s$^{-1}$ which is $\sim$200 km s$^{-1}$ less than previous results based upon far fewer galaxies. 

We utilized SDSS g and r band catalogs to map the red sequence. The cluster is elongated and bimodal along a southeast to northwest axis in agreement with the X-ray emission. We placed subcluster priors on the respective BCGs in the northwest and southeast and ran one and two subcluster models with the GMM. The two halo model was preferred according to the BIC score, and the results suggest a plane of the sky merger with the an unequal mass ratio.

The radio relic in A2443 is unlike other relics in our sample. It is the only ``roundish'' relic \citep{Feretti:2012} in our sample, and the ultra-steep spectrum and proximity to the cluster center suggests it is a radio phoenix class relic. These greatly differ from the larger scale relics such as the Toothbrush and Sausage relics found in 1RXSJ0603 and CIZAJ2242, respectively. Phoenixes appear to be more closely linked with the more localized radio plasma of  AGN. This explains why the relic is positioned much closer to the center of the cluster than others in our sample. This also suggests that the shock is much closer to the center of the cluster than other systems, which implies that A2443 is a young merger. The small projected offset between subclusters supports this picture.

%% file: A2744.tex
\subsubsection{Literature Review}

A2744 is an extremely well observed and analyzed cluster, and is known to be a complex system with multiple merging subclusters \citep[see e.g.][]{Merten:2011}. The first indications of merger activity came from the detection of a radio halo and radio relic \citep{Giovannini:1999,Govoni:2001a,Govoni:2001b}. The relic is situated to the northeast of the halo and has a linear extent of $\sim$2 Mpc. \citet{Kempner:2004} presented the first detailed X-ray analysis with Chandra and concluded a major merger is occurring north to south with a large line of sight velocity component. Additionally, a smaller northwest component is infalling creating a possible bow-shock. \citet{Owers:2011} generally agreed with this picture and with deeper Chandra observations detect a shock to the southeast of the cluster center. Note that no shock is identified to lie coincident with the radio relic. \citet{Owers:2011} also presents a detailed galaxy analysis with $\sim$300 spectroscopic cluster members. They present evidence for three subclusters undergoing a major north to south merger. They also find an interloping subcluster associated with the northwest ICM component identified in the X-ray observations. The main two components (north and south) have a large velocity difference, which substantiates the large line of sight component to the merger axis. More recently, \citet{Eckert:2016} detected a shock coincident with the radio relic using XMM-Newton and Suzaku data. Most recently, \citet{George:2017} used the Murchison Widefield Array to study the low frequency radio emission of the cluster between 88 MHz and 215 MHz. The radio halo and relic were detected at all observed frequencies. The spectral index between 88 MHz and 1.4 GHz was measured to be -1.09$\pm$0.05 and -1.01$\pm$0.07 for the halo and relic, respectively. 

A2744 is a member of the Frontier Fields \citep{Lotz:2014} sample of clusters, and thus the deep HST observations have allowed for very detailed strong lensing analyses of the core of the mass distribution. \citet{Merten:2011} presented the first joint strong/weak lensing analysis along with X-ray and galaxy analyses. It was found that A2744 is a complex merger with at least four subclusters including an enhanced ram-pressure slingshot \citep[see e.g.][]{Hallman:2004}. \citet{Merten:2011} estimated the self interaction cross section of dark matter based on offsets between X-ray and lensing peaks resulted in one of the first upper limits of $\sigma_{DM}<$ 3$\pm$1 cm$^{2}$g$^{-1}$.

Several lens models have been calculated for A2744 \citep[][with more associated with the HST Frontier Fields]{Lam:2014,Johnson:2014,Jauzac:2015,Kawamata:2016, Mahler:2017}. These strong lensing analyses mainly probe the cluster center where a large line of sight velocity difference between at least two subclusters has been identified. Updated strong and weak lensing analyses over the wider field have confirmed a massive ($M_{200}=\text{2.06}\pm\text{0.42}\times\text{10}^{\text{14}}\,\text{M}_\odot$) and complex merging system \citep{Jauzac:2016,Medezinski:2016}. Interestingly, the results in \citet{Medezinski:2016} confirm large offsets between mass, galaxy, and X-ray components ($\sim$70--80$\arcsec$). 

Additional evidence relating to the merger includes the discovery of four jellyfish type galaxies undergoing recent starburst \citep{Owers:2012}. Further study of the star formation in cluster galaxies have found evidence of shock induced star formation \citep{Rawle:2014}. The number of massive halos have raised the question of compatibility with $\Lambda$CDM structure formation models. \citet{Jauzac:2016} searched for analogs in the MXXL simulation \citep{MXXL}, and \citet{Schwinn:2017} use extreme value statistics to estimate that a volume ten times the MXXL volume would be required to find an analog. 

\subsubsection{Results} 

We utilized archival spectroscopy from NED and B and R band images from Subaru SuprimeCam. The spectroscopic catalog of 380 galaxies is well fit by a single Gaussian despite the clear multi-modality in the red sequence luminosity map. The global cluster redshift and velocity dispersion are 0.30590$\pm$0.00035 and 1602$\pm$61 km s$^{-1}$, respectively. The red sequence luminosity distribution has three distinct peaks. The cluster core has a line of sight bimodal distribution of redshifts \citep{Owers:2011,Mahler:2017} (see Figure \ref{fig:A15}). The northern peak in the red sequence luminosity distribution is elongated east to west and coincident with two peaks identified in lensing studies. Additionally there is a lower redshift, western peak associated with a known lensing over-density. We place five subcluster priors conservatively on these galaxy light peaks and set redshift priors based on the BCG redshifts of the respective subclusters. A2744 is the most complicated cluster of our sample, and we run 27 separate models with our GMM ranging from one to five subclusters.  

The lowest BIC model is a four component model with the northern luminosity peak merged into a single subcluster, and the main halo center (where strong lensing analyses have been completed) split into two line of sight groups with a large line of sight velocity difference, and a fourth subcluster associated with the westernmost luminosity peak (see Figure \ref{fig:A15}). There is no obvious pair of subclusters aligned to explain the radio relic in the east; however, subcluster 1 has an east to west elongation. Given the complex state of the merger, it is difficult to fully explore merger scenarios without detailed simulations. Inspecting the X-ray surface brightness (Figure \ref{fig:A15}) indicates that the northern luminosity peak is actively merging given the large displacement between cluster light and the ICM. Furthermore, subcluster 4 (in the west) appears to have disrupted the ICM suggesting it has passed from east to west to reach its present position. 

%% file: A3365.tex
\subsubsection{Literature Review} 

A3365 is a little studied system with one radio relic and one radio relic candidate discovered in NVSS and observed with VLA and WSRT at 1.4 GHz \citep{vanWeeren:2011}. They observed the cluster with INT and generated galaxy number density contours which suggest three separate galaxy peaks. One of the peaks is another cluster (RXC J0548.8-2154), which sits 9$\arcmin$ to the west of the center of A3365 at the same redshift. \citet{vanWeeren:2011} determined that together the system composed of A3365 and RXC J0548.8-2154 comprise the merging system. The X-ray peak is located closer to RXC J0548.8-2154, but is clearly disturbed. We will refer to the whole system as A3365.

\subsubsection{Results} 

We obtained spectroscopic redshifts with five Keck DEIMOS slitmasks and photometry in g and r bands with Subaru SuprimeCam. The spectroscopic survey resulted in 150 cluster members well fit with a single Gaussian of redshift 0.09273$\pm$0.00028 and velocity dispersion 981$\pm$58 km s$^{-1}$. The red sequence luminosity distribution has three peaks. Two of these have similar redshifts, and the western-most has a higher redshift with a peculiar velocity of $\sim$1900 km s$^{-1}$. We place subcluster priors on each over-dense region with redshift windows conservatively straddling the respective BCG redshifts. We run one, two, and three component models with the GMM and find the three component model best fits the data according to the BIC score.  

Inspection of the projected (RA, DEC) space distribution of the three subclusters suggests the east and middle subclusters likely merged to launch the shock associated with the radio relic. There is no clear alignment of subclusters with the radio relic candidate, which sits just north of the middle subcluster. These subclusters have similar redshifts, while the west subcluster has a large peculiar velocity. The east subcluster contains most of the mass of the cluster given its higher velocity dispersion.

Inspection of the X-ray surface brightness map substantiates the merger picture between the east and middle subclusters. The gas has mostly stayed in the east and been stripped from the middle subcluster in its apparent westward trajectory. Within the east, there are two bright galaxies along the extended ICM, which may indicate ongoing merger activity within the east. Finally, although at the edge of the XMM field of view, the western subcluster appears to have its ICM coincident with the galactic light distribution suggesting it has not yet merged.

%% file: A3411.tex
\subsubsection{Literature Review}

A3411 was included in the CIZA cluster catalog \citet{Ebeling:2002} and Planck SZ cluster catalog \citet{Planck_catalog:2011}, but \citet{vanWeeren:2013} were the first to publish a detailed analysis of this system. Their shallow (10 ks) Chandra X-ray map showed that the cluster gas has a cometary shape with the head located near the center of the northern subcluster (A3411) and the tail pointing in the direction of the southern subcluster (A3412), providing clear evidence that this is a dissociative cluster merger. Their VLA 1.4 GHz observations show diffuse emission near the northern subcluster, which they classify as a halo despite being highly elongated, and diffuse emission near the southern subcluster, which they class as a radio relic.  They find that the southern radio relic is peculiar due to its fragmented morphology and suggest that this may be due to interactions with cosmic filaments or that it may reflect the presence of electrons in fossil radio bubbles that are being re-accelerated by the passing shock. \citet{Giovannini:2013} largely confirm the \citet{vanWeeren:2013} findings and add that the average polarization of the radio relic is $\sim$20\%. 

\citet{vanWeeren:2017} identified a direct connection between a cluster member AGN and a portion of the radio relic and showed spectral aging across the radio jet portion connecting the relic to the galaxy. They then found an X-ray shock in deeper Chandra data and showed that the spectral index changed near the shock. The shock has a low Mach number ($\mathcal{M}<1.7$) and would be unable to generate enough acceleration to observe the radio relic were it not for the pre-accelerated plasma from the AGN. This was the first direct observation of an AGN/shock connection in this manner and the clearest evidence yet that radio relics may be intimately connected to the AGN population. Also presented in \citet{vanWeeren:2017} are four of five of our DEIMOS slitmasks and our g and i band SuprimeCam imaging. A bimodal galaxy distribution is identified with equal masses from velocity dispersions. The masses and redshifts of the subclusters are combined with the projected separation and input into the MCMAC dynamics code \citep{MCMAC,Dawson:2012} to estimate merger parameters. A3411 is found to have a 3D pericenter speed of $\sim$2600 km s$^{-1}$. A3411 is observed near apocenter, and within $\sim45\degr$ of the plane of the sky at 95$\%$ confidence.

\subsubsection{Results} 

We have since added an additional slitmask of spectroscopic data from Keck DEIMOS in the eastern edge of the cluster. With this mask, the total number of cluster members increases from 174 \citep[presented in][]{vanWeeren:2017} to 242. The global velocity distribution is well fit by a single Gaussian with redshift 0.16225$\pm$0.00030 and velocity dispersion 1239 km s$^{-1}$ in agreement with \citet{vanWeeren:2017}. The red sequence luminosity distribution is largely bimodal with a lower peak on the eastern edge of the north to south elongation and a significant peak to the north of the main two subclusters. 

We run the GMM with models ranging from one to three subclusters. The lowest BIC model matches the bimodal picture presented in \citet{vanWeeren:2017} with two nearly equal, large subclusters arranged north to south. We note that the merger between these two likely did not cause the shock that is associated with the AGN and analyzed by \citet{vanWeeren:2017}. Instead, that shock is likely associated with an optically poor group presently in the eastern flank of the cluster. The main merger likely created the shock that is far to the south and spawned the patchy relic morphology in the south. The X-ray surface brightness fits the picture of the main merger with a cool-core remnant in the north and a trailing wake feature back toward to the south. Lensing studies and updated dynamical modeling are required to further understand the cluster's mass distribution and merger history. Low frequency radio observations may help explain the patchy radio relic morphology and associated shock; although, since the ICM is diffuse in the south, a direct X-ray shock is unlikely to be detected without an extremely deep Chandra image.

%% file: CIZAJ2242.tex
\subsubsection{Literature Review}

CIZAJ2242 contains the famous Sausage relic. The cluster was first discovered by \citet{Kocevski:2007} in the second CIZA sample, as it is situated behind the plane of the galaxy as observed from Earth (although it is away from the bulge). This explains the high extinction $A_{v}=\text{1.382}$. This is likely the reason for the dearth of optical studies of the system prior to the Merging Cluster Collaboration's analysis \citep{Stroe:2014a, Jee:2015,Dawson:2015, Stroe:2015,Sobral:2015}. 

\citet{vanWeeren:2010} conducted the first comprehensive radio study of the system (including WSRT, GMRT, and VLA observations) finding evidence of shock acceleration and spectral aging associated with the outward-moving shock. This was later confirmed with the follow-up study of \citet{Stroe:2013}.  \citet{vanWeeren:2010} also observed that the northern relic is strongly polarized at the 50-60\% level, and used this to infer that the merger angle must be within $\sim30\degr$ of the plane of the sky. They also used the spectral index to infer a Mach number of $\sim$4.6. \citet{Stroe:2014b} observed with the system with the Arcminute Microkelvin Imager at 16 GHz and presented the first high frequency detection of diffuse radio emission associated with clusters.  They found diffuse emission at the northern radio relic but note that the detected flux is inconsistent with diffusive shock acceleration predictions. Recent simulations suggest that adjusting the pre-existing fossil population, magnetic fields, and/or turbulence can help explain this spectral steepening above 2 GHz \citep{Kang:2016,Donnert:2016,Fujita:2016}. \citet{Stroe:2014c} studied the spectral age of radio features in the cluster. They found that individual galaxies with radio lobes appeared to be traveling either north or south indicating the relative motion of the subclusters within the plane of the sky.

\citet{vanWeeren:2011b} conducted a suite of simulations studying potential analogs to the system and argue that the cluster is probably undergoing a merger in the plane of the sky (less than 10$\degr$ from edge-on) with a mass ratio of about 2:1, and an impact parameter of 400 kpc. The core passage of the clusters happened $\sim$1 Gyr ago. \citet{Kang:2012} conducted diffusive shock acceleration simulations of the Sausage and found that Mach numbers from 2.0--4.5 were supported depending on the amount of pre-existing cosmic ray electrons. However, they question the ability of the merger event to produce such an elongated shock. 

\citet{Kierdorf:2016} studied the relic at 4.85 and 8.35 GHz with the 100-m Effelsberg telescope. They observe a variable polarization fraction from $\sim$55\% at the eastern end of the northern relic to $\sim$25\% at the western end. These values are in agreement with those estimated by\citet{vanWeeren:2010} at higher resolution, which suggest the magnetic fields are ordered on scales larger than the Effelsberg beam (300 kpc). 

Three detailed X-ray analyses have been conducted, one with XMM-Newton \citep{Ogrean:2013a,Ogrean:2013b} and two with Suzaku \citep{Ohashi:2013,Akamatsu:2013}.  The results from XMM-Newton show an extreme north to south elongation of the X-ray gas largely consistent with the merger axis suggested by the radio relics. The XMM instrumental background levels prevent characterization of the surface brightness profile at the location of the northern radio relic. Suzaku observations near the southern radio relic provide evidence for a shock with Mach number $\sim$1.2-1.3. \citet{Ogrean:2013a,Ogrean:2013b} also note two interesting features of the gas. The first is a ``wall" of hot gas east of the cluster center, and while not associated with a radio relic it does extend into the region behind the northern relic; they suggest that it may be indicative a more complex merger scenario. The second feature is a ``smudge" of enhanced X-ray emission coincident with the eastern portion of the northern radio relic.  \citet{Akamatsu:2013} found evidence for temperature jump at the location of the northern radio relic corresponding to a Mach number of 3.15$\pm$0.52, while lower than that predicted by the radio \citep[4.6$\pm$1.3;][]{vanWeeren:2010} they are consistent within the 68\% confidence intervals. They did not see a jump in the surface brightness profile, but they claim that this is due to the large Suzaku PSF ($\sim$380 kpc) being much larger than the width of the relic ($\sim$50 kpc).  Note that they do not observe a significant temperature jump in the region east of the cluster.

The Merging Cluster Collaboration completed a detailed optical analysis of the system. \citet{Stroe:2014a} conducted an H$\alpha$ survey of the cluster and find an order of magnitude boost in the normalization of the galaxy luminosity function in the vicinity of the relics, even greater than that of other known mergers at the same redshift. \citet{Jee:2015} presented the first weak lensing analysis with the Subaru imaging described in \S\ref{subsec:sample_subaru_observations}. Despite the large and variable extinction, the data show a massive and bimodal system with a north to south orientation that largely agrees the red sequence galaxies. The cluster as a whole is found to be very massive ($M_{200}=11.0_{-3.2}^{+3.7}\times10^{14} M_{\sun}$ and $9.8_{-2.5}^{+3.8}\times10^{14} M_{\sun}$ for the northern and southern subclusters respectively). The galaxy luminosity and weak lensing mass peaks exhibit a $\sim$190 kpc offset at a $\sim$2$\sigma$ level. \citet{Dawson:2015} present a subcluster and redshift analysis using our Subaru SuprimeCam imaging and Keck DEIMOS spectroscopic data. The subcluster algorithm is an expectation-maximizing GMM analysis suitable for simpler bimodal systems. The north and south subclusters are found to have similar line of sight velocities ($\Delta v=69\pm190$ km s$^{-\text{1}}$). The velocity dispersions of the two subclusters indicate a 1:1 mass ratio in agreement with the lensing study. Finally, \citet{Sobral:2015} analyzed 83 H$\alpha$ emitters among the DEIMOS spectroscopy and find that star-formation is boosted in the hottest X-ray regions near the subcluster cores. Additionally, evidence is found for enhancement from passing shocks in metal-rich galaxies. 

\citet{Okabe:2015} present a weak lensing analysis using a color-color selection, which is in principle more effective than the \citet{Jee:2015} color-magnitude selection, and find the southern subcluster to be two times more massive than the north. They also find the total mass to be $\sim50\%$ of the mass as \citet{Jee:2014}, but the difference between the two masses are consistent within the uncertainty of the respective results. Note that the simulations presented in \citet{vanWeeren:2011b} found the reciprocal mass ratio (the more massive subcluster in the north) further demonstrating the uncertainty in the mass ratio.

Most recently, \citet{Molnar:2017} completed a suite of hydrodynamical simulations finding a 1.3:1 mass ratio with a total mass of $\sim1\times10^{15}$ M$_{\odot}$ best describes the observations. 

\subsubsection{Results}

Using the same spectroscopy and imaging as \citet{Dawson:2015}, we find the global redshift distribution to be well fit with a single Gaussian of redshift 0.18865$\pm$0.00032 and velocity dispersion 1184$\pm$64 km s$^{-1}$ in agreement with \citet{Dawson:2015}. The red sequence luminosity distribution is clearly bimodal. We fit one and two subcluster models with the GMM and find the two halo model. The two subclusters are similar in redshift and well aligned with the two radio relics and the extension of the ICM. CIZAJ2242 is a very clean, bimodal system with a simple geometry. Subclusters 1 and 2 have redshifts and velocities in good agreement with \citet{Dawson:2015}; although, the south subcluster has a slightly higher velocity dispersion instead of the north. These results show that our MCMC-GMM recovers the results of packaged EM-GMM codes on simple geometries.

%% file: MACSJ1149.tex
\subsubsection{Literature Review}

\citet{Ebeling:2007} were the first to report on MACSJ1149 as part of the MAssive Cluster Survey \citep[MACS;][]{Ebeling:2001}. \citet{Ebeling:2014} followed up with Gemini GMOS and Keck DEIMOS spectroscopy. Of the MACS clusters, MACSJ1149 has the highest reported velocity dispersion of $\sim\text{1800}\,\text{km s}^{-\text{1}}$ as well as an X-ray luminosity of $\sim\text{18}\times\text{10}^{\text{44}}\,\text{erg}\,\text{s}^{-\text{1}}$ and temperature of $\sim\,\text{9}\,\text{keV}$ \citep{Ebeling:2007}. \citet{Smith:2009} followed with a strong lensing analysis of the system based on HST/ACS imaging and Keck spectroscopic confirmation of multiply imaged galaxies finding a mass within 500 kpc of 6.7$\pm$0.4$\times$10$^{\text{14}}$M$_{\odot}$. They found the core to be composed of a dominant structure associated with the BCG along with three additional massive halos along a northwest-southeast axis. \citet{Zitrin:2009} published an independent strong lensing analysis and noted the largest known lensed image of a single spiral galaxy. This galaxy is a face-on spiral at a redshift of 1.49 and has been multiply imaged. One of these images contained a supernova, which in turn was multiple imaged by a cluster galaxy into an Einstein cross pattern. The supernova appeared in another of the images predictably and the time difference between the appearance of the supernova can be used to constrain cosmology and the lens model of MACSJ1149 \citep{Refsdal:1964, Kelly:2015, Sharon:2015, Oguri:2015, Diego:2016, Treu:2016}. \citet{Zitrin:2009} argue that a nearly uniform mass distribution over a 200 kpc radius with a surface density near the critical density is needed to explain the small amount of image distortion in this spiral. \citet{Zheng2012} used the cluster's powerful magnification to discover a $\text{z}\sim\text{10}$ galaxy. These results are largely thanks to MACSJ1149 being selected as part of the Cluster Lensing and Supernova Survey \citep[CLASH;][]{Postman:2012} and as part of the HST Frontier Fields program \citep{Lotz:2014}. CLASH studied MACSJ1149 with ground based weak lensing with Subaru/SuprimeCam \citep{Umetsu:2014}, and with a joint weak and strong lensing analysis utilizing Subaru SuprimeCam and HST ACS \citep{Umetsu:2015}, respectively. Both analyses resulted in extremely high masses for $M_{\text{200}}$: 25.4$\pm$5.2$\times$10$^{\text{14}}$ M$_{\odot}$ and 25.02$\pm$5.53$\times$10$^{\text{14}}$ M$_{\odot}$. MACSJ1149 is among the most massive clusters \citep{PlanckMass}. 

\citet{Bonafede:2012} presented the first radio analysis with evidence for three diffuse sources including a radio relic southeast and a candidate relic west of the cluster and a candidate radio halo as well. The radio halo has an extremely steep spectral index between 323 MHz and 1.4 GHz: $\alpha \approx \text{2}$ indicating the cluster merger is not recent and/or the (re)acceleration process was not efficient \citep{Brunetti:2008}. The relics are not situated symmetrically with the extended ICM and mass distribution; the southeast relic is colinear, but the western relic candidate is nearly perpendicular to the merger axis suggesting it may actually not be a shock-tracing relic. 

\citet{Ogrean:2016} presented a deep (365 ks) Chandra X-ray analysis. MACSJ1149 is found to be among the most X-ray luminous ($L_{X,\text{[0.1--2.4\,keV]}}=\text{(1.62}\pm\text{0.02)}\times\text{10}^{\text{45}}\,\text{erg}\,\text{s}^{-\text{1}}$) and hottest ($T_{X} = \text{10.73}^{+\text{0.62}}_{-\text{0.43}}\,\text{keV}$) clusters known. X-ray surface brightness appears to be relatively regular for a multi-component merger indicating that MACSJ1149 could be an old merger in agreement with the steep spectral index of the candidate radio halo. A weak cold front to the northeast of the cluster was found as well as a possible surface brightness edge that could be the leading bow shock may indicate that merger activity is still present or that sloshing may be occurring in the ICM. 

Using the archival Subaru SuprimeCam imaging and Keck DEIMOS and Gemini GMOS spectroscopy we completed a subcluster and redshift analysis \citep{Golovich:2016}. A second massive subcluster is discovered to the south that extraordinarily lacks an associated X-ray peak. We interpreted that the merger between the northern subcluster (which has been extensively studied as part of CLASH and the Frontier Fields) and the southern subcluster stripped the gas of the southern subcluster. This merger occurred mostly in the plane of the sky ($\Delta v_{LOS} \sim 300$ km s$^{-1}$). A dynamics analysis shows the merger to be similar to El Gordo \citep{Ng:2015} in terms of its age and merger velocity. A third subcluster is identified to be merging with the northern subcluster along the line of sight, which perhaps provides explanation for the cold front discovered by \citet{Ogrean:2016}. 

\subsubsection{Results}

Here we re-evaluate the substructure with our analysis scheme utilized in this paper. This is largely similar to the analysis presented in \citet{Golovich:2016} with only minor changes, and we complete our analysis routine here for completeness. This will also test that slight changes in the redshift analysis do not significantly alter results. Our shrinking aperture technique results in 260 spectroscopic cluster members within R$_{200}$ (as opposed to 278 in \citet{Golovich:2016}). The global redshift distribution is well fit by a single Gaussian with a redshift of 0.54362$\pm$0.00052 and velocity dispersion 1668$\pm$76 km s$^{-1}$, making MACSJ1149 the highest redshift system in our sample. 

We placed subcluster priors on the two bright subcluster peaks in the red sequence luminosity distribution as well as a broad, uninformative prior covering the northern extension of the red sequence luminosity distribution with a large peculiar line of sight velocity difference. This potential subcluster is evident in the DS-test with the group of red circles in the north of the cluster. We run models with one to three subclusters and identify the three subcluster model to have the lowest BIC score. The three halo model is in good agreement with the results of \citet{Golovich:2016}.

%% file: MACSJ1752.tex
\subsubsection{Literature Review}

MACSJ1752 was discovered as a bright X-ray source in ROSAT and confirmed spectroscopically by \citet{Ebeling:2001}. MACSJ1752 was first reported as a candidate radio relic cluster by \citet{Edge:2003}, although only in passing. \citet{vanWeeren:2012b} observed MACSJ1752 with the WSRT and found radio relics with sizes 1.3 and 0.9 Mpc, in the northeast and southwest respectively. They also measured integrated spectral indices of -1.16 and -1.1, suggesting shocks with Mach numbers of 3.5--4.5.  They argue that the radio relics and X-ray distribution suggest a 2:1 mass ratio binary event that occurred with an impact parameter about four times the core radius of the larger subcluster. They also note a possible radio halo associated with the system. \citet{Bonafede:2012} observed the system with GMRT and confirmed both the radio relics and radio halo.  They find spectral indices in good agreement with the \citet{vanWeeren:2012b} measurements.  \citet{Bonafede:2012} also resolve spectral steepening across each of the relics, similar to that seen in the CIZAJ2242.  They find that the NW relic has an average polarization of ~20\% with values up to 40\%, and the SW relic has an average polarization of ~10\% with values up to 40\%, with the higher polarization seen at the outer edge of the relics.  They suggest that this decrease towards the inner edge could be due to turbulent motions developing after the shock randomizing the magnetic field.  The northeast relic is among the brightest relics known, with similar brightness and morphology as the northern radio relic of PSZ1G108 \citep{deGasperin:2015}. 

\subsubsection{Results}

We observed MACSJ1752 with four Keck DEIMOS slitmasks and with Subaru SuprimeCam in g, r, and i bands. The spectroscopic survey resulted in 176 cluster member galaxies within R$_{200}$. The red sequence luminosity map is badly affected by a very bright star near the northeast BCG. Before a full galaxy and lensing analysis can be completed, this stellar profile must be subtracted in a manner similar to the method presented in \citet{Jee:2016}. This star also hindered the spectroscopic survey since SDSS imaging was used for targeting, and SDSS did not detect objects in the vicinity of the star. To detect the northeast BCG spectroscopically we placed a slit manually. We still detected many galaxies to the north and west of the BCG, so we proceed with a subcluster analysis here. We centered subcluster priors on the two BCGs with conservative projected position and redshift ranges. We ran one and two subcluster models, and the two subcluster model was preferred, although only slightly. We suspect more spectroscopic completion near the core of the northeast subcluster will improve the BIC of a two halo model. 

The two subclusters have similar redshifts and velocity dispersions suggesting a 1:1 merger in the plane of the sky. The X-ray surface brightness distribution shows two ``bullet-like'' gas cores that are nearly coincident with the red sequence luminosity peaks. The survival of the two cores suggests a merger with a non-zero impact parameter. However, the radio relics, galaxies, and ICM are all approximately colinear along a northeast to southwest axis.

%% file: PLCKG287.tex
\subsubsection{Literature Review}

PLCKG287 is the second most significant SZ detected cluster of the 20 new clusters in the Planck Early Release Catalog \citep{Planck_catalog:2011} (after only Abell 2163). \citet{Bagchi:2011} observed PLCK G287 with GMRT at 150 MHz and VLA at 1.4 GHz and discovered a pair of radio relics, that have a very large projected separation of 4.4 Mpc. They also identify a potential radio halo closer to the cluster center. \citet{Bagchi:2011} found evidence for PLCKG287 being a post merger system with optical imaging from the IUCAA 2m and a shallow 10 ks XMM-Newton observation. They estimate a SZ and X-ray based masses of $M_{500} = 25\pm$1.0$\times10^{14}$ and $26\pm1.0\times10^{14}$ M$_\odot$, respectively. These masses are consistent with the weak lensing based mass of \citet{Gruen:2014}: $M_{200} = 53 \pm 13\times10^{14}$ M$_\odot$ and $M_{500} = 28\pm5.0\times10^{14}$ M$_\odot$.  \citet{Gruen:2014} also noted a number of exceptional strong lens arc-like features. The two most separated features are 165$\arcsec$ apart. This is much larger than the expected Einstein radius of 37$\arcsec$ given the 68\% upper confidence NFW fit and assuming a source redshift of 5. Thus it is called into question whether these arcs are actually multiple images. \citet{Zitrin:2017} completed a strong lensing analysis with HST imaging and 20 sets of multiple images. A high-redshift dropout galaxy is identified with a photometric redshift of $\sim$7, and the critical area for $z\sim10$ source galaxies is found to be 2.58 arcmin$^{2}$ making PLCKG287 the largest known cosmic lens. Most recently, \citet{Finner:2017} analyzed our Subaru/SuprimeCam imaging in a detailed weak gravitational lensing analysis. They find a complex cluster core with multiple components. They find weak evidence for a subcluster in the south that could have merged to launch the radio relics, but the merger scenario is unclear with the lensing study alone.

\citet{Bonafede:2014} observed PLCK G287 with GMRT at 150, 325, and 610 MHz. They determined that the northern most relic identified by \citet{Bagchi:2011} is actually a radio galaxy with two lobes and some other point sources, however the northern relic closest to the cluster and the southern relic (2.8 Mpc from center) are confirmed. \citet{Bonafede:2014} make the argument that the small projected distance of the northern relic and the large projected distance of the southern relic is due to two separate passes of a merging subcluster ($\sim$10$\%$ of the mass of the main cluster). Most recently, \citet{George:2017} observed the cluster with the Murchison Widefield Array at 88, 118, 154, 188, and 215 MHz. The two radio relics and halo are confirmed at each frequency. The northwest and southeast relics have spectral indices between 88 MHz and 3 GHz measured to be $-1.19\pm0.03$ and $-1.36\pm0.04$, respectively. The radio halo was blended with several point sources, and no spectral index was estimated.

\subsubsection{Results}

We observed PLCKG287 with three Keck DEIMOS slitmasks and with Subaru SuprimeCam in g and r bands. We also obtained spectroscopy from VLT GMOS (M. Girardi, M. Nonino, private communication). These spectra will be presented in a more detailed spectroscopic and photometric analysis in a forthcoming paper (Girardi et al. in prep.). The spectroscopic survey results in 302 spectroscopic cluster members well fit by a single Gaussian with redshift 0.38321$\pm$0.00046 and velocity dispersion 1756$\pm$74 km s$^{-1}$. The red sequence luminosity distribution is dominated by a single peak with a much weaker peak sitting $\sim$2 Mpc to the south. The DS-test corroborates this peak with a collection of blue circles in the vicinity. We placed subcluster priors over the two luminosity peaks and ran one and two halo models with the GMM. The two models have similar BIC scores with the two halo model only slightly preferred. 

The main subcluster of PLCKG287 has an extreme velocity dispersion, and the second subcluster is situated near the south radio relic, which would be unexpected in a bimodal merger scenario. The relative positions of the two radio relics has been discussed with a proposal for a two core passage model by \citet{Bonafede:2014}.

The weak lensing analysis of \citet{Gruen:2014} did not detect a second massive peak in the south. The X-ray surface brightness map does not reveal obvious X-ray emission indicating a separate ICM peak associated with the subcluster in the south. This could indicate that it has been stripped, or that the subcluster is very low mass. The bimodal merger picture is complicated further by the relative position of the relic and galaxies. The shock should run ahead occur as the subcluster approaches apocenter, but the subcluster is unexpectedly far from the center of the cluster.

In the scenario proposed by \citet{Bonafede:2014}, where the south relic was launched at first core passage, and the north relic is launched during a more recent second core passage by the original infalling subcluster (note this model does not include the southern subcluster detected by our GMM), one complicating factor is the bright southern relic. The radio emission of a radio relic is believed to be short-lived \citep{Skillman:2013}. A possible explanation is a scenario similar to Abell 3411. The cluster galaxies in the south could host an AGN that has created a region of mildly relativistic fossil electrons that are now being reaccelerated by the passing shock. Deeper X-ray and radio observations in this region would be necessary to search for such a scenario.

%% file: PSZ1G108.tex
\subsubsection{Literature Review}

PSZ1G108 is a little studied system that first was discovered recently by \citet{Planck_SZ} using its SZ signal. The cluster mass was later estimated by the Planck Collaboration to be $M_{500}=7.7\pm0.6\times10^{14}\,\text{M}_\odot$ \citep{PlanckMass}. \citet{deGasperin:2015} completed the only detailed analysis of the cluster and found it hosts one of the most powerful pair of radio relics known to date in addition to a radio halo with WSRT and GMRT in several frequencies. The injection spectral index of the northern and southern relic was found to be -1.02 and -1.17, respectively. The relics notably comprise the second most powerful radio system known at 1.4 GHz ($L=44.6\pm0.3\times10^{24}$ W Hz$^{-1}$). Each relic shows 10--30\% polarized emission. ROSAT show the X-ray emission to be centered between the two relics, but the data are insufficient to detect merger activity. The estimated luminosity is $L_{X}\sim7.5\times10^{44}$ erg s$^{-1}$ based on the ROSAT emission, and using the $(L,T)$ scaling relations provided by \citet{Pratt:2009} a temperature of 6.5 keV was reported. 

\subsubsection{Results}

We observed PSZ1G108 with two slitmasks on Keck DEIMOS and in g and r bands on Subaru SuprimeCam. Unfortunately, tracking problems made the SuprimeCam images unusable, so we made use of DSS images to trace the red sequence. The 40 spectroscopic galaxies are well fit by a single Gaussian in redshift space with a global redshift of 0.33494$\pm$0.00059 and velocity dispersion 873$\pm$90 km s$^{-1}$. The red sequence distribution is elongated along the same axis as the two radio relics. The data are insufficient to complete a subcluster analysis, so we present only a single halo GMM model. 

The X-ray surface brightness map indicates a main ICM peak near the middle of the cluster equidistant from the two relics with X-ray emission streaming northward toward the bright northern relic. Deeper X-ray observations along with photometry and spectroscopy are needed to study this system in detail.

%% file: RXCJ1053.tex
\subsubsection{Literature Review}

RXCJ1053 is a little studied cluster although it appears in both SDSS and REXCESS cluster catalogs. \citet{Aguerri:2007} reported a velocity dispersion of $\sim$650 km s$^{-1}$. \citet{Rudnick:2009} made the first mention of a radio relic with an extent of 1 Mpc and a low flux.  \citet{vanWeeren:2011} presents the WSRT 1382 MHz observations and finds a weak radio relic in the west. A general bimodal picture is found with the X-ray emission between the two galaxy peaks and the radio relic along the same axis. Between the low velocity dispersion, weak radio emission ($0.2\times10^{24}$ W Hz$^{-1}$), and low X-ray luminosity ($L_{X}=0.44\times10^{44}$ erg s$^{-1}$) the system appears to be a low mass merger. 

\subsubsection{Results}

We observed RXCJ1053 with four slitmasks of Keck DEIMOS spectroscopy and in g and r bands with Subaru SuprimeCam. The spectroscopic survey resulted in 119 cluster members within R$_{200}$ very well fit by a single Gaussian with redshift 0.07217$\pm$0.00021 and velocity dispersion 653$\pm$41 km s$^{-1}$ in agreement with \citet{Aguerri:2007}. RXCJ1053 is the lowest redshift cluster in our sample as well as the cluster with the lowest velocity dispersion. The red sequence luminosity distribution is aligned along a southeast to northwest axis with the relic at the northwest end. There are three red sequence luminosity peaks with the brightest peak in the middle but the BCG in the northwest. The southeast peak is just above our 3$\sigma$ contour. 

We place subcluster priors on the three peaks with conservative redshift priors and run one, two, and three halo models with the GMM. The lowest BIC model has two subclusters over the middle and northwestern peaks. Subcluster 1 (northwest) contains the cluster BCG, but it has a very low velocity dispersion. Meanwhile, subcluster 2 is slightly less settled with no dominant BCG, but it contains most of the mass of the cluster. The two subclusters are well aligned with the radio relic to the northwest, and the X-ray surface brightness distribution indicates a merger along the same axis. The brightest X-ray emission is associated with subcluster 1. Lower mass subclusters have been observed holding onto their gas (e.g. the Bullet Cluster, ZwCl0008), and often indicate a cool-core remnant. Deep X-ray observations would be necessary to fully classify RXCJ1053 as a low mass ``Bullet-like'' cluster. 

%% file: RXCJ1314.tex
\subsubsection{Literature Review}

RXCJ1314 is a double relic cluster. \citet{Valtchanov:2002} obtained 37 spectroscopic redshifts of cluster members and found that there is a significant bimodal distribution in redshift space ($\sim$1700 km/s separation) and is supported by a bimodal distribution of galaxies in projected space.  The double relics have been observed on VLA and GMRT \citep{Feretti:2005,Venturi:2007,Venturi:2013}, although little work has been done studying the spectral steepening or polarization.  \citet{Mazzotta:2011} briefly reported on their analysis of XMM-Newton data (106 ks), and found a shock front consistent with the leading edge of the western radio relic.  They note that there is excellent agreement between the Mach number estimates of X-ray and radio.  Interestingly this shock front is M-shaped which they suggest may indicate infalling material. Most recently, \citet{George:2017} observed the cluster with the Murchison Widefield Array at low frequencies between 88 MHz and 215 MHz. The east relic is clearly detected in each band except the lowest frequency (88 MHz), but the west relic and halo are blended for all frequencies. The east and west relics are found to have spectral indices between 118 MHz and 1.4 GHz of -1.03$\pm$0.12 and -1.22$\pm$0.09, respectively. 

\subsubsection{Results}

We observed RXCJ1314 with two Keck DEIMOS slitmasks and with Subaru SuprimeCam in g and r broadband filters as well as NB814 for H$\alpha$ study. Here we only analyze the broadband imaging. The spectroscopic survey resulted in 156 cluster members, which are well fit by a single Gaussian with redshift 0.24703$\pm$0.00035 and velocity dispersion 1094$\pm$64 km s$^{-1}$. The red sequence luminosity distribution has three peaks. Two are aligned east to west between the double radio relics, and the third is to the south. Unfortunately we do not have spectroscopic coverage for the south subcluster, but inspection of the X-ray surface brightness distribution suggests that the subcluster has yet to merge since the ICM is well aligned with the galactic light. Interestingly, there is a stream of gas suggesting the subcluster may have had a glancing interaction or that it is embedded within a gas filament that is being accreted by the main cluster to the north. 

We place subcluster priors on the east and west subclusters and allow for a conservative velocity prior. The GMM is run for one and two halo models. The two halo model is preferred according to the BIC score. The two subclusters have $\sim$1500 km s$^{-1}$ line of sight velocity differences, which is the largest among the double relic systems in our sample. The velocity dispersions of the two subclusters are similar suggesting a 1:1 mass ratio merger. 

The large line of sight velocity difference suggests a merger axis with a substantial component along the line of sight. The radio contours for this cluster (see Figure \ref{fig:A24}) are based on our VLA polarization observations. For details on the preparation of these contours see \citet{Golovich:2017, Benson:2017}, where we present analogous observations for ZwCl0008 and ZwCl2341. These radio observations will provide an upper limit on the viewing angle of the merger axis based upon the semi-analytic work of \citet{ensslin1998}, which may help contain the dynamics of this cluster. The dynamics of RXCJ1314 will be studied in a forthcoming paper. 

%% file: ZwCl0008.tex
\subsubsection{Literature Review}

\citet{vanWeeren:2011c} first identified ZwCl 0008 as a double radio relic system while searching the 1.4 GHz NVSS, 325 MHz WENSS, and 74 MHz VLSS surveys. They carried out a radio survey of ZwCl 0008 with GMRT observations at 241 MHz and 640 MHz and WSRT observations at 1.3--1.7 GHz in full polarization mode. Two radio relics were identified, with the eastern relic larger than the western relic. Spectral index maps show a steepening trend toward the cluster center for both relics indicating motion away from the center. The spectral indices at the front of the relics were reported to be $-\text{1.2}\pm\text{0.20}\,\text{and}\,-\text{1.0}\pm\text{0.15}$ for the east and west relics respectively. Taking these as the injection spectral indices, Mach numbers ($\mathcal{M}$) of $\text{2.2}_{-\text{0.1}}^{+\text{0.2}}$ and $\text{2.4}_{-\text{0.2}}^{+\text{0.4}}$ were reported for the east and west relics. In addition, the polarization was measured at $5-25\%$ for the east relic and $5-10\%$ for the west relic. 

ZwCl 0008 was studied in a follow up simulation analysis by \citet{Kang:2012}, whose diffusive shock acceleration simulations showed that $\mathcal{M}=\text{2}$ explains the relics in ZwCl 0008 regardless of the level of pre-existing relativistic electrons. They also find a projection angle between 25 and 30$^{\circ}$ to model the spectral index and radio flux.

\citet{Kierdorf:2016} studied the eastern relic at 4.85 and 8.35 GHz with the Efelsberg 100-m telescope. The relic is detected at both frequencies, but only the southern portion in the high frequency band. Polarized emission is detected at 4.85 GHz; whereas, \citet{vanWeeren:2011c} detected only patchy polarized emission at 1.4 GHz. A maximum polarization fraction of 26$\pm$7\% and 22$\pm$4\% is estimated for the 8.35 and 4.85 GHz images, respectively, which is in areement with the value estimated by \citet{vanWeeren:2011c}. 

Most recently, \citet{Golovich:2017} present a wide range of observations including HST ACS+WFC, Subaru SuprimeCam, Keck DEIMOS, Chandra, and JVLA imaging. The HST and Subaru imaging are used to complete a weak lensing analysis, which finds a $\sim$4:1 mass ratio between two mass peaks. The lower mass peak is associated with a cool-core remnant analogous to the ``bullet'' in the Bullet Cluster. The JVLA observations were used to estimate the polarization of the radio relics, which were measured to be as high as 40$\%$. The spectroscopy and photometry indicate a bimodal merger as well. Interestingly, there are large offsets between the BCGs and lensing peaks; although, the uncertainties in the lensing centroids are large due to the low mass and redshift of this system. Finally, a dynamics analysis shows the cluster to be near apocenter having passed pericenter with a relative velocity of $\sim$2000 km s$^{-1}$ $\sim$1 Gyr before the observed state. 

\subsubsection{Results}

Here we re-evaluate the galaxy substructure with our MCMC-GMM. In \citet{Golovich:2017}, we utilized a simpler GMM without MCMC functionality. Our shrinking aperture method results in 116 spectroscopic cluster members, and the red sequence luminosity distribution based on our Subaru SuprimeCam g and r observations is bimodal along an east to west axis. The cluster redshift distribution is well fit by a single Gaussian with redshift 0.10383$\pm$0.00027 and velocity dispersion 791$\pm$56 km s$^{-1}$ in agreement with \citet{Golovich:2017}. 

We place subcluster priors on the two BCGs with conservative redshift ranges. The two halo model is preferred over the single halo model although only slightly. Based on the preponderance of evidence for a bimodal merger, we only study the two halo model. The main difference between this subcluster model and the results of \citet{Golovich:2017} is the mass ratio based on the velocities. Subcluster 1 (east) is more massive according to the lensing analysis; however, the velocity dispersion of subcluster 2 was larger in \citet{Golovich:2017}. Here, while still similar, the velocity dispersions suggest subcluster 1 is more massive, in agreement with the lensing results; although, with more similar masses than the 4:1 mass ratio suggested by the lensing analysis. Finally, the results here are in agreement with a low line of sight velocity difference between the two subclusters (90$\pm$130 km s$^{-1}$).

%% file: ZwCl1447.tex
\subsubsection{Literature Review}

ZwCL1447 is a single relic and halo system that is part of the \citet{Dressler:1999} sample of 10 Distant Rich Clusters of Galaxies. \citet{Giovannini:2009} found a radio relic candidate in the north, however when \citet{Govoni:2012} investigated the system with more radio observations they found that this northern relic candidate was actually a radio halo; additionally, they detected a southern radio relic.  Both the radio halo and southern radio relic have considerable point source contamination.  \citet{Giovannini:2009} find that both the ROSAT X-ray measured gas distribution and the SDSS measured cluster galaxy distribution are elongated along the axis connecting the radio halo and relic, typical with post pericenter systems.

\subsubsection{Results}

We observed ZwCl1447 with two slitmasks of Keck DEIMOS and in g, r, and i bands with Subaru SuprimeCam. The spectroscopic survey results in 116 cluster members well fit with a single Gaussian of redshift 0.37574$\pm$0.00057 and velocity dispersion 1382$\pm$90 km s$^{-1}$. The red sequence luminosity distribution has three peaks; however, a bright star in the south has affected the contours. This star will need to be modeled out in more detailed analyses, but this is beyond the scope of this study. 

Two of the red sequence luminosity peaks are aligned with the radio relic, and the third is situated to the south, just east of the relic. The radio halo is slightly offset from the main subcluster to the north suggesting the ICM could be displaced away from the center of the cluster. ZwCl 1447 is the only cluster in our sample without Chandra or XMM-Newton data. We present the X-ray map from RASS in Figure \ref{fig:A26}, which shows structure along the supposed merger axis, but is too low resolution to comment further. 

We placed subcluster priors on the two luminosity peaks aligned with the radio relic. We left out the third southernmost peak due to our lack of spectroscopic coverage in this region. We run one and two halo models with the GMM, and the one halo model is preferred with a lower BIC. This is likely due to our sparse spectroscopic sampling in the south due to the bright star. We used SDSS imaging for target selection, and the photometric catalog is incomplete in this region. There are many galaxies within the stellar aureole that are unrecovered from SDSS. 

We note that the cluster velocity dispersion is high suggesting a massive system; however, stellar subtraction, X-ray observations, and more spectroscopic observations in the south are necessary to fully analyze this system. 

%% file: ZwCl1856.tex
\subsubsection{Literature Review}

ZwCl1856 (also known as PSZ1 G096.89+24.17) is a double relic system identified as such through a search of Planck SZ selected clusters \citep{Planck_SZ} in NVSS, WENSS, VLSS surveys carried out by \citet{deGasperin:2014}. The Planck Collaboration estimates the mass to be $M_{500}=4.4^{+0.45}_{-0.48}\times10^{14}$ M$_\odot$ \citep{PlanckMass}. The RASS emission corresponds to a X-ray luminosity of $L_{X}=3.7\times10^{44}$ erg s$^{-1}$. Using $(L,T)$ scaling relations from \citet{Pratt:2009}, this corresponds to $T_{X}\sim4$ keV \citep{deGasperin:2014}.

\citet{deGasperin:2014} carried our WSRT 21 cm observations sufficient to study the polarization of the relics. The two relics are symmetrically located along a north--south axis separated by $\sim$2 Mpc. The polarization is found to be $\sim$10$\%$ for each relic. Finally, a radio halo is marginally detected. \citet{deGasperin:2014} suggests lower frequency observations could confirm or refute the presence of a halo. 

\subsubsection{Results}

We observed ZwCl1856 with two slitmasks on Keck DEIMOS and with Subaru SuprimeCam in g and r bands. The spectroscopic survey utilized Digitized Sky Survey (DSS) imaging for targeting, and as a result of the poor seeing the majority of detected objects were stars. The survey resulted in 48 cluster member redshifts that are well fit by a single Gaussian of redshift 0.30362$\pm$0.00061 and velocity dispersion 990$\pm$106 km s$^{-1}$. As with PSZ1G108, the SuprimeCam imaging is mostly unusable due to tracking issues.  Two of the six exposures in g band and all exposures in r band were lost to this issue. In order to define the color of detected objects, we observed PSZ1G108 with Gemini-N GMOS in g and r bands. We utilized a 2$\times$1 mosaic covering the north and south portions of the cluster field including the two radio relics. Each pointing is composed of 5$\times$110 s exposures. The data were reduced with the standard Gemini IRAF package, and the images were coadded with the same methodology as our SuprimeCam imaging. 

The red sequence luminosity distribution is bimodal along a north to south axis symmetrically placed between the two radio relics. We placed subcluster priors on the two luminosity peaks and allowed for a conservative redshift range for our GMM. The two models are nearly equal according to the BIC score with the two halo model slightly preferred. We study only the two halo model due to the clear evidence of a major merger from the double radio relics and the X-ray surface brightness distribution, which is elongated along the same axis (see Figure \ref{fig:A27}). The two resulting subclusters have similar redshifts (line of sight velocity difference of 189$\pm$266 km s$^{-1}$) and velocity dispersions indicating a 1:1 mass ratio merger in the plane of the sky. In summary, ZwCl1856 appears to be among the cleanest bimodal systems in our sample; however, further observations will be necessary to study this system.

%% file: ZwCl2341.tex
\subsubsection{Literature Review}

\citet{Bagchi:2002} were the first to discover the diffuse radio emission associated with the radio relics in the 1.4 GHz NVSS survey, and concluded that the emission was the first evidence of cosmic-ray particle acceleration taking place at cosmic shocks. They also noted that ZwCl2341 is likely in the process of ongoing structure formation, so in many ways they were the first to directly observe and realize the association between radio relics and merging clusters.  \citet{vanWeeren:2009} followed-up the system with GMRT 157, 241, and 610 MHz observations, confirming this interpretation. They were able to measure radio spectral indices of -0.49$\pm$0.18 and -0.76$\pm$0.17 for the northern and southern relics, respectively.  In their analysis they also presented Chandra X-ray observations and found that the ICM of the system is highly disturbed and elongated along the axis of the two relics. \citet{Giovannini:2010} obtained follow-up 1.4 GHz VLA imaging of the system and confirmed the existence of the relics.  Interestingly they also found evidence for polarized (11\%) diffuse emission along an optical filament of galaxies between the two relics, which they posit may be a giant and highly elongated radio halo or the merging of two clusters both hosting a central radio halo.  \citet{Akamatsu:2013} observed the system with Suzaku in hopes of detecting temperature or density jumps associated with the radio relic locations. While they observe a decrease in both, the angular resolution of Suzaku was not sufficient to check for the presence of a sharp discontinuity. \citet{Ogrean:2014} did not detect a shock in the location of the radio relic in XMM-Newton and Chandra observations. 

\citet{Boschin:2013} conducted a spectroscopic survey of the system, obtaining 142 redshifts, 101 of which were cluster members.  Analysis of the velocity distribution of cluster members yielded two distinct groups, and the possibility of a third.  Spatial distribution of the galaxies shows an elongated shape to the cluster in the SSE-NNW direction with four over densities (three very significant), matching the elongated X-ray gas distribution. Their color-magnitude selection of cluster members showed a more complex structure, yielding eight galaxy density peaks. 

\citet{Benson:2017} present our Subaru SuprimeCam and Keck DEIMOS observations (see also Paper 1). \citet{Benson:2017} find a three subcluster model best describes the data using a similar MCMC-GMM subclustering analysis to the one used in this paper. \citet{Benson:2017} interpret the three subclusters to be involved in two mergers with the plane of the sky merger forming the two radio relics with a secondary line of sight merger within the north between two low mass subclusters. Using our Subaru imaging, \citet{Benson:2017} complete the first weak lensing analysis finding a total mass of  $4.49\pm1.72\times10^{14}$ M$_\odot$, which is substantially lower than previous estimates based on the velocity dispersion of subclusters \citep{Boschin:2013}. 

\subsubsection{Results}

Here we use our analysis scheme to study the same data presented in \citet{Benson:2017}. Namely, ZwCl2341 was observed with three slitmasks on Keck DEIMOS resulting in 224 cluster members within R$_{200}$. The redshift distribution is well fit with a single Gaussian with redshift 0.26960$\pm$0.00029 and velocity dispersion 1036$\pm$49 km s$^{-1}$. The red sequence was mapped with g and r band images from Subaru SuprimeCam. The distribution has three distinct peaks aligned between the radio relics from northwest to southeast. There is an additional extension of red sequence luminosity toward the east of the northern peak.

\citet{Benson:2017} elected to fit no more than three halos with the GMM. The preferred model was a three halo model associated with the northern and southern luminosity peaks with two in the north splitting in velocity space. Here we place the same three halos but with an additional halo associated with the middle luminosity peak, which was less distinct in the \citet{Benson:2017} luminosity map. 

We run 12 models ranging from one to four halos. The preferred model is analogous to \citet{Benson:2017} with a slightly lower velocity dispersion for the southernmost subcluster (subcluster 3 in Figure \ref{fig:A28}). Subclusters 2 and 3 are well aligned with the radio relics and at similar redshifts suggesting a merger in the plane of the sky. The southern subcluster appears much more massive based on the velocity dispersion. Subcluster 1 has a higher redshift and may be involved in a secondary merger with subcluster 2. 

Inspection of the X-ray surface brightness map shows general elongation between the radio relics but also a clumpy morphology in the northeast indicating dynamical activity. Further X-ray observations and deeper imaging for more precise lensing will be necessary to fully understand this complex cluster.